\documentclass[runningheads]{llncs}

\usepackage{graphicx} 
\usepackage{xcolor} 

\usepackage[utf8]{inputenc}
\usepackage[T1]{fontenc}
\IfFileExists{luximono.sty}{
  \usepackage[scaled=0.9]{luximono}
}{
  \usepackage[scaled=0.81]{beramono}
}
\usepackage{bold-extra} 
\usepackage{microtype}

\usepackage{amsmath} 
\usepackage{amssymb} 
\usepackage{stmaryrd} 
\usepackage{multirow}

\usepackage[inline]{enumitem} 
  \setlist*[enumerate]{label=(\arabic*)}

\usepackage{wrapfig}
\usepackage{placeins} 

\usepackage{xspace} 
\usepackage{suffix} 
\usepackage{xparse}
\usepackage{letltxmacro} 
\usepackage{etoolbox}
\usepackage{xifthen}

\newcommand*{\ListingsFontSize}{\fontsize{8.0}{8.3}\selectfont}
\newcommand*{\SilverFont}{\ttfamily\ListingsFontSize}
\usepackage{silver}

\usepackage[nounderscore]{syntax} 

  \renewcommand{\litleft}{\begin{math}}
  \renewcommand{\litright}{\end{math}}

\usepackage{adjustbox}  
\usepackage{caption}    

\usepackage{hyperref} 

\usepackage{prooftree} 
\DeclareMathAlphabet{\cmcal}{OMS}{cmsy}{m}{n}

\newcommand*{\VoilaFont}{\SilverFont}
\newcommand*{\VoilaFontSize}{\ListingsFontSize}
\newcommand*{\InlineVoilaFontSize}{\small}

\lstdefinelanguage{voila}{
  language=,
  sensitive=true,
  morecomment=[l]{//},
  morecomment=[s]{/*}{*/},
  morekeywords=[1]{ 
    struct,
    void, int, bool, null, true, false,
    procedure, returns, do, while, CAS,
    parallel
  },
  morekeywords=[2]{ 
    id, Set, set, frac, Int, Nat,
    region, guards, interpretation, state, actions,
    unique, duplicable, manual,
    abstract_atomic,
    interference, in, on, requires, ensures, invariant,
    make_atomic, update_region, open_region, use_atomic, using, with
  },
  morekeywords=[3]{ 
    lemma, use,
    fold, unfold,
    assume, assert, inhale, exhale
  },
  basicstyle={\VoilaFont\VoilaFontSize},
  commentstyle={\color[HTML]{747678}\textit},
  keywordstyle={[1]\color[HTML]{0005FF}},
  keywordstyle={[2]\color[HTML]{671800}},
  keywordstyle={[3]\color[HTML]{EC008C}},
  mathescape=true,
  escapechar=§,
  moredelim=**[is][\normalfont\itshape]{'}{'} 
}

\lstdefinelanguage{plaintext}{
  language=,
  keywordstyle=\bfseries,
  ndkeywordstyle=\bfseries,
  morecomment=[l]{//},
  mathescape=true
}

\lstnewenvironment{voila}[1][]{
  \lstset{
    language=voila,
    floatplacement={tbp},
    captionpos=b,
    frame=lines,
    xleftmargin=8pt,
    xrightmargin=8pt,
    #1
  }
}{}

\newcommand*{\inlvoila}{%
  \lstinline[%
    language=voila,%
    columns=fixed,%
    basicstyle={\VoilaFont\InlineVoilaFontSize},%
    commentstyle={\color{black}\InlineVoilaFontSize},
    keywordstyle={[1]\color{black}\InlineVoilaFontSize},
    keywordstyle={[2]\color{black}\InlineVoilaFontSize},
    keywordstyle={[3]\color{black}\InlineVoilaFontSize},
    keywordstyle={[4]\color{black}\InlineVoilaFontSize}
  ]%
}
\newcommand*{\vl}{\inlvoila}

\LetLtxMacro\oldttfamily\ttfamily
\DeclareRobustCommand{\ttfamily}{\oldttfamily\InlineVoilaFontSize}

\newcommand*{\MaybeFloatBarrier}{\FloatBarrier}

\newcommand*{\Figref}[1]{Fig.~\ref{fig:#1}}
\newcommand*{\figref}{\Figref}
\newcommand*{\Secref}[1]{Sec.~\ref{sec:#1}}
\newcommand*{\secref}{\Secref}
\newcommand*{\Appref}[1]{App.~\ref{sec:#1}}
\newcommand*{\appref}{\Appref}
\newcommand*{\Lineref}[1]{Line~\ref{line:#1}}
\newcommand*{\lineref}[1]{line~\ref{line:#1}}
\WithSuffix{\newcommand*}\Figref*[1]{Fig.~#1}
\WithSuffix{\newcommand*}\figref*{\Figref*}
\WithSuffix{\newcommand*}\Secref*[1]{Sec.~#1}
\WithSuffix{\newcommand*}\secref*{\Secref*}
\WithSuffix{\newcommand*}\Lineref*[1]{Line~#1}
\WithSuffix{\newcommand*}\lineref*[1]{line~#1}
\WithSuffix{\newcommand*}\Lines*[1]{Lines~{#1}}
\WithSuffix{\newcommand*}\lines*[1]{lines~{#1}}
\WithSuffix{\newcommand*}\Linerange*[1]{Lines~{#1}}
\WithSuffix{\newcommand*}\linerange*[1]{lines~{#1}}

\newcommand*{\ie}{i.e.\xspace}
\newcommand*{\Eg}{E.g.\xspace}
\newcommand*{\eg}{e.g.\xspace}

\newcommand*{\cf}{cf.\xspace}

\newcommand*{\wrt}{w.r.t.\xspace}

\newcommand*{\fgc}{fine-grained concurrency\xspace}

\newcommand*{\poc}{proof outline checker\xspace}

\newcommand*{\pol}{proof outline language\xspace}

\newcommand*{\termacronym}{}%
\NewDocumentCommand{\term}{O{} m}{%
  \textit{#2}%
  \renewcommand*{\termacronym}{}%
  \ifdefmacro{#1}{%
    \ifdefvoid{#1}{}{\renewcommand*{\termacronym}{#1}}%
  }{%
    \ifthenelse{\equal{#1}{}}{}{\renewcommand*{\termacronym}{#1}}%
  }%
  \ifdefvoid{\termacronym}{}{ (\termacronym)}%
  \xspace%
}

\newcommand*{\coqishtools}{proof checkers\xspace}

\newcommand*{\caperishtools}{automated verifiers\xspace}
\newcommand*{\caperishtool}{automated verifier\xspace}

\newcommand*{\voilaishtools}{proof outline checkers\xspace}

\newcommand*{\MakeAtomic}{\textsc{MakeAtomic}\xspace}
\newcommand*{\UpdateRegion}{\textsc{UpdateRegion}\xspace}
\newcommand*{\OpenRegion}{\textsc{OpenRegion}\xspace}
\newcommand*{\UseAtomic}{\textsc{UseAtomic}\xspace}
\newcommand*{\Substitution}{\textsc{Substitution}\xspace}

\newcommand*{\Consequence}{\textsc{Consequence}\xspace}
\newcommand*{\AExists}{\textsc{AExists}\xspace}
\newcommand*{\Exists}{\textsc{Exists}\xspace}
\newcommand*{\TripleWeak}{\textsc{AWeakening1}\xspace}
\newcommand*{\InterferenceWeak}{\textsc{AWeakening2}\xspace}
\newcommand*{\LvlWeak}{\textsc{AWeakening3}\xspace}
\newcommand*{\ProcCall}{\textsc{FunctionCall}\xspace}

\newcommand{\ctext} [1] {\texttt{#1}}
\newcommand{\many} [1] {\overline{#1}} 

\newcommand{\alt}{~~|~~}

\newcommand{\level} [0] {\ctext{level}}
\newcommand{\alevel} [0] {\ctext{alevel}}

\newcommand{\actxtDomain} [0] {\ctext{update}}

\newcommand{\foreachSymb}{\textit{foreach}}
\newcommand{\foreach}[2]{\textit{\foreachSymb}~{#1}~\textit{in}~{#2}}
\newcommand{\foreachStart}{\textit{do}}
\newcommand{\foreachEnd}{\textit{end}}

\newcommand{\foreverySymb}{\textit{forall}}
\newcommand{\forevery}[2]{\textit{\foreverySymb}~{#1}~\textit{in}~{#2}}

\newcommand{\foroneSymb}{\textit{exists}}
\newcommand{\forone}[2]{\textit{\foroneSymb}~{#1}~\textit{in}~{#2}}

\newcommand{\actionset}[1]{\metafunc{Actions}({#1})}
\WithSuffix{\newcommand}\actionset*[1]{{\normalfont\actionset{#1}}}

\newcommand{\persistentVars} [1] {\many{P}}
\newcommand{\nonPerisitentVars} [1] {\many{P}}

\newcommand*{\qdot}{\mathbin{\cdot}}

\newcommand*{\slstar}{\mathbin{\ast}} 
\newcommand*{\pointsto}{\mapsto} 

\NewDocumentCommand{\region}{s m e{_} e{^} >{\SplitList{,}}d()}{%
  \texttt{\InlineVoilaFontSize#2}%
  \ensuremath{%
    \IfNoValueTF{#3}{%
      \IfBooleanF{#1}{%
        _r%
      }%
    }{%
      _#3%
    }%
    \IfNoValueF{#4}{^#4}%
  }%
  \def\listsep{\def\listsep{,\xspace}}%
  \IfNoValueTF{#5}{%
  }{%
    (\ProcessList{#5}{\regionputarg})%
  }%
  \xspace%
}
\newcommand*{\regionputarg}[1]{%
  \listsep\texttt{#1}%
}

\newcommand*{\tadaforall}{\raisebox{-0.035ex}{\includegraphics{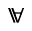}}\xspace}

\NewDocumentCommand{\atriple}{s o r<> m o r<>}{%
  \def\atripleargtwo{#2}%
  \def\atripleargfive{\exists #5 \qdot}%
  \IfNoValueT{#2}{%
    \IfBooleanTF{#1}{%
      \def\atripleargtwo{}%
    }{%
      \def\atripleargtwo{x \in X}%
    }%
  }%
  \IfNoValueT{#5}{%
    \IfBooleanTF{#1}{%
      \def\atripleargfive{}%
    }{%
      \def\atripleargfive{\exists y \in Y \qdot}%
    }%
  }%
  \tadaforall 
  \atripleargtwo \qdot 
  \langle#3{}\rangle
  \ \text{\vl!#4!}\ 
  \atripleargfive 
  \langle#6{}\rangle%
}

\newcommand*{\guardname}[1]{\textsc{#1}}

\NewDocumentCommand{\guard}{s r[] e{_}}{%
  \def\guardthree{_#3}%
  \IfNoValueT{#3}{%
    \IfBooleanTF{#1}{%
      \def\guardthree{}%
    }{%
      \def\guardthree{_r}%
    }%
  }%
  \left[\guardname{#2{}}\right]\guardthree%
}

\newcommand*{\atctx}{\cmcal{A}} 
\newcommand*{\interp}{I} 

\newcommand*{\transT}{\cmcal{T}} 

\NewDocumentCommand{\trans}{s o o m m}{%
  \def\transargthree{}%
  \IfNoValueF{#3}{%
    \def\transargthree{\forall #3 \qdot}%
  }%
  \IfBooleanF{#1}{%
    \guardname{#2}:
  }%
  \transargthree #4 \rightsquigarrow #5
}

\NewDocumentCommand{\trc}{E{_}{{R}} r()}{%
  \transT_#1(#2)^{\ast}
}

\newcommand*{\trackdiamond}{\blacklozenge}

\NewDocumentCommand{\trackres}{o d() d<>}{%
  \def\trackresone{#1}%
  \IfNoValueT{#1}{%
    \def\trackresone{r}%
  }%
  \def\trackresres{\trackdiamond}%
  \IfNoValueF{#2}{%
    \def\trackresres{(#2)}%
  }%
  \trackresone \Mapsto \trackresres%
}

\newcommand*{\macro}[1]{\texttt{\MakeUppercase{#1}}}

\newcommand{\codenote}[1]{%
  \renewcommand*{\ctext}[1]{\texttt{\scriptsize ##1}}%
  \textsf{\scriptsize #1}%
}

\newcommand*{\metafunc}[1]{\textsc{#1}}
\WithSuffix{\newcommand}\metafunc*[1]{{\normalfont\metafunc{#1}}}

\makeatletter

\newskip \point

\def \premisespacing{\quad}

\def \RulePremisesNewlineMore[#1]#2.#3#4{\@ifnextchar\bgroup{\RulePremisesNewlineMore[#1]{#2}.{#3\premisespacing#4}}{\@ifnextchar.{\RulePremisesNewline[#1]{{\begin{array}{c}#2\\#3\premisespacing#4\end{array}}}}{\RuleMultiPremise[#1]{{\begin{array}{c}#2\\#3\end{array}}}{#4}}}}

\def \RulePremisesNewline[#1]#2.#3{\@ifnextchar\bgroup{\RulePremisesNewlineMore[#1]{#2}.{#3}}{\@ifnextchar.{\RulePremisesNewline[#1]{{\begin{array}{c}#2\\#3\end{array}}}}{\RuleMultiPremise[#1]{#2}{#3}}}}

\def \RuleMultiPremise[#1]#2#3{\@ifnextchar\bgroup{\RuleMultiPremise[#1]{#2\premisespacing#3}}{\@ifnextchar.{\RulePremisesNewline[#1]{#2\premisespacing#3}}{\prooftree #2\justifies#3 \using{#1}\endprooftree}}}

\def \RuleWithName[#1]#2{\@ifnextchar\bgroup {\RuleMultiPremise[#1]{#2}}{\@ifnextchar.{\RulePremisesNewline[#1]{#2}}{\prooftree \justifies #2 \using{#1} \endprooftree}}}

\def \RuleWithInfo[#1]{\@ifnextchar[{\RuleWithNameAndCondition[#1]}{\RuleWithName[(#1)]}}

\def \RuleWithNameAndCondition[#1][#2]{\RuleWithName[(#1)^{#2}]}

\def \Inf{\proofrulebaseline=2ex \abovedisplayskip12\point\belowdisplayskip12\point \abovedisplayshortskip8\point\belowdisplayshortskip8\point \@ifnextchar[{\RuleWithInfo}{\RuleWithName[ ]}}

 \def \InfBox{\@ifnextchar[{\InfBoxWidth}{\InfBoxWidth[0]}}
 \newcount\argW
 \newcount\dargW
 \newcount\boxW
 \newcount\sboxW
 \def \InfBoxWidth[#1]#2{\@ifnextchar\bgroup{\DBwname[#1]{#2}}{\DBwname[#1]{}{#2}}}
 \def \DBwname[#1]#2#3{\setlength {\unitlength}{.5\proofrulebaseline}
\boxW#1 \multiply\boxW2 \sboxW\boxW \advance\sboxW-3
\setbox123=\hbox{$#3$}\dargW\wd123\divide\dargW\unitlength
\argW\dargW \divide\argW2
\if 0#1
  \boxW\argW \sboxW\boxW \advance\boxW3
\fi
\begin {picture}(\dargW,7.8115)(-\argW,0)
    \put(.5,2.6)    {\begin{picture}(0,0)
    \put(0,6)   {\line(-1,0)    {\boxW}}
    \put(0,6)   {\line(1,0) {\boxW}}
    \put(-\boxW,6)  {\line(1,-2)    {3}}
    \put(\boxW,6)   {\line(-1,-2)   {3}}
    \put(0,0)   {\line(-1,0)    {\sboxW}}
    \put(0,0)   {\line(1,0) {\sboxW}}
    \put(0,2)   {\makebox(0,2)  {\mbox{$#2$}}}
    \put(0,-2)  {\makebox(0,0){$#3$}}
    \end{picture}}
    \end {picture}}

\makeatother

\begin{document}


\title{Concise Outlines for a Complex Logic:\\A Proof Outline Checker for TaDA (Full Paper)}
\titlerunning{Concise Outlines for a Complex Logic}

\author{Felix A. Wolf \and 
Malte Schwerhoff \and
Peter Müller
}
\institute{Department of Computer Science, ETH Zurich\\
\email{\{felix.wolf,malte.schwerhoff,peter.mueller\}@inf.ethz.ch}}

\maketitle

\begin{abstract}
Modern separation logics allow one to prove rich properties of intricate code, \eg functional correctness and linearizability of non-blocking concurrent code. However, this expressiveness leads to a complexity that makes these logics difficult to apply. Manual proofs or proofs in interactive theorem provers consist of a large number of steps, often with subtle side conditions. On the other hand, automation with dedicated verifiers typically requires sophisticated proof search algorithms that are specific to the given program logic, resulting in limited tool support that makes it difficult to experiment with program logics, \eg when learning, improving, or comparing them. 
Proof outline checkers fill this gap. Their input is a program annotated with the most essential proof steps, just like the proof outlines typically presented in papers. The tool then checks automatically that this outline represents a valid proof in the program logic. In this paper, we systematically develop a proof outline checker for the TaDA logic, which reduces the checking to a simpler verification problem, for which automated tools exist. Our approach leads to proof outline checkers that provide substantially more automation than interactive provers, but are much simpler to develop than custom~automatic~verifiers. 

\end{abstract}


\section{Introduction}
\label{sec:introduction}

Standard separation logic enables the modular verification of heap-manipulating sequential~\cite{OHearnRY01,Reynolds02} and data-race free concurrent programs~\cite{OHearn04,Brookes04}. More recently, numerous separation logics have been proposed that enable the verification of fine-grained concurrency by incorporating ideas from concurrent separation logic, Owicki-Gries~\cite{OwickiG76}, and rely-guarantee~\cite{Jones83}. Examples include 
CAP~\cite{DinsdaleYoungDGPV10},
iCAP~\cite{SvendsenB14},
CaReSL~\cite{TuronDB13},
CoLoSL~\cite{RaadVG15},
FCSL~\cite{SergeyNB15},
GPS~\cite{TuronVD14},
RSL~\cite{VafeiadisN13}, and
TaDA~\cite{PintoDG14}
(see Brookes et~al.~\cite{BrookesO16} for an overview).
These logics are very expressive, but challenging to apply because they often
comprise many complex proof rules.
\Eg our running example (\figref{lock_tada_proof_outline}) consists of two statements, but requires
over 20
rule applications in TaDA, many of which have non-trivial instantiations and subtle side conditions. This complexity seems inevitable for challenging verification problems involving, \eg fine-grained concurrency or weak memory.

The complexity of advanced separation logics makes it difficult to develop proofs in these logics. It is, thus, crucial to have tools that check the validity of proofs and automate parts of the proof search. One way to provide this tool support is through \term{proof checkers}, which take as input a nearly complete proof and check its validity. They typically embed program logics into the higher-order logic of an interactive theorem prover such as Coq. Proof checkers exist, \eg for RSL~\cite{VafeiadisN13} and FCSL~\cite{SergeyNB15}. Alternatively, \term{automated verifiers} take as input a program with specifications and devise the proof automatically. They typically combine existing reasoning engines such as SMT solvers with logic-specific proof search algorithms. Examples are Smallfoot~\cite{BerdineCO05} and Grasshopper~\cite{PiskacWZ14} for traditional separation logics, and Caper~\cite{DinsdaleYoungPAB17} for fine-grained concurrency.

Proof checkers and automated verifiers 
strike different trade-offs in the design space. Proof checkers are typically very expressive, enabling the verification of complex programs and properties, and produce foundational proofs. However, existing proof checkers 
offer little automation. Automated verifiers, on the other hand, significantly reduce the proof effort, but compromise on expressiveness and require substantial development effort, especially, to devise custom proof search algorithms.

It is in principle possible to increase the automation of proof checkers by developing proof tactics, or to increase the expressiveness of automated verifiers by developing stronger custom proof search algorithms. However, such developments are too costly for the vast majority of program logics, which serve mostly a scientific or educational purpose. As a result, adequate tool support is very rare, which makes it difficult for developers of such logics, lecturers and students, as well as engineers to apply, and gain experience with, such logics.

To remedy the situation, several tools took inspiration from the idea of \term{proof outlines}~\cite{Owicki75,AptBO09}, formal proof skeletons that contain the key proof steps, but omit most of the details. Proof outlines are a standard notation to present program proofs in publications and teaching material. \term{Proof outline checkers} such as Starling~\cite{WindsorDSP17} and VeriFast~\cite{JacobsSPVPP11} take as input a proof outline and then check automatically that it represents a valid proof in the program logic. They provide automation for proof steps for which good proof search algorithms exist, and can support expressive logics by requiring annotations for complex proof steps. Due to this flexibility, proof outline checkers are especially useful for experimenting with a logic, in situations where foundational proofs are not essential.

In this paper, we present Voila, a \poc for TaDA~\cite{PintoDG14}, which goes beyond existing 
proof outline checkers and automated verifiers 
by supporting a substantially more complex program logic, handling fine-grained concurrency, linearizability, abstract atomicity, and other advanced features. We believe that our systematic development 
of Voila 
generalizes to other complex logics. Our contributions are as follows:
\begin{itemize}
\item
  The Voila \term{proof outline language}, which supports a large subset of TaDA and enables users to write 
  proof 
  outlines 
  very similar to those used by the TaDA authors~\cite{PintoDG14,Pinto16} (\secref{voila_language_overview}). 

\item
 A systematic approach to automate the expansion of a proof outline into a full \emph{proof candidate} via a normal form and heuristics (\secref{proof_candidate}). Our approach automates most proof steps (20 out of 22 in the running example from \figref{lock_tada_proof_outline}).

\item
  An encoding of the proof candidate into Viper~\cite{MuellerSS16}, which checks its validity without requiring any TaDA-specific proof search algorithms (\secref{viper_encoding}).

\item
  The Voila \poc, the \emph{first} tool that supports specification for linearization points, provides a high degree of automation, and achieves good performance (\secref{evaluation}). Our submission artifact with the Voila tool ready-to-use can be found at \cite{VoilaArtifact}, and the Voila source repository is located at \cite{VoilaRepository}.
\end{itemize}

\paragraph{Outline.} \Secref{overview} gives an overview of the TaDA logic and illustrates our approach.
\secref{voila_language_overview} presents the Voila proof outline language, and \secref{proof_workflow} summarizes how we verify proof outlines. We explain how we automatically expand a proof outline into a proof candidate in \secref{proof_candidate} and how we encode a proof candidate into Viper in \secref{viper_encoding}. 
In \secref{evaluation}, we evaluate our technique by verifying several challenging examples, 
discuss related work in \secref{related_work}, and conclude in \secref{conclusion}.

The appendix contains
many further details, including: the full version and Viper encoding of our running example, with TaDA levels (omitted from this paper, but supported by Voila) and nested regions; additional inference heuristics; general Viper encoding scheme; encoding of a custom guard algebra; and a substantial soundness sketch.

\section{Running Example and TaDA Overview}
\label{sec:overview}

\begin{figure}
  \begin{minipage}[t]{0.52\textwidth}\vspace{0pt}
    \includegraphics[
      width=1.0\linewidth,
      trim={6.25cm 14.95cm 6.45cm 2.7cm},
      clip
    ]{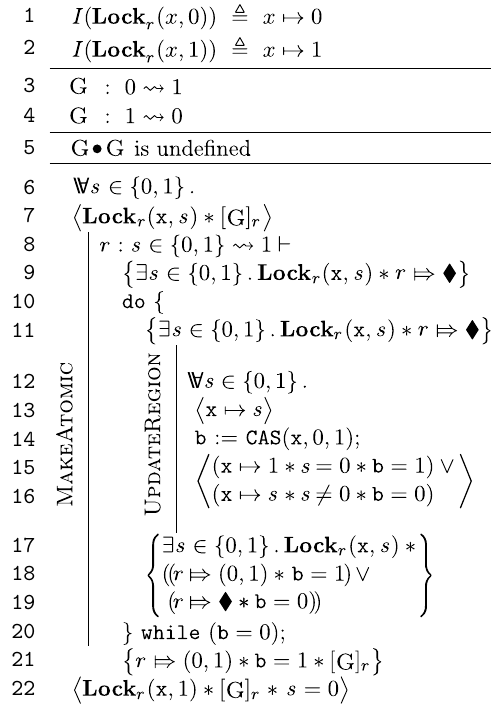}
  \end{minipage}\hfill
  \begin{minipage}[t]{0.45\textwidth}\vspace{0pt}
    \caption{
      TaDA spinlock example with shared region \region*{Lock}; 
      adapted with only minor changes 
      from TaDA~\cite{PintoDG14}.
      The lock region (\linerange*{1--2}) comprises a single memory location, whose value is either 0 (available) or 1 (acquired).
      Guard \guardname{G} allows locking and unlocking (\linerange*{3--4}), and is unique (\lineref*{5}).
      The proof outline (\linerange*{6--22}) shows a CAS-based \vl!lock! operation with~atomic~specifications.      
      An enclosing region (\region*{CAPLock} in da Rocha Pinto et~al.~\cite{PintoDG14},
      verifiable by Voila and shown in \appref{full_lock_caplock_tada_voila}) then
      establishes the usual lock semantics.
      Levels (denoted by $\lambda$ in TaDA) are omitted from the discussion in this paper, but supported by Voila and included in \appref{full_lock_caplock_tada_voila}.
    }
    \label{fig:lock_tada_proof_outline}
  \end{minipage}
\end{figure}

\Figref{lock_tada_proof_outline} shows our running example: a TaDA proof outline for the \vl!lock! procedure of a spinlock. As in the original publication~\cite{PintoDG14}, the outline shows only two out of 22 proof steps and omits most side conditions.
We use this example to introduce the necessary TaDA background, explain TaDA proof outlines, and illustrate the corresponding Voila proof outline.

\subsection{Regions and Atomicity}
\label{sec:shared_regions_atomic_triples}

TaDA targets shared-memory concurrency with sequentially consistent memory.
TaDA programs manipulate \term{shared regions}, data structures that are concurrently modified according to a specified \term{protocol}
(as in rely-guarantee reasoning~\cite{Jones83}).
A shared region such as \region{Lock}_r(x, $s$) is an abstraction over the region's content, analogous to abstract predicates~\cite{ParkinsonB05} in traditional separation logic.
In our example (\linerange*{1--2}), the lock owns memory location $x$ (denoted by separation logic's points-to predicate $x \pointsto \_$),
and its \term{abstract state} $s$ is 0 or 1, indicating whether it is unlocked or locked. Here, the abstract state and the content of the memory location coincide, but they may differ in 
general.
The subscript $r$ uniquely identifies a region instance. 
Note that 
TaDA's region assertions are duplicable, such that multiple threads may obtain an instance of the \region{Lock} resource and invoke operations on the lock.

Lines 3--5 define the protocol for modifications of a lock as a labeled transition system. The labels are \term{guards} -- abstract resources that restrict when a transition may be taken. Here, guard \guardname{G} allows both locking and unlocking (\lines*{3-4}), and is unique (\lineref*{5}). Most lock specifications use duplicable guards to allow multiple threads to compete for the lock; 
in this example, the usual lock semantics is 
established by an enclosing region (\region*{CAPLock}~\cite{PintoDG14}; see \appref{full_lock_caplock_tada_voila}).

Lines 6--22 contain the proof outline for the \vl!lock! procedure, which updates a lock \vl!x! from an undetermined state -- it can seesaw between locked and unlocked due to environment
interference -- to the locked state. Importantly, this update appears to be atomic to clients of the spinlock. These properties are expressed by the \term{atomic TaDA triple} (\lines*{6, 7, and 22})
\begin{displaymath}
  \atriple*
    [s \in \{0,1\}]
    <\region{Lock}(x,$s$) \slstar \guard[G]>
    {lock(x)}
    <\region{Lock}(x,1) \slstar \guard[G] \slstar s = 0>
\end{displaymath}

\noindent
Atomic triples 
(angle brackets)
express that their statement is linearizable~\cite{HerlihyW90}. The abstract state of shared regions occurring in pre- and postconditions of atomic triples is interpreted relative to the linearization point, \ie
the moment in time when the update becomes visible to other threads (here, when the CAS
operation on \lineref*{14} 
succeeds).
The \term{interference context} $\tadaforall s \in \{0,1\}$ is a special binding for the abstract region state that forces
callers to guarantee that the environment keeps the lock state in $\{0,1\}$ until the linearization point is reached (a vacuous restriction in this case).

The precondition of the triple states that an instance of guard $G$ for region $r$, $\guard[G]$, is required to execute \vl!lock(x)!. The postcondition expresses that, at the linearization point, the lock's abstract
state was changed from unlocked ($s = 0$) to locked (\region{Lock}(x,1)). In general, callers must assume that a region's abstract state may have been changed by the environment after the linearization point was reached;
here, however, the presence of the unique guard $\guard[G]$ enables the caller of \vl!lock! to conclude (by the transition system) that the lock remains locked.

\subsection{TaDA Proof Outline}
\label{sec:tada_proof_outline}

\begin{figure}[t]
\begin{center}
\includegraphics[
  width=1.0\textwidth,
  trim={0.1cm 3.9cm 2.7cm 0.2cm},
  clip
]{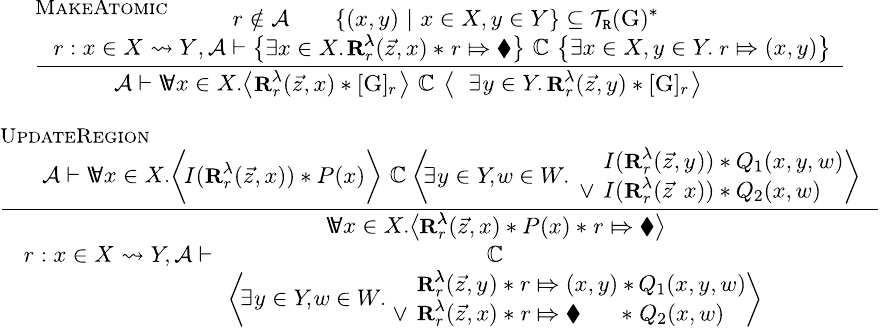}
\caption{
  Simplified versions of two key TaDA rules used in \figref{lock_tada_proof_outline}.
  \MakeAtomic establishes an atomic triple (conclusion) for a linearizable block of code (premise), which includes checking that a state update complies with the region's transition system: $\trc(G)$ is the reflexive, transitive closure of the transitions that \guardname{G} allows.
  \UpdateRegion identifies a linearization point, for instance,
  a CAS statement. If successful, the diamond tracking resource $\trackres<D>$ is exchanged for the witness tracking resource $\trackres(x,y)$ to record the performed state update; otherwise, the diamond resource is kept, such that the operation can be attempted again. 
\vspace{-4mm}
}
\label{fig:simplified_tada_key_rules}
\end{center}
\end{figure}

Lines 6--22 of the proof outline in \figref{lock_tada_proof_outline} show the main proof steps; \figref{simplified_tada_key_rules} shows simplified versions of the applied key TaDA rules.
\MakeAtomic establishes an atomic triple by checking that a block of code is atomic \wrt a shared region abstraction (hence the change from non-atomic premise triple, written with curly braces, to an atomic conclusion triple).
\UpdateRegion identifies the linearization point inside this code block.
Rule \MakeAtomic 
requires that the \term{atomicity context}, a set $\atctx$ of \term{pending updates}, of the premise triple includes any region updates performed by the statement of the triple
(there can be at most one such update per region).
In the proof outline, this requirement is reflected on \lineref*{8}, which shows the intended update of the lock's state: $r: s \in \trans*{\{0,1\}}{1}$
(following TaDA publications, we omitted the tail of the atomicity context from the outline).
\MakeAtomic checks that the update is allowed by the region's transition system with the available guards (the rule's second premise in \figref{simplified_tada_key_rules}), but the
check is omitted from the proof outline.
Then \MakeAtomic temporarily exchanges the corresponding guard $\guard[G]$ for the \term{diamond tracking resource} $\trackres<D>$ (\lineref*{9}), which serves as evidence that the intended update was not yet performed.

Inside the loop, an application of \UpdateRegion identifies the CAS (\lineref*{14}) as the linearization point. The rule requires the diamond resource in its precondition (\lineref*{11}), modifies the shared region (\linerange*{12--16}), and case-splits in its postcondition: if the update failed (\lineref*{19}) then the diamond is kept for the next attempt; otherwise (\lineref*{18}), the diamond is exchanged for the 
\term{witness tracking resource} $\trackres(0,1)$, which indicates that the region was updated from abstract state 0 to 1.
At the end of \MakeAtomic (\lines*{21--22}), the witness resource is consumed and the desired abstractly atomic postcondition is established, stating that the shared region was updated from 0 to 1 at the linearization point.

\subsection{Voila Proof Outline}
\label{sec:incr_voila_proof_outline}

\Figref{incr_voila} shows the \emph{complete} proof outline of our example 
in the Voila \pol, which closely resembles the TaDA outline from \figref{lock_tada_proof_outline}. In particular, the \vl!region! declaration defines a region's interpretation, abstract state, and transition system, just like the initial declarations in \figref{lock_tada_proof_outline}. The subsequent proof outline for procedure \vl!lock! annotates the same two rule applications as the TaDA outline and a very similar loop invariant. The Voila proof outline verifies automatically via an encoding into Viper, but the outline is expressed completely in terms of TaDA concepts; it does not expose any details of the underlying verification infrastructure.
This means that our tool automatically infers the additional 20 rule applications, and all omitted side conditions, thereby closing the gap between the user-provided proof outline and a corresponding full-fledged proof.

\begin{figure}[t]
\begin{voila}
struct cell { int val; }

region Lock(id r, cell x)
  interpretation { x.val |-> ?v && (v == 0 || v == 1) }
  state { v }
  guards { unique G; }
  actions { G: 0 ~> 1; G: 1 ~> 0; }

abstract_atomic procedure lock(id r, cell x)
  interference ?s in Set(0, 1);
  requires Lock(r, x, s) && G@r;
  ensures  Lock(r, x, 1) && G@r && s == 0;
{
  bool b;
  make_atomic using Lock(r, x) with G@r {
    do
      invariant Lock(r, x);
      invariant !b ==> r |=> <D>;
      invariant  b ==> r |=> (0, 1);      
    {
      update_region using Lock(r, x) {
        b := CAS(x, 0, 1);
      }
    } while (!b);
  }
}
\end{voila}
\vspace{-2mm}
\caption{
  The Voila proof outline of our example, strongly resembling the TaDA proof outline from \figref{lock_tada_proof_outline}.
  \vl!id! is the type of region identifiers; primitive types are passed by value, structs by reference. Logical variables are introduced using a question mark; \eg \vl!x.val $\pointsto$ ?v! binds the logical variable \vl!v! to the value of the location \vl!x.val!. \vl!&&! denotes separating conjunction. 
}
\label{fig:incr_voila}
\end{figure}

\section{Proof Outline Language}
\label{sec:voila}
\label{sec:voila_language_overview}

Proof outlines annotate programs with rule applications of a given program logic. These annotations indicate where to apply rules and how to instantiate their meta-variables.
The goal of a proof outline is to convey the essential proof steps; ideally, consumers of such outlines can then construct a full proof with modest effort. Consumers may be human readers~\cite{Owicki75}, or tools that automatically check the validity of a proof outline~\cite{JacobsSPVPP11,MooijW05,WindsorDSP17}; our focus is on the latter.

The key challenge of designing a proof outline language is to define annotations that accomplish this goal with low annotation overhead for proof outline authors. To approach this challenge systematically, we classify the rules of the program logic (here: TaDA) into three categories:
(1)~For some rules, the program prescribes where and how to apply them,
\ie they
do not require any annotations. We call such rules \term{syntax-driven rules}. An example in 
standard 
Hoare logic is the assignment rule, where the assignment statement prescribes how to manipulate 
the 
adjacent assertions.
(2)~Some rules can be applied and instantiated in many meaningful ways. For such rules, the author of the proof outline needs to indicate where or how to apply them through suitable annotations. Since such rules often indicate essential proof steps, we call them \term{key rules}. 
In proof outlines for standard Hoare logic, 
the while-rule typically requires an annotation \emph{how} to apply it, namely the loop invariant. The rule of consequence typically requires an annotation \emph{where and how} to apply it, \eg to strengthen the precondition of a triple or to weaken its postcondition.
(3)~The effort of authoring a proof outline can be greatly reduced by applying some rules heuristically, based on information 
already present in the outline.
We call such rules \term{bridge rules}. Heuristics reduce the annotation overhead, but may lead to incompleteness 
if they
fail; a proof outline language may provide annotations to complement the heuristics in such situations, slightly blurring the distinction between key and bridge rules. \Eg the Dafny verifier~\cite{Leino10} applies heuristics to guess termination measures for 
loops,
but also offers an annotation to provide a measure manually, 
if
necessary.

The rule classification depends on the proof search capabilities of the verification tool that is used to check the proof outline.
We use Viper~\cite{MuellerSS16}, which provides a high degree of automation for standard separation logic and, thus, allows us to focus on the specific aspects of TaDA\@.

In the rest of this section, we give an overview of the Voila proof outline language and, in particular, discuss which TaDA rules are supported as syntax-driven, key, and bridge rules.
Voila's
grammar can be found in \appref{voila_grammar},
showing that Voila strongly resembles TaDA, 
but requires fewer technical details.

\paragraph{Expressions and Statements.} 

Voila supports all of TaDA's programming language constructs, including variables and heap locations, primitive types and operations thereon, atomic heap reads and writes, loops, and procedure calls. Consequently, Voila supports the corresponding syntax-driven TaDA rules.

\paragraph{Background Definitions.}

Voila's syntax for declaring regions and transitions closely resembles TaDA, but \eg subscripts are replaced by additional parameters, such as the region identifier \vl!r!.
A region declaration defines the region's content via an \vl!interpretation! assertion, and its value via a \vl!state! function. The latter may refer to region parameters, as well as values bound in the interpretation, such as \vl!v! in the example from \figref{incr_voila}.
The region's transition system is declared by introducing the guards and the permitted \term{actions}, \ie transitions. Voila includes several built-in guard algebras (adopted from Caper~\cite{DinsdaleYoungPAB17}); additional ones can be encoded, see \appref{encodability_main}.
A region declaration introduces a corresponding region predicate, which has an additional out-parameter that yields the region's abstract state (\eg \vl!s! in the precondition of 
procedure
\vl!lock! in \figref{incr_voila}), as defined by the \vl!state! function. We omit this out-parameter when its value is irrelevant.

\paragraph{Specifications.}

Voila proof outlines require specifications for procedures, and invariants for loops; we again chose a TaDA-like syntax for familiarity. 
Explicit loop invariants are required by Viper, but also enable us to automatically instantiate certain bridge rules (see framing in \secref{proof_candidate}).

Recall that specifications in TaDA are written as atomic or non-atomic triples, and include an interference context and an atomicity context. Voila simplifies the notation significantly by requiring these contexts only for abstractly-atomic procedure specifications; for all statements and rule applications, they are determined automatically, despite changing regularly during a proof.
For procedures with abstractly-atomic behavior (modifier \vl!abstract_atomic!), the interference context is declared through the \vl!interference! clause. \Eg for procedure \vl!lock! from \figref{incr_voila}, it corresponds to TaDA's interference context $\tadaforall s \in \{0,1\}$.

\paragraph{Key Rules.}

In addition to procedure and loop specifications, Voila requires user input only for the following fundamental TaDA rules: \UpdateRegion, \MakeAtomic, \UseAtomic, and \OpenRegion; applications of all other rules are automated. Since they capture the core ideas behind TaDA, these rules are among the most complex rules of the logic and admit a vast proof search space. Therefore, their annotation is essential, for both human readers~\cite{PintoDG14,Pinto16} and automatic checkers.
As seen in \figref{incr_voila}, the annotations for these key rules include only the used region and, for updates, the used guard; all other information present in the corresponding TaDA rules is derived automatically.

\paragraph{Bridge Rules.}

All other TaDA rules are applied automatically, and thus have no 
Voila counterparts. 
This includes
all structural rules for manipulating triple atomicity (\eg \TripleWeak, \AExists), interference contexts (\eg \Substitution, \InterferenceWeak), and levels (\eg \LvlWeak). Their applications are heuristically derived from the program, applications of key rules, and adjacent triples.
TaDA's frame rule is also automatically applied by leveraging Viper's built-in support for framing, combined with additional encoding steps to satisfy TaDA's frame stability side condition.
Finally, TaDA entailments are bridge rules when they can be automated by the used verification tool. For Viper, this is the case for standard separation logic entailments, which constitute the majority of entailments to perform.
To support TaDA's \term{view shifts}~\cite{DinsdaleYoungBGPY13,Pinto16} -- entailments similar to the classical rule of consequence, but involving arbitrary definitions of regions and guard algebras -- Voila provides specialized annotations.

\section{Proof Workflow}
\label{sec:proof_workflow}

Our approach, and corresponding implementation, 
enables the following workflow: users provide a proof outline and possibly some annotations 
for
complex entailments, but never need to insert any other
rule.
Hence, if the outline summarizes a valid proof, verification is automatic, without a tedious process of manually applying additional rules. 
If the outline is invalid, 
our tool reports which specification
(\eg loop invariant)
it could not prove or which key rule application it could not verify, and why 
(\eg missing guard).

Achieving this workflow, however, is challenging: by design, proof outlines provide the important proof steps, but are not complete proofs. Consider, \eg the TaDA and Voila outlines from \figref{lock_tada_proof_outline} and \figref{incr_voila}, respectively. Applying \UpdateRegion produces an atomic triple in its conclusion, whereas the while-rule requires a non-atomic triple for the loop body. A complete proof needs to perform the necessary adjustment through additional applications of bridge rules, which are not present in the proof outlines, and thus need to be inferred.

Our workflow is enabled by first expanding proof outlines into \term{proof candidates}, in two main steps: step~1 automatically inserts the applications of all syntax-driven rules; step~2 expands further by applying heuristics to insert bridge rule applications. The resulting proof candidate contains the applications of all rules of the program logic. Afterwards, we check that the proof candidate corresponds to a valid proof, by encoding it as a Viper program that checks whether all proof rules are applied correctly.
Our actual implementation deviates slightly from this conceptual structure, \eg because Viper does not require one to make the application of syntax-driven rules, framing, and entailment checking explicit.

\section{Expanding Proof Outlines to Proof Candidates}
\label{sec:proof_candidate}

Automatically expanding a proof outline is ultimately a proof search problem, with a vast search space in case of complex logics such as TaDA. Our choice of key rules (and corresponding annotations) reduces the search space, but it remains vast, due to TaDA's many structural rules that can be applied to almost all triples.
To further reduce the search space, 
without introducing additional annotation overhead, we devised 
(and enforce) 
a \term{normal form} for 
proof candidate 
triples. 
 Our normal form 
allows us to define \term{heuristics} for the application of bridge rules \emph{locally}, based only on adjacent rule applications, without having to inspect larger proof parts. This locality reduces the search space substantially, and enables us to automatically close the gap between user-provided proof outline and finally verified proof candidate. 
In our running example, our heuristics infer 20 out of 22~rule~applications.

It might be helpful to consider an analogy with standard Hoare logic:
its rule of consequence can be applied to each Hoare triple. A suitable normal form could restrict proofs to use the rule of consequence only at the beginning of the program and for each loop (as in a weakest-precondition calculus). A heuristic can then infer the concrete applications, in particular, the entailments used in the rule application, treating the rule as a bridge rule.

\paragraph{Normal Form.}

Our normal is established by a combination of syntactic checks and proof obligations in the final Viper encoding. Its main restrictions are as follows:
\begin{enumerate*}

\item All triples are either exclusively atomic or non-atomic, which enables us to infer the triple kinds from statements and key rule applications. Due to this restriction, Voila cannot express specifications that combine atomic and non-atomic behaviors.
However, such specifications do not occur frequently (see \Secref*{5.2.3 in \cite{Pinto16}} for an example) and could be supported via additional annotations.

\item All triple preconditions, as well as the postconditions of non-atomic triples, are \term{stable}, \ie cannot be invalidated by  (legal) concurrent operations. 
In contrast, TaDA requires stability only for certain assertions. Our stronger requirement enables us to \emph{rely} on stability at various points in the proof instead of having to \emph{check} it -- most importantly, when Viper automatically applies its frame rule. To enforce this restriction, we eagerly stabilize assertions through suitable weakening steps.

\item In atomic triples, the state of every region 
is bound by exactly one interference quantifier ($\tadaforall$), which simplifies the manipulation of interference contexts, \eg for procedure calls. To the best of our knowledge, this restriction does not limit the expressiveness of Voila proofs.

\item Triples must hold for a \emph{range} of atomicity contexts $\atctx$, rather than just a single context. This stronger proof obligation rules out certain applications of \MakeAtomic\ -- which we have seen only in contrived examples -- but it increases automation substantially and improves procedure modularity.
\end{enumerate*}

By design, our normal form prevents Voila from constructing certain TaDA proofs. However, the only practical limitation is that Voila does not support TaDA’s combination of atomic and non-atomic behavior in a single triple. As far as we are aware, all other normal form restrictions do not limit expressiveness for practical examples, or can be worked around in systematic ways.

\paragraph{Heuristics.}

We employ 
five main heuristics: 
to determine when to change triple atomicity,
to ensure stable frames by construction,
to compute atomicity context ranges,
to compute levels,
and to compute interference contexts in procedure body proofs.
All heuristics are based on inspecting adjacent rule applications and their proof state. 
We briefly discuss the first three heuristics here, and refer readers to \appref{additional_details_heuristics} for the remaining two heuristics. There, we give a more detailed explanation, and illustrate our heuristics in the context of our running example.
(1)~%
Changing triple atomicity corresponds to an application of (at least) TaDA rule \TripleWeak, necessary when a non-atomic composite statement (\eg the \vl!while! statement in \figref{lock_tada_proof_outline}) has an abstract-atomic sub-statement (\eg the atomic CAS in \figref{lock_tada_proof_outline}). We infer all applications of this rule.
(2)~%
A more complex heuristic is used in the context of framing: TaDA's frame rule requires the \term{frame}, \ie, the assertion preserved across a statement, to be stable. 
For simple statements such as heap accesses, it is sound to rely on Viper's built-in support for framing.
For composite statements with arbitrary user-provided \term{footprints} (assertions such as a loop invariant describing which resources the composite statement may modify), 
we greedily infer frame rule applications that attempt to preserve all information outside the footprint. The inferred applications are later encoded in Viper such that the resulting frame is stable, by applying suitable weakening steps.
(3)~%
Atomicity context ranges are heuristically inferred from currently owned tracking resources and level information. Atomicity contexts are not manipulated by a specific TaDA rule, but they need to be instantiated when applying 
rules: most importantly, TaDA's procedure call rule, but also \eg \MakeAtomic and \UpdateRegion (see \figref{simplified_tada_key_rules}).

In our experience, our heuristics fail \emph{only} in two scenarios: the first are contrived examples, concerned with TaDA resources in isolation, 
not properties of actual code -- where they fail to expand a proof outline into a valid proof.
More relevant is the second scenario, where our heuristics yield a valid proof that Viper then fails to verify because it requires entailments 
that Viper cannot discharge automatically.
To work around such problems when they occur, Voila allows programmers to provide additional annotations 
to indicate where to apply 
complex entailments. 

Importantly, a failure of our heuristics does not compromise soundness: if they infer invalid bridge rule applications, \eg whose side conditions do not hold, the resulting invalid proof candidates are rejected by Viper in the final validation.

\section{Validating Proof Candidates in Viper}
\label{sec:viper_encoding}

Proof candidates 
-- \ie the user-provided program with heuristically inserted bridge rule applications -- 
do not necessarily represent valid proofs, \eg when users provide incorrect loop invariants. To check whether a proof candidate actually represents a valid proof, we need to verify (1)~that each rule is applied correctly, in particular, that its premises and side conditions hold, and (2)~that the property shown by the proof candidate entails the intended specification. To validate proof candidates automatically, we use the existing Viper tool~\cite{MuellerSS16}.
In this section, we give a high-level overview of how we encode proof candidates into the Viper language.

\paragraph{Viper Language.}

Viper uses a variation of separation logic~\cite{SmansJP09,ParkinsonSummers12} whose assertions separate access permissions from value information: separation logic's points-to assertion \vl!x.f $\pointsto$ v!
is expressed as \vl!acc(x.f) && x.f == v!,
and separation logic predicates~\cite{ParkinsonB05} are similarly split into a predicate (abstracting over permissions) and a heap-dependent function (abstracting over values). Well-definedness checks ensure that the heap is accessed only under sufficient permissions.
Viper provides a simple imperative language, which includes in particular two statements to manipulate the verification state: \vl!exhale $A$! asserts all logical constraints in assertion $A$, removes the permissions in $A$ from the current state (or fails if the permissions are not available) and assigns non-deterministic values to the corresponding memory locations (to reflect that the environment could now modify them); \vl!inhale $A$! analogously assumes constraints and adds permissions.

\paragraph{Regions and Assertions.}

\begin{figure}[t]
\begin{center}
\setlength{\tabcolsep}{5pt}
\begin{minipage}[b]{0.5\textwidth}
\begin{voila}[language=silver]
$\llbracket$region R(r: id, $\overline{\texttt{p: t}}$)
  interpretation $I$
  state $S$
  guards $G$
  actions $A$$\rrbracket$ -> 
predicate R(r: Ref, $\overline{\texttt{p:} \llbracket \texttt{t} \rrbracket}$) { $\llbracket I\rrbracket$ }
  
function R_State(r: Ref, $\overline{\texttt{p:} \llbracket \texttt{t} \rrbracket}$): $T$
  requires  R(r,$\overline{\texttt{p}}$)
{ unfolding R(r,$\overline{\texttt{p}}$) in $\llbracket S\rrbracket$ }
  
$\textit{foreach}$ g($\overline{\texttt{p': t'}}$) $\in G$:
  predicate R_g(r: Ref, $\overline{\texttt{p':} \llbracket \texttt{t'} \rrbracket}$)
$\textit{end}$

field diamond: Bool
\end{voila}
\end{minipage}\hfill
\begin{minipage}[b]{0.5\textwidth}
\begin{voila}[language=silver]
field val: Int

predicate Lock(r: Ref, x: Ref) { 
  acc(x.val) && 
  (x.val == 0 || x.val == 1)
}

function Lock_State
             (r: Ref, x: Ref): Int
  requires Lock(r, x) 
{ unfolding lock(r, x) in x.val }

predicate Lock_G(r: Ref)

field diamond: Bool
\end{voila}
\end{minipage}
\vspace{-2mm}
\caption{
  Excerpt of the Viper encoding of regions; general case (left), and for the 
  lock 
  region from \figref{incr_voila} (right). 
  The encoding function is denoted by double square brackets; overlines denote lists; \textit{foreach} loops are expanded statically. 
  Type $T$ is the type of the state expression $S$,
  which is inferred.
  Actions $A$ do not induce any global declarations.
  The elements of struct types and type \vl!id! are encoded as Viper references (type \vl!Ref!). The \vl!unfolding! expression temporarily unfolds a predicate into its definition; it is required by Viper's backend verifiers.
  The struct type \vl!cell! from \figref{incr_voila} is encoded as a Viper reference with field \vl!val!
  (in Viper, all objects have all fields declared in the program).
}
\label{fig:region_enc}
\end{center}
\end{figure}

TaDA's regions introduce various resources such as region predicates and guards. We encode these into Viper permissions and predicates as summarized in \figref{region_enc} (left). Each region \vl!R! gives rise to a 
corresponding predicate,
which is defined by the region interpretation.
A region's abstract state may be accessed by a Viper function \vl!R_State!, which is defined based on the region's \vl!state! clause,
and depends on the region predicate.
Moreover, we introduce an abstract Viper predicate \vl!R_g! for each guard \vl!g! of the region.

These declarations allow us to encode most TaDA assertions in a fairly straightforward way. \Eg the assertion 
\region{Lock}_r(x, $s$) from \figref{lock_tada_proof_outline} is encoded as a combination of a region predicate and the function yielding its abstract state: \vl!Lock($r$,x) && Lock_State($r$,x) == $s$!. We encode region identifiers as references in Viper, which allows us to use the permissions and values of designated fields to represent resources and information associated with a region instance. \Eg we use the permission \vl!acc($r$.diamond)! to encode the TaDA resource $\trackres<D>$.

\paragraph{Rule Applications.}

\begin{wrapfigure}[4]{r}{0.195\textwidth}
\newcommand{\triple}[3]{\{\;#1\;\}\;#2\;\{\;#3\;\}}
\makeatletter 
\DeclareRobustCommand
  \myvdots{\vbox{\baselineskip3.5\p@ \lineskiplimit\z@
    \hbox{.}\hbox{.}\hbox{.}}}
\makeatother
\scalebox{0.78}{
\Inf{
  \Inf[R]{
    \Inf
      {\myvdots\vspace{-3pt}}
      {\triple{P_p}{s}{Q_p}}        
  }{\triple{P_c}{s}{Q_c}}
  }{{\myvdots}}
}
\end{wrapfigure}

Proof candidates are tree structures, where each premise of a rule application $R$ is established as the conclusion of another rule application, as illustrated on the right. To check 
the 
validity of a candidate, we check 
the 
validity of each rule application. For rules 
that are 
natively supported by Viper (\eg the assignment rule), Viper performs all necessary checks. Each other rule application is checked via an encoding into the following sequence of Viper instructions:
(1)~Exhale the precondition $P_c$ of the conclusion to check that the 
required assertion holds.
(2)~Inhale the precondition $P_p$ of the premise
since it may be assumed
when proving
the premise.
(3)~After the code $s$ of the premise, exhale the postcondition $Q_p$ of the premise to check that 
it was established by the proof for the premise.
(4)~Inhale the postcondition $Q_c$ of the conclusion.
Steps~2 and~3 are performed for each premise of the rule. 
Moreover, we assert the side conditions of each rule.
If a proof candidate is invalid,
\eg composes incompatible rules,
one of the checks above fails and 
the candidate is rejected.

Using this encoding of rule applications as building blocks, we can assemble entire procedure proofs as follows: for each procedure, we inhale its precondition, encode the rule application for 
its
body, and then exhale 
its postcondition.

\paragraph{Example: Stabilizing Assertions.}

Recall that an assertion $A$ is stable if and only if the environment cannot invalidate $A$ by performing 
any 
legal region updates. In practice, this means that the environment cannot hold a guard that allows it to change the state of a region in a way that violates $A$.
The challenge of \emph{checking} stability as a side-condition is to \emph{avoid higher-order quantification} over region instances and guards, which is hard to automate.
We address this challenge by eagerly \term{stabilizing} assertions in the Viper encoding, \ie we weaken Viper's verification state such that the remaining information about the state is stable. We achieve this effect by first assigning non-deterministic values to the region state and then constraining these to be within the states permitted by the region's transition system, taking into account the guards the environment~could~hold.
The Viper code for stabilizing instances of \vl!Lock! can be found in \appref{enc_outline_steps_ext_disc}.

\section{Evaluation}
\label{sec:evaluation}

We evaluated Voila on nine benchmark examples from Caper's test suite, with the
Treiber's stack~\cite{Treiber86} variant \texttt{BagStack} being the most
complex example, and report verification times and annotation overhead.
Each example has been verified in two versions: a version with Caper's
comparatively \emph{weak} non-atomic specifications, and another version with
TaDA's \emph{strong} atomic specifications; see \secref{related_work} for a
more detailed comparison of Voila and Caper.
An additional example, \texttt{CounterCl}, demonstrates the encoding of a custom
guard algebra not supported in Caper (see \appref{supported_tada_ingredients}).
To evaluate performance stability, we seeded four 
examples with 
errors in the loop invariant, procedure postcondition, code, and region
specification, respectively.
Our benchmark suite is relatively small, but each example involves nontrivial specifications. 
To the best of our knowledge, no other (semi-)automated tool is able to verify similarly strong specifications.

\paragraph{Performance.}

\begin{figure}[t]
\begin{center}
\setlength{\tabcolsep}{5pt}
\begin{minipage}[t]{0.47\textwidth}
\scalebox{0.8}{
\begin{tabular}{l|c|r|r|r}
Program                   & LOC & Stg    & Wk    & Cpr  \\
\hline
\texttt{SLock}            & 15  & 2.6    & 2.1   & 1.4   \\ 
\texttt{TLock}            & 23  & 21.8   & 8.1   & 2.4   \\ 
\texttt{TLockCl}          & 16  & 2.9    & 2.6   & 0.5   \\ 
\texttt{CASCtr}           & 25  & 3.9    & 2.7   & 1.5   \\ 
\texttt{BoundedCtr}       & 24  & 8.1    & 5.1   & 63.1  \\ 
\texttt{IncDecCtr}        & 28  & 4.2    & 3.1   & 2.9   \\ 
\texttt{ForkJoin}         & 16  & 2.1    & 1.3   & 1.0   \\ 
\texttt{ForkJoinCl}       & 28  & 2.9    & 2.3   & 1.6   \\ 
\texttt{BagStack}         & 29  & 29.9   & 18.0  & 211.6  \\ 
\texttt{CounterCl}        & 45  & \multicolumn{1}{c|}{-} & 5.8 & \multicolumn{1}{c}{-} \\ 
\end{tabular}
}
\end{minipage}\hfill
\begin{minipage}[t]{0.47\textwidth}
  \scalebox{0.75}{
\begin{tabular}{l|l|r|r|r}
Program & Err & Stg & Wk & Cpr \\ \hline
\multirow{4}{*}{\texttt{CASCtr}}
        & L   & 1.5 & 1.9 & 1.5 \\
        & P   & 2.5 & 1.9 & 11.2 \\
        & C   & 1.5 & 1.2 & 0.5 \\
        & R   & 1.2 & 1.1 & 0.3 \\
\hline
\multirow{4}{*}{\texttt{TLock}}
        & L   & 3.9   & 7.2 & 2.0 \\
        & P   & 7.2   & 3.4 & 2.4 \\
        & C   & 15.6  & 1.8 & 0.6 \\
        & R   & 4.1   & 1.8 & 0.7 \\
\hline
\multirow{4}{*}{\texttt{TLockCl}}
        & P   & 2.9 & 2.6 & 143.4 \\
        & C   & 2.5 & 2.5 & 115.5 \\
        & R   & 1.8 & 1.7 & 5.0 \\
\hline
\multirow{4}{*}{\texttt{BagStack}}
        & L   & 26.5 & 17.8 & \multicolumn{1}{c}{> 600} \\
        & P   & 27.9 & 17.7 & \multicolumn{1}{c}{> 600} \\
        & C   & 26.3 & 17.8 & \multicolumn{1}{c}{> 600} \\
        & R   & 14.4 & 9.2  & 216.6
\end{tabular}
  }
\end{minipage}
\caption{
  Timings in seconds for successful (left table) and failing (right table) verification
  runs; lines of code (LOC) are given for Voila programs and exclude proof
  annotations. 
  \emph{Stg}/\emph{Wk} denote strong/weak Voila specifications; \emph{Cpr}
  abbreviates Caper.
  Programs include spin and ticket locks, counters (\emph{Ctr}), and client programs
  (\emph{Cl}) using the proven specifications.
  Errors (\emph{Err}) were seeded in loop invariants (\emph{L}), postconditions
  (\emph{P}), code (\emph{C}), and region specifications (\emph{R}).
\vspace{-4mm}
}
\label{fig:performance}
\end{center}
\end{figure}

\figref{performance} shows the runtime for each example in seconds. All measurements were
carried out on a Lenovo W540 with an Intel Core i7-4800MQ and 16GB of RAM,
running Windows 10 x64 
and Java HotSpot
JVM 18.9 x64; Voila was compiled using Scala 2.12.7. We used a recent checkout
of Viper and Z3 4.5.0 x64 (we failed to compile Caper against newer versions of
Z3).
Each example was verified ten times (on a continuously-running JVM); after
removing the highest and lowest measurement, the remaining eight values were
averaged. Caper (which compiles to native code) was measured analogously.

Overall, Voila's verification times are good; most examples verify in under five seconds. Voila is slower than Caper and its logic-specific symbolic execution engine, but it exhibits stable performance for successful and failing runs, which is crucial in the common case that proof outlines are developed interactively, such that the checker is run frequently on incorrect versions. As demonstrated by the error-seeded versions of 
\texttt{TLockCl} and \texttt{BagStack}, Caper's performance is~less~stable.

Another interesting observation is that strong specifications typically do not take significantly longer to verify,
although only they require the full spectrum of TaDA ingredients and make use of
TaDA's most complex rules, \MakeAtomic and \UpdateRegion.
Notable exceptions are: \texttt{BagStack}, where only the strong specification
requires sequence theory reasoning; and \texttt{TLock} and
\texttt{BoundedCtr}, whose complex transition systems with many disjunctions
significantly increase the workload when verifying atomicity rules such as
\MakeAtomic.

\paragraph{Automation.}

Voila's annotation overhead, averaged over the programs with \emph{strong}
specifications from \figref{performance}, is 0.8 lines of proof annotations
(not counting declarations and procedure specifications; neither for Caper) 
per line of code, which demonstrates the high degree of automation Voila achieves. 
Caper has an average 
annotation overhead of 0.13 for its programs from
\figref{performance}, but significantly weaker specifications.
Verifying only the latter in Voila does not reduce annotation overhead
significantly since Voila was designed to support TaDA's strong specifications.
The overhead reported for encodings into interactive theorem provers 
such as Coq~\cite{DokoV17,KaiserDDLV17,KleinEHACDEEKNSTW09,VafeiadisN13}
is typically much higher, ranging between~10~and~20.

\section{Related Work}
\label{sec:related_work}

We compare Voila to three groups of tools: \caperishtools,
focusing on automation;
\coqishtools, focusing on expressiveness; and \voilaishtools, designed to strike
a balance between automation and expressiveness.
Closest to our work in the kind of supported logic is the \caperishtool Caper~\cite{DinsdaleYoungPAB17}, from which we drew inspiration,
\eg for how to specify region transition systems.
Caper supports an improved version of CAP~\cite{DinsdaleYoungDGPV10}, a
predecessor logic of TaDA\@.
Caper's symbolic execution engine achieves an impressive degree of automation, which, for more complex examples, is higher than Voila's. Caper's
automation also covers slightly more guard algebras than Voila. 
However, the automation comes at the price of expressiveness, compared to Voila:
postconditions are often significantly weaker because the logic does not support
linearizability (or any other notion of abstract atomicity). \Eg Caper cannot
prove that the spinlock's \vl!unlock! procedure actually releases the lock.
As was shown in \secref{evaluation}, Caper is typically faster than Voila, but
exhibits less stable performance when a program or its specifications are wrong.

Other \caperishtools for \fgc reasoning are
SmallfootRG~\cite{CalcagnoPV07}, which can prove memory safety, but not
functional correctness, and CAVE~\cite{Vafeiadis10}, which can prove
linearizability, but cannot reason about non-linearizable code (which TaDA and
Voila can).
VerCors~\cite{OortwijnBGHZ17} combines a concurrent separation logic
with process-algebraic specifications; special program annotations are used to relate concrete program operations to terms in the abstract process algebra model. Reasoning about the resulting term sequences is automated via model checking, but is non-modular.
Summers et~al.~\cite{SummersM18} present an automated verifier for the RSL family of logics~\cite{VafeiadisN13,DokoV16,DokoV17} for reasoning about weak-memory concurrency. Their tool also 
encodes
into Viper and requires very few annotations because proofs in the RSL logics are more stylized than in TaDA\@.

A variety of complex separation
logics~\cite{VafeiadisN13,NanevskiLSD14,TuronVD14,SergeyNB15,DokoV16,DokoV17,FruminKB18,KrebbersJ0TKTCD18,JungKJBBD18}
are supported by proof checkers, typically via 
Coq encodings.
As discussed in the introduction, such tools strike a different trade-off than 
proof 
outline checkers: they provide foundational proofs, but typically offer little automation,
which hampers experimenting with logics.

Starling~\cite{WindsorDSP17} is a proof outline checker and closest to Voila in terms of the overall design, but it 
focuses on proofs that are \emph{easy} to automate. 
To achieve 
this, it
uses a simple instantiation of the Views meta-logic~\cite{DinsdaleYoungBGPY13} as its logic. Starling's logic does not enable the kind of strong,
linearizability-based postconditions that Voila can prove (see the discussion of Caper above). Starling generates proof obligations 
that
can be discharged by an SMT solver, or by
GRASShopper~\cite{PiskacWZ14} if the program requires heap reasoning.
The parts of 
an outline
that involve the heap must be written in GRASShopper's input language. In contrast, Voila does not expose the underlying 
system, 
and
users can work on the abstraction~level~of~TaDA.

VeriFast~\cite{JacobsSPVPP11} can  be seen as  an outline checker 
for a separation logic with impressive features
such as higher-order functions and predicates. It has no dedicated support for \fgc, but the developers manually encoded examples such as
concurrent stacks and queues.
VeriFast favors expressiveness over automation: proofs often require non-trivial specification adaptations and substantial
amounts of ghost code, but the results typically~verify~quickly.

\section{Conclusion}
\label{sec:conclusion}

We introduced Voila, a novel proof outline checker that supports most of TaDA's features, and achieves a high degree of automation and good performance. This enables concise proof outlines with a strong resemblance of TaDA\@.

Voila is the first deductive verifier that can reason automatically about a procedure's effect at its linearization point, which is essential for a wide range of concurrent programs. Earlier work either proves much weaker properties (the preservation of basic data structure invariants rather than the functional behavior of procedures) or requires substantially more user input (entire proofs rather than concise outlines).

We believe that our systematic approach to developing Voila can be generalized to other complex logics. In particular, encoding proof outlines into an existing verification framework allows one to develop proof outline checkers efficiently, without developing custom proof search algorithms. Our work also illustrates that an intermediate verification language such as Viper is suitable for encoding a highly specialised program logic such as TaDA. 
During the development of Voila, we uncovered and fixed several soundness and modularity issues in TaDA, which the original authors acknowledged and had partly not been aware of. We view this as anecdotal evidence of the benefits of tool support that we described in the introduction.

Voila supports the vast majority of TaDA's features; most of the others can be supported with additional annotations.
The main exception are TaDA's hybrid assertions, which combine atomic and non-atomic behavior. Adding support for those is future work. Other plans include an extension of the supported logic, \eg to handle 
extensions of TaDA~\cite{PintoDGS16,DOsualdoFGS19}.

\paragraph{Acknowledgements.}

We thank the anonymous referees of this paper, and earlier versions thereof, for suggesting many improvements to the explanation of our work. We are also thankful to Thomas Dinsdale-Young and Pedro da Rocha Pinto for instructive discussions about their work, TaDA\@, and for feedback on Voila.


\bibliographystyle{splncs04}
\bibliography{paper_full_arxiv}


\newpage
\appendix

\MaybeFloatBarrier
\section{TaDA Key Proof Rules}
\label{sec:tada_key_proof_rules}

\begin{figure}
\begin{center}
\includegraphics[
  width=1.0\textwidth, 
  trim={-0.2cm 0cm 0cm -0.2cm},
  clip
]{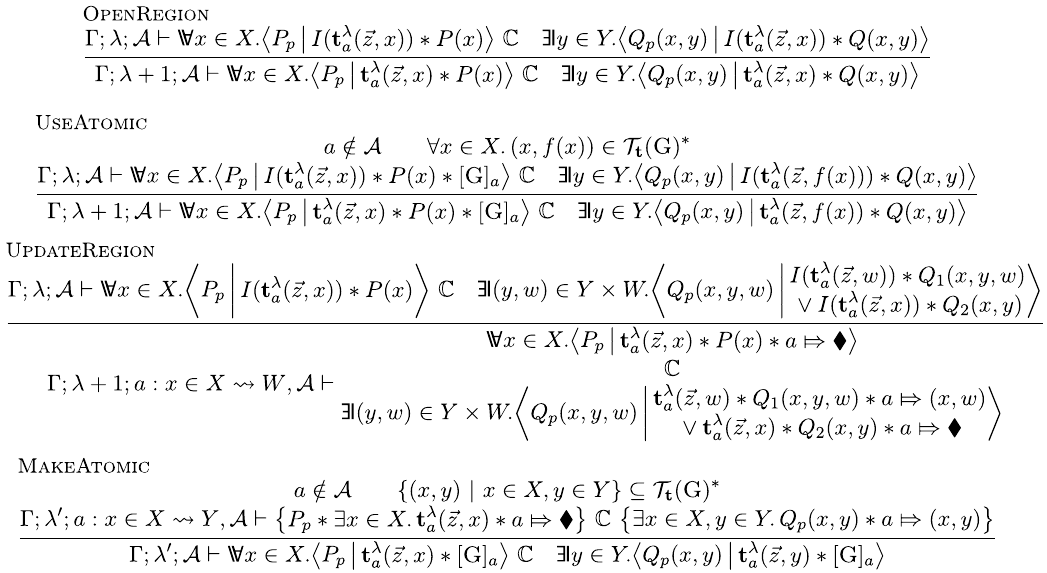}
\caption{
  \figref{tada_key_proof_rules} shows TaDA's key proof rules as they are presented
  in da Rocha Pinto's thesis~\cite{Pinto16}, including public and private assertions,
  and region levels. Levels are omitted from the discussion in this paper,
  but supported by Voila. The combination of public and private assertions in a rule triple is currently not supported by Voila.
}
\label{fig:tada_key_proof_rules}
\end{center}
\end{figure}

\MaybeFloatBarrier
\section{Supported TaDA Ingredients}
\label{sec:supported_tada_ingredients}

\Figref{voila_features} provides an overview of Voila's features, \wrt TaDA ingredients and Caper guard algebras~\cite{DinsdaleYoungPAB17}.
The left column lists TaDA features and to which extent their incur annotation overhead.
\emph{None} means that the ingredient does not surface at all in a Voila program.
\emph{Once} means that there is a one-time annotation per Voila program, typically in the form of a background declaration such as a region.
In contrast, \emph{proc} means that the feature requires a one-time annotation per Voila procedure, typically as part of a procedure specification.
Next, \emph{low} means that the feature may result in more than one annotation per procedure: for regions, these are new-region statements (one per newly created region instance), in addition to region declarations. Tracking resources, on the other hand, typically appear in invariants of loops that repeat until an update succeeded.
Finally, View shifts incur a \emph{medium} annotation overhead: most standard view shifts are automated by Voila and do not require annotations, but for complex, manually encoded examples, additional annotations may be required. See also \appref{encodability_main}.

Most of Caper's guard algebras are supported by Voila, and as such, do not incur any additional overhead (the guards themselves must still be mentioned, \eg in specifications). Only counting and sum guards are not directly supported by Voila; they can be encoded, which will require additional annotations. See also \appref{encodability_main} for an example of a manually encoded guard algebra.

\begin{figure}[h]
\begin{center}
\setlength{\tabcolsep}{5pt}
\begin{minipage}[t]{0.47\textwidth}
\begin{tabular}{l|c}
Ingredient             & Annotations \\ \hline
Regions                & low      \\
Transition systems     & once     \\
Triple kinds           & proc     \\
Interference contexts  & proc     \\
Atomicity contexts     & none     \\
Levels                 & proc     \\
Tracking resources     & low      \\
Private vs. public     & --       \\
View Shifts            & medium   \\
Stability              & none     \\
Framing                & none     \\
\end{tabular}
\end{minipage}\hfill
\begin{minipage}[t]{0.47\textwidth}
\begin{tabular}{l|c}
Guard Algebra  & Support \\ \hline
Trivial        & built-in \\
All-or-nothing & built-in \\
Counting       & encodable \\
Indexed        & built-in \\
Product        & built-in \\
Permissions    & built-in \\
Sum            & encodable \\
  \\
  \\
  \\
  \\
\end{tabular}
\end{minipage}
\end{center}
\caption{
  Supported TaDA ingredients, with a classification of the incurred annotation overhead, and Caper guard algebras~\cite{DinsdaleYoungPAB17}, with a classification of their support.
  TaDA's combination of public and private assertions in a rule triple is currently not supported by Voila.
}
\label{fig:voila_features}
\end{figure}

\MaybeFloatBarrier
\newpage
\section{Voila Grammar}
\label{sec:voila_grammar}

This section gives an overview of Voila's grammar, and shows that Voila strongly resembles TaDA, but requires fewer technical details in its annotation language.

\begin{figure}
  \newcommand*{\br}[1]{\texttt{\{}#1\texttt{\}}}
  \newcommand*{\pr}[1]{\texttt{(}#1\texttt{)}}
  \centering
  \begin{align*}
  t &::= \texttt{id} \alt \texttt{bool} \alt \texttt{int} \alt \texttt{frac} \alt S 
  \\
  e &::= x \alt \texttt{?}x \alt l \alt 
        e\ \texttt{\&\&}\ e \alt e\ \texttt{||}\ e \alt \texttt{!}e \alt e \Rightarrow e \alt
        e\ \mathit{op}\ e
  \\
  a &::= e \alt
        x.f \pointsto e \alt
        a\ \texttt{\&\&}\ a \alt e \Rightarrow a \alt
        R\pr{r, \overline{e}} \alt
        G(\overline{e})\texttt{@}r \alt
        r \Mapsto \trackdiamond \alt r \Mapsto \pr{e,e}
  \end{align*}
  \caption{
    Voila's core syntax for types $t$, expressions $e$, and assertions $a$).
    Types $t$ include the type of region identifiers \texttt{id}, fractions \texttt{frac}, and struct types $S$.
    Expressions $e$ include variables $x$, literals $l$, fields $f$,  and the usual expression operators, \eg relational ones. They also include variable binders $\texttt{?}x$, which are allowed in only two places: the right-hand side of points-to assertions and the last parameter of a region instance, binding the region's abstract state.
    Assertions $a$ include, besides the usual separation logic assertions, region instances $R(r, \overline{e})$, where $R$ denotes a region name, $r$ a region identifier, and $\overline{e}$ the region's abstract state; the last argument may be omitted when the region state is unspecified. As usual, overlines denote lists. Moreover, assertions include guards $G(\overline{e})@r$, where $G$ denotes a guard name and $\overline{e}$ the guard arguments (guards without arguments are written as  $G@r$), and TaDA's two tracking resources. 
    For brevity, we omitted levels, collection data types (\ie sets, sequence, maps, and tuples), and more complex guards  (but see also \appref{supported_tada_ingredients} and \appref{encodability_main}).
  }
  \label{fig:grammar1}
\end{figure}

\begin{figure}
  \newcommand*{\br}[1]{\texttt{\{}#1\texttt{\}}}
  \newcommand*{\pr}[1]{\texttt{(}#1\texttt{)}}
  \newcommand*{\ass}{\texttt{:=}}
  \newcommand*{\using}{\texttt{using}}
  \newcommand*{\with}{\texttt{with}}
  \newcommand*{\grd}{G(\overline{e})\texttt{@}r}
  \newcommand*{\rgn}{R\pr{r, \overline{e}}}
  \centering
  \begin{minipage}[t]{0.40\textwidth}\flushleft
  \[\begin{array}{rcl}
  ns &::= & x\ \ass\ e \\
     &\alt& \overline{x}\ \ass\ P\pr{\overline{e}} \\
     &\alt& \texttt{if}\ (e)\ \br{s}\ \texttt{else}\ \br{s} \\
     &\alt& \texttt{while}\ \pr{e}\ \texttt{invariant}\ a\ \br{s} \\
     &\alt& s\texttt{;}\ s \\
  \\
  s  &::= & t\ x \alt ns \alt as \\
  \end{array}\]
  \end{minipage}
  \hfil
  \begin{minipage}[t]{0.58\textwidth}  
  \[\begin{array}{rcl}
  as &::= & x\text{.}f\ \ass\ e \\
     &\alt& y\ \ass\ x\text{.}f \\
     &\alt& \overline{x}\ \ass\ P\pr{\overline{e}} \\
     &\alt& \texttt{use\_atomic}\ \using\ \rgn\ \with\ \grd\ \br{as} \\
     &\alt& \texttt{make\_atomic}\ \using\ \rgn\ \with\ \grd\ \br{s} \\
     &\alt& \texttt{open\_region}\ \using\ \rgn\ \br{as} \\
     &\alt& \texttt{update\_region}\ \using\ \rgn\ \br{as} \\
  \end{array}\]
  \end{minipage}
  \caption{
    Voila's core syntax for statements $s$, atomic statements $\mathit{as}$, and non-atomic statements $\mathit{ns}$.
    Statements $s$ comprise variable declarations as well as atomic and non-atomic statements; the categorization of the latter follows TaDA\@.
    Atomic statements $\mathit{as}$ include field reads and writes, invocations of abstract-atomic procedures, and key rule statements. Following TaDA, rule statements other than \texttt{make\_atomic} may only nest atomic statements.
    Non-atomic statements $\mathit{ns}$ are local variable assignments, invocations of non-atomic procedures, and compound statements.
    For brevity, statements for creating struct and region instances have been omitted, as have ghost statements useful for encoding, \eg complex guard algebras (see \appref{encodability_main}).
  }
  \label{fig:grammar2}
\end{figure}

\begin{figure}
  \newcommand*{\br}[1]{\texttt{\{}#1\texttt{\}}}
  \newcommand*{\pr}[1]{\texttt{(}#1\texttt{)}}
  \addtolength{\jot}{-3pt} 
  \begin{minipage}[t]{0.09\textwidth}\flushleft
  \begin{align*}
  & \texttt{struct}\ S\ \br{\overline{t\ f}}
  \end{align*}
  \end{minipage}
  \hfil
  \begin{minipage}[t]{0.29\textwidth}
  \begin{align*}
  & \texttt{region}\ R\pr{\texttt{id}\ r, \overline{t\ x}} \\
  & \quad \texttt{interpretation}\ \br{a} \\
  & \quad \texttt{state}\ \br{e} \\
  & \quad \texttt{guards}\ \br{\overline{mod\ G(\overline{t\ x})}} \\
  & \quad \texttt{actions}\ \br{\overline{G(\overline{e})\colon e \leadsto e}} \\
  \end{align*}
  \end{minipage}
  \hfil
  \begin{minipage}[t]{0.40\textwidth}
  \begin{align*}
  & \texttt{abstract\_atomic procedure}\ P\pr{\overline{t\ x}}\ \\
  & \qquad \texttt{returns}\ \pr{\overline{t\ y}} \\
  & \quad \texttt{interference}\ \texttt{?}x\ \texttt{in}\ e \\
  & \quad \texttt{requires}\ a \\
  & \quad \texttt{ensures}\ a \\
  & \texttt{\{} \overline{s} \texttt{\}}
  \end{align*}
  \end{minipage}
\vspace{-3mm}
  \caption{
    Voila's core syntax for struct, region, and procedure declarations.
    Structs declare fields and induce homonymous types.
    Region declarations include name $R$, identifier $r$ and further formal arguments $\overline{t\ x}$. A region's interpretation and state are an assertion and expression, respectively. Each region may declare guards $G(\overline{t\ x})$, with formal arguments $\overline{x}$ and modifier \texttt{unique} or \texttt{duplicable}, and actions that describe possible state changes.
    Abstract-atomic procedure declarations include an interference clause that corresponds to TaDA's interference context.
    More complex guard and action definitions are omitted for brevity, as are non-atomic and lemma procedures (but see also \appref{full_lock_caplock_tada_voila} and \appref{encodability_main}).
  }
  \label{fig:grammar3}
\end{figure}

\MaybeFloatBarrier
\section{Full Lock and CAPLock Example}
\label{sec:full_lock_caplock_tada_voila}

This section complements our running example by showing TaDA outline and Voila code for
\begin{enumerate*}
\item region \region*{Lock} and procedure \vl!lock!, but with the previously omitted levels, and
\item region \region*{CAPLock} and procedure \vl!acquire!, which build on the former and provide the expected lock semantics.
\end{enumerate*}

\begin{figure}
\begin{center}
\includegraphics[
  width=1.0\textwidth,
  trim={0.1cm 16.7cm 2.2cm 0.8cm},
  clip
]{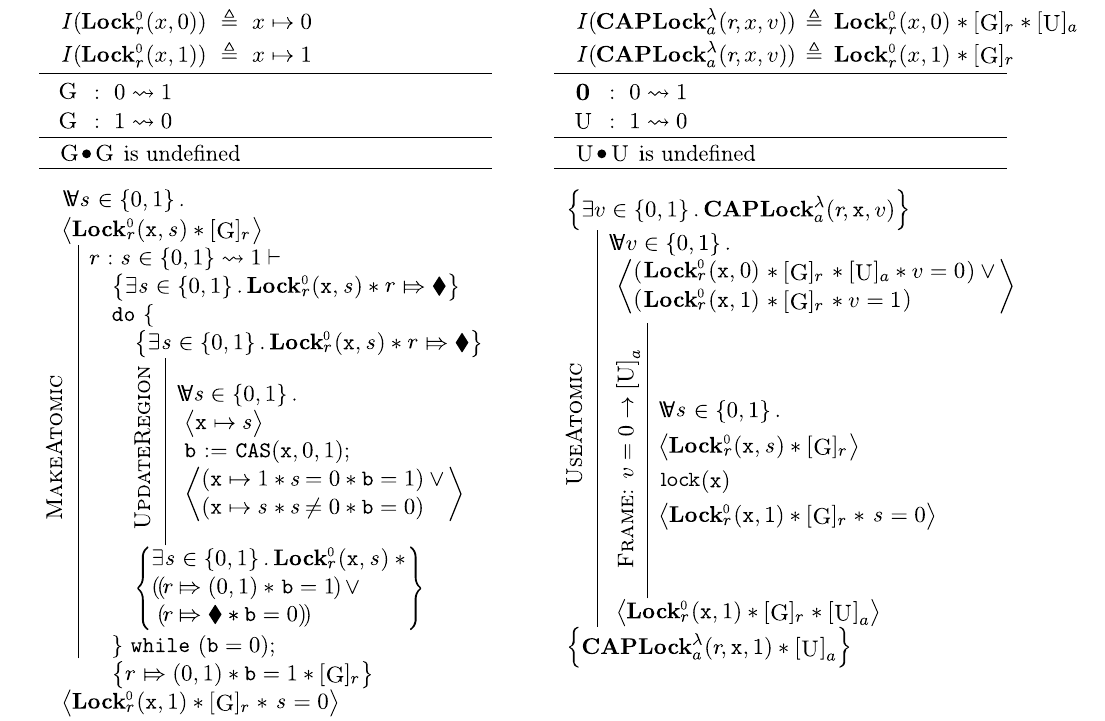}
\caption{
  TaDA declarations and proof outlines adapted from~\cite{Pinto16}: the left column repeats our running example (region \region*{Lock}, proof outline for procedure \vl!lock!), but with levels included. The right column shows the \region*{CAPLock} region and a proof outline for procedure \vl!acquire!, which build on \region*{Lock} and \vl!lock!, respectively.
  The \region*{CAPLock} abstraction provides the expected lock semantics, via its guards and actions: the vacuous zero guard \textbf{\ctext{0}} allows arbitrarily many clients to compete for the lock (\ie call \vl!acquire!), but only the holder of the unique \guardname{U} guard can release the lock again.
  Proof outlines of the latter procedures (\vl!release!/\vl!unlock! for \region*{CAPLock}/\region*{Lock}) are straightforward and have been omitted.
}
\label{fig:caplock_tada}
\end{center}
\end{figure}

The TaDA triple proved by the proof outline in the left (body of procedure \vl!lock!) and right (body of procedure \vl!acquire!) column, respectively, are the following:
\begin{gather*}
  \atctx \vdash
  \atriple*
    [s \in \{0,1\}]
    <\region{Lock}^0(x,$s$) \slstar \guard[G]>
    {lock(x)}
    <\region{Lock}^0(x,1) \slstar \guard[G] \slstar s = 0>
\\
  \lambda; \atctx \vdash
  \{\exists v \in \{0,1\} \cdot \region{CAPLock}_a^\lambda($r$,x,$v$)\}
  \text{\ \vl!acquire(x)!\ }
  \{\region{CAPLock}_a^\lambda($r$,x,1) \slstar \guard[U]_a\}
\end{gather*}

\begin{figure}[h]
  \begin{voila}
region CAPLock(id a, int lvl, id r, cell x)
  guards { 
    duplicable Z; 
    unique U;
  }
  interpretation {
    0 < lvl &&
    Lock(r, 0, x, ?v) && G@r && (v == 0 || v == 1) &&
    (v == 0 ==> U@a)
  }
  state { v }
  actions {
    Z: 0 ~> 1;
    U: 1 ~> 0;
  }

procedure acquire(id a, int lvl, id r, cell x) 
  requires CAPLock(a, lvl, r, x) && Z@a;
  ensures  CAPLock(a, lvl, r, x, 1) && U@a;
{
  use_atomic using CAPLock(a,lvl, r, x) with Z@a {
    lock(r, 0, x);
  }
}

// Repetition of our running example, but with previously omitted levels

struct cell {
  int val;
}

region Lock(id r, int lvl, cell x)
  guards { unique G; }
  interpretation {
    x.val |-> ?v && (v == 0 || v == 1)
  }
  state { v }
  actions {
    G: 0 ~> 1;
    G: 1 ~> 0;
  }

abstract_atomic procedure lock(id r, int lvl, cell x)
  interference ?s in Set(0, 1);
  requires Lock(r, lvl, x, s) && G@r;
  ensures Lock(r, lvl, x, 1) && G@r && s == 0;
  \end{voila}
\caption{
  The Voila proof outline of TaDA's CAPLock~\cite{PintoDG14}, building on our lock running example \figref{incr_voila}.
  Voila does not yet support TaDA's zero guard; instead, we use a duplicable guard \guardname{Z}.
  Following \figref{caplock_tada}, \region*{CAPLock} uses \region*{Lock} with a fixed level of 0, but Voila also verifiers a more general version, where \region*{Lock}'s level is any level smaller than \region*{CAPLock}'s.
  Also following the TaDA source, \region*{Lock}'s identifier \vl!r! is exposed as an argument to \region*{CAPLock}. An alternative would be to existentially quantify it; in Voila, this can be modeled via a ghost field of \region*{CAPLock}.
}
\label{fig:caplock_voila}
\end{figure}

\MaybeFloatBarrier
\section{Extended Discussion of our Normal Form}
\label{sec:additional_details_normal_form}

Recall from \secref{proof_candidate} that we impose a normal form on the rule triples of our proof candidate, with four main restrictions: triples are exclusively atomic or non-atomic; all triple preconditions, as well as the postconditions of non-atomic triples, are stable; in atomic triples, the state of every region in the precondition is bound by exactly one interference quantifier; and triples must hold for a range of atomicity contexts. 
The normal form is required to hold for the premises and conclusions of all syntax-driven and key rules, which allows our heuristics to exploit the restrictions when inserting applications of bridge rules. Bridge rules themselves may violate the normal form, which increases completeness.

Next, we provide additional details on the last normal form restriction: triples must hold for a heuristically determined \emph{range} of atomicity contexts $\atctx$, rather than just a single context. This stronger proof obligation rules out certain applications of \MakeAtomic\ -- which we have seen only in contrived examples -- but it increases automation substantially: most importantly, by enabling modular procedure specifications, which, as confirmed by the TaDA authors, was not possible in the original logic.

TaDA proofs require a suitable instantiation of the atomicity context $\atctx$, \ie, the set of pending region updates. Choosing a set that is too small provides weak stability guarantees and thus, leads to unnecessary weakening of assertions, whereas a set that is too large prevents certain applications of the \MakeAtomic rule. In both cases, the proof may fail even for correct programs. Moreover, for procedure specifications, it is virtually impossible to chose a single atomicity $\atctx$ that allows all possible clients to call the procedure, since each client would have to establish \emph{exactly} $\atctx$,

To overcome these problems, Voila proves triples for \emph{all} atomicity contexts within certain bounds. These bounds are inferred by proof state already present in the proof candidate, by partitioning the set of currently held region instances into two sets: the first contains all regions that the triple's statement may update; this set corresponds to the lower bound, and is manipulated according to \MakeAtomic. The second set contains all regions that the code to verify cannot update anyway; it corresponds to the upper bound, and is determined based on level information.

\MaybeFloatBarrier
\section{Extended Discussion of our Heuristics}
\label{sec:additional_details_heuristics}

\begin{figure}[t]
  \begin{center}
  \includegraphics[
    width=1.0\textwidth,
    trim={2.3cm 10.3cm 8cm 2.4cm}, 
    clip
  ]{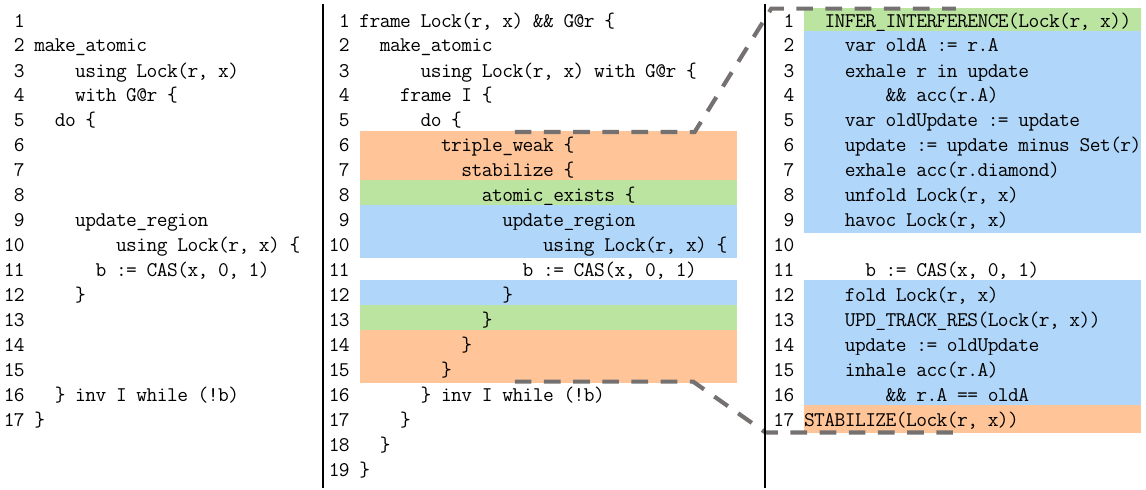}%
  \end{center}
\vspace{-5mm}
  \caption{
    Left to right: the core of our running example's \vl!lock! procedure (same
    as \figref{incr_voila}), the proof candidate with inferred bridge rules, and
    an excerpt of its Viper encoding. Colors link operations of the proof candidate to their encoding.
    \vl!I! abbreviates the loop invariant from \figref{incr_voila}.
    The encoding uses macros such as \macro{Stabilize} (more details later in this appendix) to abstract over Viper details.
  }
  \label{fig:lock_enc_overview_ext_disc}
\end{figure}

Recall from \secref{proof_candidate} that our heuristics infer bridge rule applications locally, by inspecting only adjacent rule applications that are to be composed, and their proof state. 
We employ five main heuristics: to determine when to change triple atomicity, to ensure stable frames by construction, to compute atomicity context ranges, to compute levels, and to compute interference contexts in procedure body proofs. 
The first three heuristics have been described briefly in \secref{proof_candidate}; here, we provide additional details and illustrate some of the heuristics on our running example. 
\Figref{lock_enc_overview_ext_disc} shows the Voila outline, the proof candidate, and the Viper encoding (discussed later). We visualize proof candidates by adding steps for \emph{inferred} bridge rule instantiations (\eg \vl!triple_weak!, denoting TaDA rule \TripleWeak), analogous to the user-provided key rule instantiations. For simplicity, some of the inferred steps are omitted.

\paragraph{Changing Triple Kinds.}

Atomicity changes of a triple are necessary when a non-atomic composite statement has an abstract-atomic sub-statement, such as the \vl!while! statement in \figref{caplock_tada} with its atomic body. In such cases, we apply \vl!triple_weak! (\lineref*{6} in \figref{lock_enc_overview_ext_disc}) to obtain a non-atomic triple from an atomic one. The corresponding TaDA rule \TripleWeak requires that the postcondition is stable, which we achieve via stabilization, that is, by applying a specialized TaDA entailment that weakens the postcondition to satisfy stability constructively. We denote this step with a \vl!stabilize! annotation (\lineref*{7}) in the proof candidate.

\paragraph{Framing.}

TaDA's frame rule requires the \term{frame}, \ie, the assertion preserved across a statement, to be stable. We infer frames greedily, that is, we (actually, Viper) frame as much information around a statement as soundly possible. For simple statements such as heap accesses, this approach automatically leads to stable frames. For composite statements with (arbitrary) user-provided \term{footprints} (assertions such as loop invariants describing which resources are taken into the composite statement), we need to ensure explicitly that our greedy approach does not produce an unstable frame. For this purpose, we insert explicit \vl!frame! bridge steps (\lineref*{4}) around composite statements; all other resources are then framed across, and our encoding will ensure that these frames are stable.
In our case, such composite statements are loops (invariants), calls (pre- and postconditions) and \vl!make_atomic! (using-clauses). In each case, a step \vl!frame $F$! is inserted, indicating that ``everything but footprint $F$'' will be framed across and must thus be stable.

\paragraph{Interference Contexts.}

TaDA's rules for opening a region and calling a procedure
(\OpenRegion and \ProcCall; both not necessary for our running example)
require that the state of each involved region in the precondition is bound by exactly one interference context (\tadaforall). This is not guaranteed in arbitrary TaDA proofs (where a region's state might, \eg not be bound at all), but it is in Voila, due to our normal form. 
As a consequence, no additional step is necessary before opening a region; before calling a procedure (not used by our running example), a \vl!substitution! step is inserted to check compatibility of the caller's and the callee's interference contexts.
However, the frequently necessary atomicity triple changes from non-atomic to atomic triples violate the single binder restriction of our normal form since non-atomic triples have no interference contexts; similarly, opening a region may violate the restriction since the state of nested regions is typically not bound.
To re-establish the normal form, we insert \vl!atomic_exists! steps in both cases, which automatically determine suitable interference contexts for unbound region state, \eg on \lineref*{8}, where the preceding \vl!triple_weak! changes triple atomicity.

\paragraph{Levels.}

Region levels have been omitted from the core paper, but are supported by Voila. Levels are essentially an order on region instances, and are used to prevent circular reasoning when nesting TaDA's duplicable regions. 
When a region is opened or updated, or when a procedure is called, instantiating the corresponding rule requires a specific triple level. \Eg to open a region (see rule \OpenRegion from \figref{tada_key_proof_rules}), the current level (conclusion) must be one higher than the level of the region to open.
To meet such requirements, we infer suitable instances of \LvlWeak, to changes the triple level, for every rule -- with specific level requirements -- already present in the proof candidate. Inferring and instantiating \LvlWeak is relatively straightforward, and we believe that our heuristic never fails to infer a suitable application, if one is possible.

\MaybeFloatBarrier
\section{Extended Discussion of Validating Proof Candidates in Viper}
\label{sec:additional_details_proof_candidate_validation}

Recall from \secref{viper_encoding} that proof candidates -- \ie the user-provided program with heuristically inserted bridge rule applications -- do not necessarily represent valid proofs, and that we check validity of a proof candidate by encoding it into Viper, and verifying the resulting encoding. If the candidate is invalid, the latter will fail.
In this section, we provide additional -- but still somewhat high-level -- details about the encoding of our running example. Later sections of this appendix build on this, and refine the encoding further, to provide more and more technical details.

\subsection{Primer on Viper}
\label{sec:enc_viper_primer_ext_disc}

Viper uses implicit dynamic frames~\cite{SmansJP09}, a dialect of separation logic~\cite{ParkinsonSummers12} where a points-to assertion such as \vl!x.f $\pointsto$ v! is separated into access permission \vl!acc(x.f)! and heap-dependent expressions \vl!x.f == v!.
Similarly, a separation logic predicate~\cite{ParkinsonB05} is typically represented by a Viper predicate that denotes permissions to a data structure, complemented by a heap-dependent mathematical function that abstracts over the values in the data structure.

Viper provides a simple, object-based, imperative language, which includes all statements necessary to represent TaDA programs, and makes this part of the encoding trivial. 
In addition, Viper provides two statements to manipulate assertions. For an assertion $A$, \vl!inhale $A$! adds all permissions denoted $A$ to the current state and assumes all logical constraints in $A$. Conversely, \vl!exhale $A$! asserts all logical constraints in $A$ and checks that the permissions in $A$ are available in the current state (verification fails if either check does not succeed). Moreover, it removes these permissions and assigns non-deterministic values to the corresponding memory locations (to reflect that other program components may now hold the permissions and modify the memory locations). In contrast to exhaling $A$, asserting $A$ only checks that an assertion holds (and fails otherwise), but does not~remove~permissions.

\subsection{Regions and Assertions}
\label{sec:enc_regions_assertions_ext_disc}

\begin{figure}[t]
\begin{center}
\setlength{\tabcolsep}{5pt}
\begin{minipage}[b]{0.5\textwidth}
\begin{voila}[language=silver]
$\llbracket$region R(r: id, $\overline{\texttt{p: t}}$)
  interpretation $I$
  state $S$
  guards $G$
  actions $A$$\rrbracket$ -> 
predicate R(r: Ref, $\overline{\texttt{p:} \llbracket \texttt{t} \rrbracket}$) { $\llbracket I\rrbracket$ }
  
function R_State(r: Ref, $\overline{\texttt{p:} \llbracket \texttt{t} \rrbracket}$): $T$
  requires  R(r,$\overline{\texttt{p}}$)
{ unfolding R(r,$\overline{\texttt{p}}$) in $\llbracket S\rrbracket$ }
  
$\textit{foreach}$ g($\overline{\texttt{p': t'}}$) $\in G$:
  predicate R_g(r: Ref, $\overline{\texttt{p':} \llbracket \texttt{t'} \rrbracket}$)
$\textit{end}$
  
field diamond: Bool
field R_from: $T$
field R_to: $T$

field R_X: Set[$T$]
field R_A: Set[$T$]
\end{voila}
\end{minipage}\hfill
\begin{minipage}[b]{0.5\textwidth}
\begin{voila}[language=silver]
field val: Int

predicate Lock(r: Ref, x: Ref) { 
  acc(x.val) && 
  (x.val == 0 || x.val == 1)
}

function Lock_State
             (r: Ref, x: Ref): Int
  requires Lock(r, x) 
{ unfolding lock(r, x) in x.val }


predicate Lock_G(r: Ref)


field diamond: Bool
field Lock_from: Int 
field Lock_to: Int 

field Lock_X: Set[Int]  
field Lock_A: Set[Int]  
\end{voila}
\end{minipage}
\vspace{-2mm}
\caption{Repetition of \figref{region_enc} from \secref{viper_encoding}, showing the Viper encoding of regions in the general case (left), and for the lock region from \figref{incr_voila} (right). The encoding function is denoted by double square brackets; overlines denote lists. The \textit{foreach} loop is expanded statically. Type $T$ is the type of the state expression $S$, which is inferred.
Actions $A$ do not induce any global declarations.
The elements of struct types and type \vl!id! are encoded as Viper references (type \vl!Ref!). The \vl!unfolding! expression temporarily unfolds a predicate into its definition; it is required by Viper's backend verifiers. The struct type \vl!cell! from \figref{incr_voila} is encoded as a Viper reference with field \vl!val! (in Viper, all objects have all fields declared in the program).}
\label{fig:region_enc_ext_disc}
\end{center}
\end{figure}

TaDA's regions introduce various resources such as region predicates and guards. We encode those into Viper's permissions and predicates as summarized in \figref{region_enc_ext_disc} (left). Each region \vl!R! gives rise to a predicate with the same name and parameters, which is defined by the region interpretation.
A region's abstract state may be accessed by a Viper function \vl!R_State!, which is defined based on the region's \vl!state! clause, and depends on the region predicate since the function may refer to values to which the predicate provides permissions.
Moreover, we introduce an abstract Viper predicate \vl!R_g! for each guard \vl!g! of the region; their uniqueness properties are reflected in the encoding of proof steps such as \vl!stabilize! in \figref{lock_enc_overview_ext_disc}.

These declarations allow us to encode most TaDA assertions in a fairly straightforward way. For instance, the assertion \region{Lock}_r(x, $s$) from \figref{caplock_tada} is encoded as a combination of a region predicate and the function yielding its abstract state: \vl!Lock($r$,x) && Lock_State($r$,x) == $s$!. We encode region identifiers as references in Viper, which allows us to use the permissions and values of designated fields to represent resources and information associated with a region instance. For instance, we use the permission to the \vl!diamond! field to encode the TaDA resource $\trackres<D>$. Similarly, the permissions to the fields \vl!R_from! and \vl!R_to! represent TaDA's $\trackres(x,y)$ resource, while the fields' values reflect the arguments $x$ and $y$. Therefore, $\trackres(0,1)$ from \figref{caplock_tada} is encoded as 
\ctext{acc($r$.Lock\_from) \&\& acc($r$.Lock\_to) \&\& $r$.Lock\_from == 0 \&\& $r$.Lock\_to == 1}.

Besides assertions, TaDA judgments include an interference context and an atomicity context.
An interference context of the form $\tadaforall s \in X$, associated with a region \vl!R($r$, $\ldots$)!, is represented by a field \vl!$r$.R_X!, which stores the set of values to which the environment may set the region's abstract state.
The encoding of an atomicity context $\atctx$, which tracks pending updates and prevents multiple such updates for the same region instance, is more involved. As explained in \appref{additional_details_normal_form}, we check the proof outline for all atomicity contexts within a lower and an upper bound.
The lower bound is represented by a set-typed variable \vl!update!, local to each procedure (see \figref{lock_enc_overview_ext_disc}); its value is the set of all regions currently being updated. This set is modified by \vl!make_atomic! and read by \vl!update_region!, to account for the side conditions of the corresponding TaDA rules.
The upper bound, stored in variable \vl!alevel! (omitted from \figref{lock_enc_overview_ext_disc} for simplicity), is used for verifying procedure calls: specifically, to ensure that there is not already a pending update for a region the callee might update as well.

Theoretically, procedure specifications could include the set of regions the procedure might update, but this would require additional overhead. Moreover, such a specification would in general have to include \emph{all} potentially updated regions, including those nested in other regions (which, for recursively defined regions, could be arbitrarily many).
As confirmed by the TaDA authors in personal communication, TaDA currently does not address this problem in a modular way: instead, when a proved procedure triple is used, the proof tree essentially needs to be inlined at call site, to recheck atomicity context side conditions.

For Voila, we devised a modular solution that piggybacks on TaDA's levels to overapproximate the set of regions a procedure can update: first, we determine the highest level $\lambda_\text{max}$ of all regions syntactically occurring in a procedure's precondition; this will be the procedure's level (each TaDA procedure specification triple has one) and the initial upper bound of a procedure body's, stored in \vl!alevel!. Now, due to the level-related side conditions of TaDA's proof rules, the procedure cannot update any region with a level \emph{higher} than $\lambda_\text{max}$.
During procedure body verification, \vl!alevel! is updated (by \vl!make_atomic!) to always reflect the \emph{lowest} level of any region for which an update is pending. When a call is encountered, it now suffices to check that the caller's current atomicity level (\vl!alevel!) is higher than the callee's level ($\lambda_\text{max}$) -- this guarantees that the callee will not update any region for which an update is already pending.

Lastly, the domain of an update $\atctx($\vl!r!$)$ is encoded with a set-typed field \vl!r.R_A!. Its value influences assertion stabilization: while an update is pending (\ie, inside \vl!make_atomic!), the environment may not take the region value out of \vl!r.R_A!; the latter is set to \vl!r!'s interference context (\vl!r.R_X!) when \vl!make_atomic! is entered.

\subsection{Rule Applications}
\label{sec:enc_outline_steps_ext_disc}

Recall from \secref{viper_encoding} that our proof candidates are tree structures (analogous to proof trees in standard Hoare logic), and that we check the validity of a proof candidate by checking the validity of each rule application in it. For that, we (among other things) check that the necessary triple preconditions hold, and that the executed code establishes the necessary postconditions.

\paragraph{Example.} We illustrate our encoding scheme on the body of the loop in our running example, see \figref{lock_enc_overview_ext_disc}. We discuss the proof top-down in the Hoare logic, that is, inside-out in the proof candidate and Viper encoding, starting with the \vl!CAS! statement. The \vl!CAS! statement itself is encoded as a Viper method whose specification provides the semantics of the operation. 

The proof candidate wraps the \vl!CAS! statement inside an application of the \UpdateRegion rule (the blue part in the middle column of \figref{lock_enc_overview_ext_disc}; the rule itself is shown in \figref{simplified_tada_key_rules}). \Lines*{2-6} of the Viper encoding (right column) deal with the atomicity context; we omit a detailed explanation for brevity, but recall \appref{additional_details_heuristics}. The subsequent \vl!exhale! and \vl!unfold! encode steps~1 and~2 of the rule application: instead of exhaling the entire precondition of the conclusion (step~1) and inhaling the precondition of the premise (step~2), the encoding represents only the \emph{net effect} of these two operations. Therefore, it exhales the diamond resource $\trackres<D>$. Going from the conclusion to the premise, \UpdateRegion replaces the region predicate (here, \vl!Lock($r$,x)!) by its interpretation. Given the region encoding from \figref{region_enc_ext_disc}, this is exactly what Viper's unfold operation does. Note that we instantiate the conjunct $P(x)$ in the \UpdateRegion rule to represent all other resources and properties that hold in the prestate of the rule application. Hence, it does not show up in the encoding. The subsequent \vl!havoc! operation assigns a non-deterministic value to the state of all held, \emph{still folded} \vl!Lock(r,x)! predicates. This step is necessary because TaDA region predicates are duplicable. $P(x)$ thus could contain such predicate instances (in addition to the unfolded one), and we must prevent Viper from using those instances to frame old region state around the \vl!CAS! statement, which would be unsound. As confirmed by the authors in personal communication, the latter problem is actually currently not addressed in TaDA.

The first two Viper statements after the \vl!CAS! statement (right column, \lines*{12-13}) encode steps~3 and~4 of the rule application: the fold operation replaces the interpretation of the \vl!Lock! predicate by the predicate itself. \macro{Upd\_Track\_Res} is an encoding macro (macro definitions are shown in \appref{macro_definitions_general}), which inhales, depending on the success of the \vl!CAS! operation, one of the tracking resources $\trackres(0,1)$ or $\trackres<D>$. Analogously to $P(x)$ in the precondition, we take $Q_1$ and $Q_2$ to  represent all other resources and properties that hold in the poststate of the rule application in these two cases. Since they occur in both postconditions, there is no net effect of inhaling and then exhaling them, and we can omit them from the encoding. The final two instructions (\lines*{14-16}) in the blue part of the encoding maintain the atomicity context.

Besides \UpdateRegion, the loop body contains three additional rule applications. \vl!atomic_exists! (green section) establishes the interference context, which we encode via macro \macro{Infer\_Interference}. \vl!triple_weak! (orange) weakens an atomic triple in its premise to a non-atomic triple in its conclusion. Since our encoding does not track the triple kind explicitly, \vl!triple_weak! is not directly reflected in the encoding. However, its conclusion -- like all non-atomic triples -- must be stable. This side condition is enforced in the encoding via the \macro{Stabilize} macro. We explain both stability and our treatment of interference contexts next.

\paragraph{Stability and Interference Context Inference.}

Recall that an assertion $A$ is stable if and only if the environment cannot invalidate $A$ by performing any legal region updates. In practice, this means that the environment cannot hold a guard that allows it to change the state of a region in a way that violates $A$. The challenge of checking stability as a side-condition is to avoid higher-order quantification over region
instances and guards, which is hard to automate. We address this challenge by actively \term{stabilizing} assertions in the Viper encoding. That is, we remove information from Viper's verification state such that the remaining information about the state is stable. We achieve this effect by first assigning non-deterministic values to the region state, and then constraining these to be within the states permitted by the region's transition system, taking into account the guards the environment could hold.

\begin{figure}
\begin{voila}[language=silver]
INTERFERENCE_PERMITTED(Lock(r, x), from, to) ->
     (none < perm(r.diamond) ==> Lock_State(r, x) in r.Lock_A)
  && (   from == 0 && to == 1 && ENV_MAY_HOLD(Lock_G(r))
      || from == 1 && to == 0 && ENV_MAY_HOLD(Lock_G(r)) )
   
ENV_MAY_HOLD(Lock_G(r)) -> perm(Lock_G(r)) == none

STABILIZE(Lock(r, x)) ->
  label pre_havoc
  havoc Lock(r, x)
  inhale INTERFERENCE_PERMITTED(Lock(r, x), 
      old[pre_havoc](Lock_State(r, x)), Lock_State(r, x))
      
INFER_INTERFERENCE(Lock(r, x)) ->
  havoc r.Lock_X
  inhale forall s: Int :: s in r.Lock_X 
     <==> INTERFERENCE_PERMITTED(Lock(r, x), Lock_State(r, x), s)
\end{voila}
\vspace{-3mm}
\caption{
  Encoding of stabilization and interference inference for the \vl!Lock! example.
  Viper labels enable referring to the verification state at a particular point
  in the program (\ie, they generalize \vl!old! expressions, which refer to the prestate of a method).
  We assume that symbols introduced by macros, \eg label \vl!pre_havoc!, are always fresh and never result in name clashes.
  The Viper expression \vl!perm($\rho$)! denotes the permission currently held to a resource $\rho$.
}
\label{fig:lock_enc_stabilize_and_interference_ext_disc}
\end{figure}

\figref{lock_enc_stabilize_and_interference_ext_disc} shows the encoding of stabilization
for instances of our \vl!Lock! region (macro \macro{stabilize}).
First, the region state is havocked, \ie, all information about the state is thrown away. 
Afterwards, the new region state is assumed to be any state reachable by the environment from the old state. 
We encode this property of reachability by the environment in two steps:
\macro{env\_may\_hold} yields whether a guard may be held by the environment. The
encoding depends on the guard kind: 
the environment can hold the unique guard \vl!G! only if it is not already present in the proof state.
In contrast, duplicable guards may always be held by the environment, in which case \macro{env\_may\_hold} would be defined as \vl!true!.
Building on \macro{env\_may\_hold}, \macro{interference\_permitted} encodes 
the actual reachability property: 
the environment may perform a state transition if it holds at least the guard that is required for this transition by the transition system. 
Furthermore, the transition has to stay within the atomicity context 
if an update is still pending, which is TaDA's interference rely-guarantee. 
To avoid computing the transitive closure, 
Voila requires (and checks) transition systems
to be transitively closed. 

The encoded reachability (macro \macro{interference\_permitted}) is also
essential for the inference of interference contexts.
Intuitively, the
smallest interference context, at a given program point, corresponds to the set
of states that the environment could transition to, which is exactly
the set we already need for stabilization.
Therefore, as shown in macro \macro{infer\_interference}, 
we can obtain a suitable interference context
by constraining \vl!r.Lock_X! to be the set of all states reachable by the
environment.

\subsection{Application of Built-in Viper Rules}
\label{sec:enc_framing_entailments_ext_disc}

Viper provides and automates several structural proof rules, especially the rule of consequence and the frame rule. Soundness of our encoding requires that these Viper rules are used only where permitted by TaDA\@.

TaDA's entailment rule requires entailments to be justified by view shifts~\cite{DinsdaleYoungBGPY13,Pinto16}, whereas Viper's rule of consequence may be applied for any valid entailment. We must, thus, ensure that Viper's entailment steps are indeed permitted by TaDA's entailment rule. This is the case because TaDA's view shifts impose extra requirements only on entailments that involve region and guard assertions, which are encoded as predicates in Viper. Since Viper  does not automatically (un)fold predicate instances, it cannot automatically establish entailments between region assertions. Similarly for guards: encoded as abstract predicates, Viper treats them as uninterpreted resources from which no additional information can be deduced.

Viper automatically frames information about its verification state around all statements. To ensure soundness, we explicitly remove information from the state that would otherwise be framed unsoundly, as we have illustrated with the \vl!havoc! operation in \figref{lock_enc_overview_ext_disc}.

\MaybeFloatBarrier
\section{Encoding a Custom Guard Algebra}
\label{sec:encodability_main}

Voila provides a high degree of automation, as demonstrated by our evaluation in \secref{evaluation}. For concepts not directly supported and automated, it provides various features, such as ghost code, to encode them manually. Crucially, all of these features operate on the level of Voila; programmers do not need to understand (or even be aware of) the encoding into Viper. In this section, we demonstrate Voila's support for manual encodings by an example that uses a custom guard algebra.

Specifically, we chose a TaDA-adaptation~\cite{Pinto16} of Owicki-Gries' classical parallel-increment example: given multiple threads that successively increment a shared counter in parallel, prove that the final counter state equals the sum of the local increments. 
To achieve the latter, the TaDA proof uses the custom guard algebra defined in \figref{client_guard_algebra}, which defines resources (as guards) for tracking increments, and laws that govern their use and allow relating local and total increments.
The example is included in our evaluation (\texttt{CounterCl}), and, to the best of our knowledge, cannot be encoded in any comparable tool.

\begin{figure}
\begin{gather}
\textsc{Inc}(n_1 + n_2, \pi_1 + \pi_2) = \textsc{Inc}(n_1, \pi_1) \; \bullet \; \textsc{Inc}(n_2, \pi_2) \label{def:inc_split}\\
\textsc{Total}(m) \; \bullet \; \textsc{Inc}(n, 1) \implies n = m \label{def:total_inc_equality} \\
\textsc{Total}(m) \; \bullet \; \textsc{Inc}(n, \pi) = \textsc{Total}(m+d) \; \bullet \; \textsc{Inc}(n+d, \pi) \label{def:total_inc_incr}
\end{gather}
\caption{
  Custom guard algebra (an instance of Iris' authoritative monoid~\cite{JungSSSTBD15}) used by the TaDA adaptation of Owicki-Gries' classical parallel-increment example.
  Guard \guardname{Inc} counts local increments, and can be split and merged, similar to fractional permissions~\cite{Boyland03}, in which case the local increments are split/merged as well. Guard \guardname{Total}, in contrast, is exclusive and counts the overall increments.
  Composing the whole \guardname{Inc} instance with \guardname{Total} allows concluding that the sum of the local increments equals the total count. Lastly, both values can only be changed in lockstep.  
}
\label{fig:client_guard_algebra}
\end{figure}

\figref{client_voila} shows the Voila declaration of region \vl!CClient!, whose manipulation is governed by aforementioned guard algebra.
Guards \vl!INC! and \vl!TOTAL! are declared as \vl!manual! to indicate that they are not part of a guard algebra that Voila automates (see also \appref{supported_tada_ingredients}). In particular, this means that Voila will not make any uniqueness assumptions about these guards, \eg when stabilizing region state.
The laws of the guard algebra are encoded as lemma procedures such as \vl!INC_split!, which encodes the left-to-right direction of definition \ref{def:inc_split}.
Region \vl!CClient! abstracts over the shared \vl!Counter(r,n,x)!, whose value \vl!n! corresponds to the total increment count; guard \vl!G!, declared by region \vl!Counter! (see \figref{client_sequential_voila}), is needed to increment that value.
The region's \vl!actions! clause demonstrates Voila's most general syntax for specifying region transitions, and declares that the region state can be incremented from any \vl!n! to any larger \vl!m!, by anybody holding a non-zero fraction of \vl!INC! (regardless of the latter's local increments value \vl!k!).
\figref{client_voila} also shows the specification of procedure \vl!single_client!, whose implementation (shown in \figref{client_sequential_voila}) loops until it made \vl!v! successive increments to the shared counter. 
Note that \vl!single_client! could be parametric in the permission amount required for \vl!INC! (currently fixed to \vl!1/2!), which would allow arbitrarily many parallel instances (\eg \vl!1/t! for a statically-unknown number of \vl!t! threads).

\begin{figure}
\begin{voila}
region CClient(id s, id r, cell x)
  guards { 
    manual INC(int, frac); 
    manual TOTAL(int);
  }
  interpretation {
    Counter(r, x, ?n) && G@r && TOTAL(n)@s
  }
  state { n }
  actions {
    ?n, ?m, ?k, ?p | 0f < p && n < m | INC(k, p): n ~> m;
  }
  
lemma INC_split(id s, int k1, int k2, frac p1, frac p2)
  requires INC(k1 + k2, p1 + p2)@s;
  requires 0f < p1 && 0f < p2;
  ensures  INC(k1, p1)@s && INC(k2, p2)@s;

procedure single_client(id s, id r, cell x, int m)
  requires CClient(s, r, x, _) && INC(0, 1/2)@s;
  ensures  CClient(s, r, x, _) && INC(m, 1/2)@s;
\end{voila}
\caption{
  Example declarations from the Voila encoding of TaDA's counter-client example, including the \vl!CClient! region, and the signature of procedure \vl!single_client! (see also \figref{client_sequential_voila}), which is executed by each thread.
  Lemma procedure \vl!INC_split! encodes the left-to-right direction of definition \ref{def:inc_split} from \figref{client_guard_algebra} by means of pre- and postconditions. The remaining algebra laws are encoded analogously, and omitted for brevity.  
}
\label{fig:client_voila}
\end{figure}

\figref{client_parallel_voila} shows the central part of the verified code: first, guard \vl!INC(0,0)! is split into two equal fractions by using lemma procedure \vl!INC_split!; afterwards, two calls to \vl!single_client! are run in parallel.
Upon termination, lemma procedure \vl!INC_merge! (whose straightforward declaration we omitted), corresponding to the right-to-left direction of guard algebra definition \ref{def:inc_split}, is used to combine the \vl!INC! guards obtained from the postconditions of \vl!single_client! into a single instance \vl!INC(20,1f)!.
Subsequent ghost code then opens (unfolds) region \vl!CClient! to bind the -- at this point unknown -- value of the counter to the logical variable \vl!n!.
Finally, lemma procedure \vl!TOTAL_INC_equality!, corresponding to guard algebra definition \ref{def:total_inc_equality}, is used to learn that \vl!n!'s value is equal to \vl!INC!'s value, \ie \vl!20!.
Note that the lemma application would (here) fail for values other than \vl!20!, and that it is possible to work with statically unknown values, \eg \vl!m1!, \vl!m2! and \vl!m1 + m2! instead of constants \vl!9!, \vl!11! and \vl!20!.

\begin{figure}
\begin{voila}
// Allocate memory and create region instances ...

use INC_split(s, 0, 0, 1/2, 1/2);

parallel {
  single_client(s, r, x, 9);
  single_client(s, r, x, 11);
}

use INC_merge(s, 9, 11, 1/2, 1/2);

unfold CClient(s, r, x);
assert Counter(r, x, ?n);

use TOTAL_INC_equality(s, n, 20);
assert n == 20;

// ... destroy region instances and deallocate memory
\end{voila}
\caption{
  The central part of the Voila encoding of TaDA's Owicki-Gries adaptation: We use a lemma method to split the guard \vl!INC! before the parallel execution of two calls to \vl!single_client!. After the calls, another lemma method is used to recombine \vl!INC! and to sum up the local increments. Finally, we assert the equality between local and total increments.
}
\label{fig:client_parallel_voila}
\end{figure}

\begin{figure}
\begin{voila}
struct cell {
  int f;
}

region Counter(id r, cell x)
  guards { unique G; }
  interpretation { x.f |-> ?n }
  state { n }
  actions { ?n, ?m | n < m | G: n ~> m; }

abstract_atomic procedure incr(id r, cell x)
  interference ?n in Int;
  requires Counter(r, x, n) && G@r;
  ensures Counter(r, x, n + 1) && G@r;

procedure single_client(id s, id r, cell x, int m)
  requires CClient(s, r, x, _) && INC(0, 1/2)@s;
  ensures  CClient(s, r, x, _) && INC(m, 1/2)@s;
{
  int i := 0;

  while (i < m)
    invariant CClient(s, r, x, _);
    invariant INC(i, 1/2)@s;
  {
    use_atomic
      using CClient(s, r, x, ?v) with INC(i, 1/2)@s;
    {
      incr(r, x);
      use TOTAL_INC_inc(s, v, i, 1/2);
    }

    i := i + 1;
  }
}
\end{voila}
\caption{
  Further Voila code from the parallel counter example: the \vl!Counter! region and its \vl!incr! procedure, and the implementation of the \vl!single_client! procedure that is executed by each thread. We slightly simplified the loop invariant by omitting obvious properties.
  The body of \vl!incr!, omitted for brevity, is similar to procedure \vl!lock! from our running example in \figref{incr_voila}: a loop around a \vl!CAS! that attempts to increment the counter by one.
}
\label{fig:client_sequential_voila}
\end{figure}

In addition to lemma procedures, Voila provides several ghost operations for manipulating its verification state, including: in-/exhale statements for gaining/giving up resources; unfold/fold statements for opening/closing regions; but also region ghost fields, \eg for witnessing existentials.
All of these can be used to encode TaDA proof steps that are beyond what Voila automates, and to experiment with potential extensions.
Ghost operations are always applied on the Voila level such that users do not need to be aware of the encoding into Viper.

\MaybeFloatBarrier
\section{Soundness}
\label{sec:soundness_sketch_main}

In this section, we briefly explain how to show that our approach is sound \wrt TaDA. A comprehensive proof sketch is available in \appref{soundness}.
Recall our high-level approach: 
we take a Voila procedure proof outline with precondition, body, and postcondition, expand it into a proof candidate by adding further rule applications, encode the proof candidate into Viper, and verify the resulting Viper program. To prove soundness, we need to show that for each proof outline successfully verified in this approach, there exists a derivation in TaDA for a TaDA triple whose precondition, statement, and postcondition correspond to those in Voila. For this proof, we assume that the Viper verification backend verifiers are sound; showing their soundness is an orthogonal concern. 

We establish soundness in two main steps: we first prove a lemma that relates the execution of Viper statements and states to TaDA judgments and derivations. In a second step, we instantiate this lemma to show that, for a verified Voila procedure, there indeed exists a corresponding TaDA derivation.

Intuitively, our lemma expresses the following property: given a Viper prestate, a sequence of Viper statements corresponding to an encoded Voila statement, and the Viper poststate determined by executing the sequence of Viper statements, we can derive a valid TaDA triple. The proof goes by induction on the (program and rule) statements of the proof candidate. It maps Viper states to TaDA assertions, and uses the statements and rule applications in the proof candidate to construct a TaDA proof.
To enable the mapping from Viper states to TaDA assertions, we prove that our encoding maintains several invariants on Viper states.
Some of these invariants are due to Voila's normal form, for instance, that Viper states correspond to stable TaDA assertions for pre- and non-atomic postconditions.
Others are due to global properties of TaDA: \eg that a region under update (diamond resource is held) is always in the current atomicity context, which is a prerequisite for a successful TaDA proof.
Yet other invariants are technicalities enabling the mapping from Viper states to TaDA assertions, such as having either no or full permission (rather than arbitrary fractions) to certain Viper fields and predicates.

The above lemma relates the Viper encoding to a derivation of a TaDA triple. What remains to be shown is that this triple actually corresponds to the Voila procedure we encoded.
For the triple's precondition and statement, this correspondence is ensured by construction, since we obtain them from the Voila proof candidate. The latter also implies that the initial Viper state (corresponding to the Voila precondition) satisfies aforementioned invariants.
For the triple's postcondition (and to conclude the soundness proof), we need to show that the user-provided Voila postcondition is implied by the TaDA postcondition that we obtained from the final Viper state. This is also ensured by construction, since the last Viper statement asserts the Voila postcondition.

\MaybeFloatBarrier
\section{Macro Definitions for our Running Example}
\label{sec:macro_definitions}

In \appref{additional_details_proof_candidate_validation}, an overview of our encoding was given, which utilized macros to structure the encoding, and to abstract over the generated Viper code. If a verification backend other than Viper were to be chosen, the macro definitions would most likely have to be adapted, but (probably) not how these macros are combined.

In this section, we present the definitions of several core macros, \ie
the Viper code they expand to:
macros \macro{action\_permitted} and \macro{less} are concerned with checking
that a state transition is valid and enabled by held guards;
\macro{interference\_permitted} and \macro{stabilize} simulate environment
interference and ensure stable assertion, respectively;
and \macro{infer\_interference} is crucial for reducing user-required
annotations, by inferring all internal interference contexts.

The definitions shown in this section have been instantiated for our running
example (the lock region), and the corresponding explanations refer to the
running example to build up intuition. Subsequently, section
\appref{macro_definitions_general} show all macros, in their general form.

\subsection{Transition System Compliance}
\label{sec:macro_definitions_compliance}

Recall that \MakeAtomic (\cf \figref{simplified_tada_key_rules} and
\figref{tada_key_proof_rules}; likewise for \UseAtomic) requires checking that a
region state change is permitted by the region's transition system, using a
particular guard; in our example, the update from 0 to 1 (and vice versa) using
guard \guardname{G}.
In general, checking compliance of a state change requires showing that there
exists a region transition
\begin{enumerate*}
\item
  that can be instantiated such that its pre- and poststate match the performed
  state change, and
\item
  that is enabled by a specific guard (the one specified in the proof outline).
\end{enumerate*}
 
\figref{lock_enc_compliance} shows how macro \macro{action\_permitted} encodes
these two requirements (as previously mentioned, the shown macro definition is
specific to the running example's \region*{Lock} region):
since the number of transition options (\vl!actions!) is always finite, a
disjunction of the different options suffices.
A transition option with specified guard $g'$ is enabled by a guard $g$ if,
according to the guard algebra, guard $g$ entails $g'$; this is encoded by
macro \macro{less}.
Encoding this guard entailment for the algebra of guard \guardname{G}
from our running example is straightforward: \guardname{G} is entailed by
itself, potentially combined with other guards (\eg in larger examples).
In general, the definition of \macro{less} is more involved since Voila supports
more complex guard algebras (recall \figref{voila_features}), but the general
encoding is similar: \eg given a fractional guard algebra, $g$ entails $g'$ if
$g$'s fraction is larger.

\begin{figure}
\begin{voila}[language=silver]
ACTION_PERMITTED(Lock, from, to, g) ->
      from == to
   || from == 0 && to == 1 && LESS(Lock_G, g)
   || from == 1 && to == 0 && LESS(Lock_G, g)

LESS(Lock_G, Lock_G && _) -> true
LESS(_     , _          ) -> false
\end{voila}
\caption{
  Definition of macros \macro{action\_permitted} and \macro{less}, instantiated
  for our running example.
  For simplicity, macro definitions utilize structural pattern matching as known
  from, \eg Haskell. 
  \macro{less} encodes guard entailment, \ie \macro{less}\texttt{($g$, $g'$)} is
  true if, according to the guard algebra, guard $g'$ entails $g$. 
  For brevity, the definition is shown modulo commutativity of guard
  composition (\ie the case for \texttt{\_ \&\&} \guardname{G} is omitted).
}
\label{fig:lock_enc_compliance}
\end{figure}

\subsection{Environment Interference and Assertion Stability}
\label{sec:macro_definitions_stabilization}

Recall from \appref{additional_details_proof_candidate_validation} that verifier state is
stabilized by simulating possible transitions that the environment is
permitted to perform.
This simulation is a constructive approach to satisfying TaDA's assertion
stability requirement: 
an assertion $A$ is stable if, for each region instance $R$ and for each guard
(in general, combination of guards) $g$ potentially held by the environment, $A$
does not contradict $R$ being in any state reachable with $g$.
Our constructive approach eliminates the need for higher-order quantifications
over region instances and guards -- which are typically not supported by
automated verification backends.

\figref{lock_enc_actions} shows the definition of macro \macro{stabilize}, which
encodes assertion stabilization, and three helper macros:
\begin{enumerate*}
  \item 
    \macro{interference\_permitted} is similar to \macro{action\_permitted}, and
    states that the environment is permitted to perform state transitions if it
    \emph{could} hold the necessary guards.
  \item
    Correspondingly, \macro{env\_may\_hold} encodes if the environment could
    hold certain guards: \eg only if not held by the current context, for unique
    guards such as \guardname{G}; and always, for duplicable guards (not used
    here).
    The Viper expression \vl!perm($\rho$)! denotes the permission amount
    currently held to a resource $\rho$.
  \item
    Finally, \macro{stabilize} stabilizes verification state by first havocking
    a region's state, followed by constraining it to be any state reachable (by
    the environment) from the pre-havoc state.
    To avoid a fixpoint computation, Voila requires (and checks) that state
    transition systems are transitively closed.
\end{enumerate*}

\begin{figure}
\begin{voila}[language=silver]
INTERFERENCE_PERMITTED(Lock(r, x), from, to) ->
     (none < perm(r.diamond) ==> Lock_State(r, x) in r.Lock_A)
  && (   from == 0 && to == 1 && ENV_MAY_HOLD(Lock_G(r))
      || from == 1 && to == 0 && ENV_MAY_HOLD(Lock_G(r)) )
   
ENV_MAY_HOLD(Lock_G(r)) -> perm(Lock_G(r)) == none

STABILIZE(Lock(r, x)) ->
  label pre_havoc
  havoc Lock(r, x)
  inhale INTERFERENCE_PERMITTED(Lock(r, x),
            old[pre_havoc](Lock_State(r, x)), Lock_State(r, x))
\end{voila}
\caption{
  Encoding of stabilization, split into three macros.
  Viper labels enable referring to the verification state at a particular point
  in the program (\ie they generalize \vl!old! expressions, which refer to the
  state in which the precondition held).
  The Viper expression \vl!perm($\rho$)! denotes the permission amount currently
  held to a resource $\rho$, here to predicate instance \vl!Lock_G(r)!.
}
\label{fig:lock_enc_actions}
\end{figure}

The first line of \macro{interference\_permitted} accounts for the TaDA's
property that, while an update is pending (\ie before the linearization point is
reached), the environment may not take a region \vl!r! outside the current
procedure's interference context \vl!r.X!.
More details about the latter are provided in subsection
\appref{macro_definitions_interference_inference}.

Lastly, note that any assertion can be checked for stability by inhaling it,
stabilizing it, and asserting it; this is done by Voila for region
interpretations, procedure specifications and loop invariants, all of which must
be stable.

\subsection{Interference Context Inference}
\label{sec:macro_definitions_interference_inference}

In TaDA, every rule is parametrized with an interference context (denoted by $X$
in the proof rules, see \eg \figref{tada_key_proof_rules}) for \emph{atomic}
triples, but not for \emph{non-atomic} ones.
As a consequence, when going from non-atomic triples to atomic triples, \eg when
sequentially composing atomic statements, a potentially different interference
context is newly required.

In Voila, users only need to specify interference contexts once (as part of a
procedure's signature), whereas all other interference contexts are inferred,
via macro \macro{infer\_interference}.
More specifically, we infer the smallest interference context (at a given
program point) that accounts for all possible environment transitions -- which
is exactly the set we already need for stabilizing Viper's verification state.
Consequently, macro \macro{infer\_interference}, shown in
\figref{lock_enc_interference_context}, determines a lock's interference context
by first havocking the corresponding field and then constraining the context to
exactly those states that the environment could reach.

Note that TaDA in principle allows arbitrarily small interference contexts, but
we have not yet found an example where our inference heuristic prevented a
successful verification. Furthermore, note that initial interference contexts
from procedure preconditions still influence (in particular, restrict) inferred
contexts, but only indirectly, via the encoding of \MakeAtomic.

In addition to inferring intermediate interference contexts, Voila also
automatically propagates interference contexts to nested regions (region
assertions occurring in another region's interpretation), which are not even
visible in procedure specifications.
To illustrate how interference contexts are propagated to nested regions,
consider a double counter region \region*{DCounter}(r,x,y) whose interpretation
contains two counters \region*{Counter}(r1, x) and \region*{Counter}(r2, y), and
whose region state is the sum of the individual counter states.
When opening the \region*{DCounter}, we constrain the interference context of
counters \vl!r1! and \vl!r2! to contain value $s_1$ and $s_2$, respectively, iff
\region*{DCounter}'s interference context contains $s_1 + s_2$.
This approach generalizes straightforwardly to more complex situations.

\begin{figure}
\begin{voila}[language=silver]
INFER_INTERFERENCE(Lock(r, x))
  havoc r.Lock_X
  inhale forall s: Int :: s in r.Lock_X 
     <==> INTERFERENCE_PERMITTED(Lock(r, x), Lock_State(r, x), s)
\end{voila}
\caption{
  Viper encoding of interference context inference: a region's interference
  context \vl!r.Lock_X! is inferred to be the set of states the environment could
  currently reach.
}
\label{fig:lock_enc_interference_context}
\end{figure}

%
%

\MaybeFloatBarrier
\section{General Macro Definitions}
\label{sec:macro_definitions_general}

This section presents all encoding macros, in their general form; we suggest to
read \appref{macro_definitions} first, to build up an intuition for the
encoding. The macros are also referenced from the the soundness sketch shown in
\appref{soundness}.

\begin{figure}
\newcommand*{\mxA}{\many{\ctext{x}_\ctext{A}}}
\begin{voila}[language=silver, mathescape=true]
ACTION_PERMITTED(R, from, to, g) ->
  from == to || ($\forone{\ctext{A}}{\actionset*{\ctext{R}}}:$
    exists $\mxA$ :: from == from$_\ctext{A}$($\mxA$) && to == to$_\ctext{A}$($\mxA$) 
      && c$_\ctext{A}$($\mxA$) && LESS(g$_\ctext{A}$($\mxA$), g)
  )
   
INTERFERENCE_PERMITTED(R(r, $\many{\ctext{p}}$), from, to) ->
       (none < perm(r.diamond) ==> to in r.R_A)
    && from == to || ($\forone{\ctext{A}}{\actionset*{\ctext{R}}}:$
          exists $\mxA$ :: from == from$_\ctext{A}$($\mxA$) && to == to$_\ctext{A}$($\mxA$) 
            && c$_\ctext{A}$($\mxA$) && ENV_MAY_HOLD(g$_\ctext{A}$($\mxA$))
       )

LESS(g, g') -> ...

ENV_MAY_HOLD(g) -> ...

@\codenote{%% IMPORTANT: Avoid hard line-wraps in \codenote's argument
  where \ctext{R} denotes a region name (\eg \ctext{Lock}), and \ctext{R(r, $\many{\ctext{p}}$)} a region instance with identifier \ctext{r} and remaining arguments $\many{\ctext{p}}$, and where \ctext{from$_\ctext{A}$($\mxA$)} denotes the expression \ctext{from$_\ctext{A}$}, but with $\mxA$ substituted for the free (quantified) variables that occur in \ctext{from$_\ctext{A}$}
}@

\end{voila}
\caption{
  General definitions of macros \macro{action\_permitted} and
  \macro{interference\_permitted} (whereas the versions shown in
  \figref{lock_enc_compliance} were instantiated for our running example).
  \macro{action\_permitted} encodes if a transition is valid, given a specific
  guard; \macro{interference\_permitted} encodes interference the environment
  could cause.
  Macros \macro{less} (is a guard entailed by another?) and
  \macro{env\_may\_hold} (could the environment hold a particular guard?) are
  specific per supported guard algebra, and have been omitted for brevity.
  \protect\actionset*{\ctext{R}} denotes the finite set of actions (transitions)
  declared by region \ctext{R}. 
  Macro function 
  ``\ctext{$\forone{\ctext{A}}{\protect\actionset*{\ctext{R}}}: \ctext{E(A)}$}''
  expands to an iterated disjunction \ctext{E(A$_0$) || E(A$_1$) || \ldots}.
  Types (\eg for $\mxA$) have been omitted for brevity, they can be
  unambiguously inferred from, \eg involved regions.
  Given an action \ctext{A}, the expressions \ctext{c$_\ctext{A}$},
  \ctext{g$_\ctext{A}$}, \ctext{from$_\ctext{A}$} and \ctext{to$_\ctext{A}$}
  denote the four components an action declaration comprises.
}
\label{fig:encoding_transition_system}
\end{figure}

\begin{figure}
\begin{voila}[language=silver, mathescape=true]
STABILIZE(R) ->
  label pre_stabilize
  havoc forall r, $\many{\ctext{p}}$ :: none < perm(R(r, $\many{\ctext{p}}$)) ==> R(r, $\many{\ctext{p}}$)
  inhale forall r, $\many{\ctext{p}}$ :: none < perm(R(r, $\many{\ctext{p}}$)) ==>
    INTERFERENCE_PERMITTED(
      R(r, $\many{\ctext{p}}$),
      old[pre_stabilize](R_state(r, $\many{\ctext{p}}$)), R_state(r, $\many{\ctext{p}}$))
  
INFER_INTERFERENCE(R) ->
  havoc forall r :: r != null ==> r.R_X
  inhale forall r, $\many{\ctext{p}}$, s :: none < perm(R(r, $\many{\ctext{p}}$)) ==> (
    s in r.R_X <==> INTERFERENCE_PERMITTED(R(r, $\many{\ctext{p}}$), R_state(r, $\many{\ctext{p}}$), s)
  )
  
LINK_INTERFERENCE(R(r, $\many{\ctext{p}}$), s) ->
  label pre_link
  havoc ($\forevery{\ctext{c}}{C}:$ c.R$_\ctext{c}$_X)
  inhale forall $\many{\ctext{m}_{{C}}}$ :: 
    ($\forevery{\ctext{c}}{C}:$ $\ctext{m}_{{c}}$ in c.R$_\ctext{c}$_X)
      <==> 
    ($\metafunc*{StateFunction}$(R(r, $\many{\ctext{p}}$), $\many{\ctext{m}_{{C}}}$) in r.R_X)
  s
  havoc ($\forevery{\ctext{c}}{C}:$ c.R$_\ctext{c}$_X)
  inhale ($\forevery{\ctext{c}}{C}:$ c.R$_\ctext{c}$_X == old[pre_link](c.R$_\ctext{c}$_X))

  @\codenote{%% IMPORTANT: Avoid hard line-wraps in \codenote's argument
    where $C$ is the finite set of region identifiers (\eg \ctext{r'}) that occur in the interpretation of \ctext{R(r, $\many{\ctext{p}}$)}, and \ctext{R$_\ctext{c}$} is the region name associated with region identifier \ctext{c}%
  }@
\end{voila}
\caption{
  General definitions of macros \macro{stabilize} and
  \macro{infer\_interference} (whereas the versions shown in
  \figref{lock_enc_stabilize_and_interference_ext_disc} where instantiated for our
  running example), and of \macro{link\_interference}.
  \macro{stabilize} accounts for potential environment interference and ensures
  that only stable facts can be deduced.
  \macro{infer\_interference} infers interference contexts and
  \macro{link\_interference} binds the interference contexts of regions nested
  in another region instance's interpretation.
  Intuitively, \macro{link\_interference} propagates constraints on a nesting
  region's interference contexts to the interference contexts of the nested
  regions.
  Recall that \ctext{r.R\_X} (\eg \ctext{r.Lock\_X}) is the interference context
  of an instance of a region \ctext{R} with identifier \ctext{r}.  
%
  Macro function ``\ctext{$\forevery{\ctext{c}}{C}: E(c)$}''
  expands to an iterated conjunction \ctext{E(c$_0$) \&\& E(c$_1$) \&\& \ldots}.
  Similarly, $\many{\ctext{m}_{{C}}}$ expands to \ctext{m$_{\ctext{c$_0$}}$, m$_{\ctext{c$_1$}}$, \ldots}, one variable \ctext{m$_\ctext{c}$} for each $\ctext{c} \in C$.
  \ctext{$\metafunc*{StateFunction}$(R(r, $\many{\ctext{p}}$), $\many{\ctext{m}_{{C}}}$)} denotes the state of \ctext{R(r, $\many{\ctext{p}}$)} where the state of each region $\ctext{c} \in C$  occurring in the region interpretation is substituted by $\ctext{m}_{{c}}$.
  \Eg consider a region \ctext{Sum(r, $\many{\ctext{p}}$)} with an interpretation \ctext{Cell(c$_1$, $\many{\ctext{p}_1}$, ?a) \&\& Cell(c$_2$, $\many{\ctext{p}_2}$, ?b)} and the state clause \ctext{a + b}. Then, \ctext{$\metafunc*{StateFunction}$(Sum(r, $\many{\ctext{p}}$), m$_{\ctext{c$_1$}}$, m$_{\ctext{c$_2$}}$)} is \ctext{m$_{\ctext{c$_1$}}$ + m$_{\ctext{c$_2$}}$}.
  For this \ctext{Sum} example, the first \ctext{inhale forall} in the definition of \macro{link\_interference} would be instantiated as \ctext{forall m$_{\ctext{c$_1$}}$, m$_{\ctext{c$_2$}}$ :: (m$_{\ctext{c$_1$}}$ in \ctext{c$_1$}.Cell\_X \&\& m$_{\ctext{c$_2$}}$ in \ctext{c$_2$}.Cell\_X) <==> ((m$_{\ctext{c$_1$}}$ + m$_{\ctext{c$_2$}}$) in r.Sum\_X)}. 
}
\label{fig:encoding_stabilization}
\end{figure}

\begin{figure}
\begin{voila}[language=silver]
ATOMIC(s) ->
  label pre_atomic
  $\foreach{\ctext{R}}{\metafunc*{Regions}} ~ \foreachStart$
    INFER_INTERFERENCE(R)
  $\foreachEnd$
  s
  $\foreach{\ctext{R}}{\metafunc*{Regions}} ~ \foreachStart$
    havoc forall r :: r != null ==> r.R_X
    inhale forall r :: r != null ==> r.R_X == old[pre_atomic](r.R_X)
    STABILIZE(R)
  $\foreachEnd$
\end{voila}
\caption{
  General definition of macro \macro{Atomic}, which encodes changing a
  non-atomic to an atomic triple and establishing the interference context
  in accordance with our normal form. 
%
  Macro function
  ``\ctext{$\foreach{\ctext{R}}{\metafunc*{Regions}} ~ \foreachStart\ \ctext{S(R)}\ \foreachEnd$}''
  expands to an iterated sequential composition of Viper statements
  \ctext{S(R$_0$); S(R$_1$); \ldots}, and \metafunc{Regions} denotes the finite
  set of regions declared by the current Voila program.
}
\label{fig:encoding_atomicity_change}
\end{figure}

\begin{figure}
\begin{voila}[language=silver]
CALL($\many{y}$ := M($\many{e}$)) ->
  label pre_call
  $\foreach{l}{\metafunc*{Levels}(\ctext{M})} ~ \foreachStart$
    assert $\level$ > $l$ && $\alevel$ > $l$
  $\foreachEnd$  
  var $\many{z}$ := $\many{e}$
  exhale Pre$_\ctext{M}$[$\many{z}/\many{x}$]
  $\foreach{\ctext{R}}{\metafunc*{Regions}} ~ \foreachStart$
    STABILIZE(R)
  $\foreachEnd$
  havoc $\many{y}$
  inhale Post$_\ctext{M}$[$\many{z}/\many{x}$][$\many{y}/\many{r}$][old[pre_call]/old]

CALL_ATOMIC($\many{y}$ := M($\many{e}$)) ->
  label pre_call
  $\foreach{Q}{\metafunc*{Inter}(\ctext{M})} ~ \foreachStart$
    assert r$_Q$.R$_Q$_X subset S$_Q$
  $\foreachEnd$
  $\foreach{l}{\metafunc*{Levels}(\ctext{M})} ~ \foreachStart$
    assert $\level$ > $l$ && $\alevel$ > $l$
  $\foreachEnd$
  var $\many{z}$ := $\many{e}$
  exhale Pre$_\ctext{M}$[$\many{z}/\many{x}$]
  $\foreach{\ctext{R}}{\metafunc*{Regions}} ~ \foreachStart$
    STABILIZE(R)
  $\foreachEnd$
  havoc $\many{y}$
  inhale Post$_\ctext{M}$[$\many{z}/\many{x}$][$\many{y}/\many{r}$][old[pre_call]/old]

@\codenote{%% IMPORTANT: Avoid hard line-wraps in \codenote's argument
  where $\many{x}$ and $\many{r}$ are procedure \ctext{M}'s formal in- and out-arguments, respectively, and where $e[a/b]$ denotes syntactic substitution of $a$ with $b$ in $e$
}@
\end{voila}
\caption{
  General definitions of macros \macro{CALL} and \macro{CALL\_ATOMIC}.
  \ctext{Pre\_M} and \ctext{Post\_M} denote \ctext{M}'s pre- and postcondition,
  respectively.
  Viper variables \ctext{level} and \ctext{alevel} track the current judgment
  and atomicity level, respectively.
  \metafunc{Levels}(\ctext{M}) denotes the set of all levels that (directly)
  occur in the precondition of procedure \ctext{M}; they effectively determine
  the level of the procedure to be called, and thus must be smaller than the
  current levels.
  \metafunc{Inter}(\ctext{M}) denotes the set of interference clauses of
  procedure \ctext{M}. Set \ctext{S$_Q$} denotes the interference set itself (\eg
  \ctext{Set(0, 1)} in our running example), and \ctext{R$_Q$} and \ctext{r$_Q$}
  denote the region name and identifier (\eg \ctext{Lock} and \ctext{r}) that
  identify the constrained region instance, respectively.
  Calls to non-atomic procedures are encoded in the expected way, aside from the
  levels check and the stabilization of the frame. Invocations of atomic
  procedures are encoded analogously, with the additional check that the
  caller's interference contexts may not allow more interference than the callee
  permits.
}
\label{fig:encoding_call_atomic}
\end{figure}

\begin{figure}
\begin{voila}[language=silver]
UPDATE_REGION(R(r, l, $\many{\ctext{p}}$), s) ->
  label pre_update

  assert $\level$ > l
  var $\level$_store := $\level$
  $\level$ := l  
  
  exhale r in $\actxtDomain$ && acc(r.R_A)
  var $\actxtDomain$_store := $\actxtDomain$
  $\actxtDomain$ := $\actxtDomain$ minus Set(r)
  
  exhale acc(r.diamond)
  
  unfold R(r, l, $\many{\ctext{p}}$)
  havoc R(r, l, $\many{\ctext{p}}$) // Havoc other instances possibly held
  LINK_INTERFERENCE(R(r, l, $\many{\ctext{p}}$), s)
  fold R(r, l, $\many{\ctext{p}}$)
  
  // Note: the following if-else statement is abbreviated as 
  // UPD_TRACK_RES in @\SilverCommentStyle\figref{lock_enc_overview_ext_disc}@
  if (R_state(r, l, $\many{\ctext{p}}$) == old[pre_update](R_state(r, l, $\many{\ctext{p}}$))) {
    inhale acc(r.diamond)
  } else {
    inhale acc(r.R_from) && r.R_from == old[pre_update](R_state(r, l, $\many{\ctext{p}}$))
    inhale acc(r.R_to) && r.R_to == R_state(r, l, $\many{\ctext{p}}$)
  }
  
  $\actxtDomain$ := $\actxtDomain$_store
  inhale acc(r.R_A) && r.R_A == old[pre_update](r.R_A)
  
  $\level$ := $\level$_store
\end{voila}
\caption{%
  General definition of macro \macro{update\_region}. Statement \vl!s! is executed as part of the expansion of \macro{link\_interference}.
  Since we are interested in the updated region's level, the pattern match in
  the macro's signature is \ctext{R(r, l, $\many{p}$)}, \ie the level \ctext{l}
  is split off from the remaining arguments \ctext{$\many{p}$} (analogous to the
  region identifier \ctext{r}).
  Viper variable \ctext{update} tracks the set of region identifiers for which
  an atomic update is pending, and Viper fields \ctext{r.R\_from} and
  \ctext{r.R\_to} record the performed update; see also macro
  \macro{make\_atomic} in \figref{encoding_make_atomic}.
  The \ctext{if-else} statement heuristically resolves an angelic choice, which
  is not supported by Viper: a region update is assumed to have happened if the
  region state changed. See also \secref{evaluation}.
  Recall that \ctext{r.R\_A} is the domain of the atomicity context for a region \ctext{R} with identifier \ctext{r}.
}
\label{fig:encoding_update_region}
\end{figure}

\begin{figure}
\begin{voila}[language=silver]
OPEN_REGION(R(r, l, $\many{\ctext{p}}$), s) ->
  label pre_open

  assert $\level$ > l
  var $\level$_store := $\level$
  $\level$ := l  
  
  unfold R(r, l, $\many{\ctext{p}}$)
  havoc R(r, l, $\many{\ctext{p}}$) // Havoc other instances possibly held
  LINK_INTERFERENCE(R(r, l, $\many{\ctext{p}}$), s)
  fold R(r, l, $\many{\ctext{p}}$)
  
  assert R_state(r, l, $\many{\ctext{p}}$) == old[pre_open](R_state(r, l, $\many{\ctext{p}}$))
  
  $\level$ := $\level$_store
\end{voila}
\caption{%
  General definition of macro \macro{open\_region}. Statement \vl!s! is executed as part of the expansion of \macro{link\_interference}.
  The definition is similar to \macro{update\_region}, but the last \vl!assert! statement checks that the region state was not changed by executing \vl!s!.
}
\label{fig:encoding_open_region}
\end{figure}

\begin{figure}
\begin{voila}[language=silver]
USE_ATOMIC(R(r, l, $\many{\ctext{p}}$), g, s) ->
  label pre_atomic

  assert g
  assert R(r, l, $\many{\ctext{p}}$)
  assert alevel > l

  assert $\level$ > l
  var $\level$_store := $\level$
  $\level$ := l  
  
  unfold R(r, l, $\many{\ctext{p}}$)
  havoc R(r, l, $\many{\ctext{p}}$) // Havoc other instances possibly held
  LINK_INTERFERENCE(R(r, l, $\many{\ctext{p}}$), s)
  fold R(r, l, $\many{\ctext{p}}$)

  ACTION_PERMITTED(R, R_state(r, l, $\many{\ctext{p}}$), old[pre_open](R_state(r, l, $\many{\ctext{p}}$), g)
  
  $\level$ := $\level$_store
\end{voila}
\caption{%
  General definition of macro \macro{use\_atomic}. Statement \vl!s! is executed as part of the expansion of \macro{link\_interference}.
  The definition is similar to \macro{update\_region} and \macro{open\_region}, but here, validity of the atomic update performed by \vl!s! is checked.
}
\label{fig:encoding_use_atomic}
\end{figure}

\begin{figure}
\begin{voila}[language=silver]
DO_WHILE(s, b, I) ->
  s
  WHILE(b, I, s)

WHILE(b, I, s) ->
  label pre_while
  exhale I
  $\foreach{\ctext{R}}{\metafunc*{Regions}} ~ \foreachStart$
    STABILIZE(R)
  $\foreachEnd$
  inhale I

  var oldUpdate := update
  var oldLevel := level
  var oldALevel := alevel
  while (b) 
    invariant I
    invariant update == oldUpdate && level == oldLevel 
        && alevel == oldALevel
    invariant forall r :: 
        r in update ==> acc(r.A) && r.A == old[pre_while](r.A)
    $\foreach{\ctext{R}}{\metafunc*{Regions}} ~ \foreachStart$
      invariant forall r :: r != null ==>
          acc(r.R_X) && r.R_X == old[pre_while](r.R_X)
    $\foreachEnd$
  {
    s
  }
\end{voila}
\caption{%
  General definitions of macros \macro{do\_while} and \macro{while}.
  The latter is encoded using a corresponding Viper loop, preceded by an explicit
  stabilization of the loop's frame.
  The additional invariants enforce that triple level, atomicity context and
  interference context are preserved inside the loop.
}
\label{fig:encoding_while}
\end{figure}

\begin{figure}
\begin{voila}[language=silver]
EXPLICIT_FRAME_OUT ->
  $\foreach{\ctext{R}}{\metafunc*{Regions}} ~ \foreachStart$
    exhale forall r, $\many{\ctext{p}}$ :: acc(R(r, $\many{\ctext{p}}$), perm(R(r, $\many{\ctext{p}}$))
  $\foreachEnd$
  $\foreach{\ctext{G}}{\metafunc*{Guards}} ~ \foreachStart$
    exhale forall r, $\many{\ctext{p}}$ :: acc(G(r, $\many{\ctext{p}}$), perm(G(r, $\many{\ctext{p}}$))
  $\foreachEnd$
  $\foreach{\ctext{f}}{\metafunc*{Fields}} ~ \foreachStart$
    exhale forall x :: x != null ==> acc(x.f, perm(x.f))
  $\foreachEnd$  

EXPLICIT_FRAME_IN(lbl) ->
  $\foreach{\ctext{R}}{\metafunc*{Regions}} ~ \foreachStart$
    exhale forall r, $\many{\ctext{p}}$ :: acc(R(r, $\many{\ctext{p}}$), perm[lbl](R(r, $\many{\ctext{p}}$))
  $\foreachEnd$
  ...
\end{voila}
\caption{%
  General definitions for macros \macro{explicit\_frame\_out} and
  \macro{explicit\_frame\_in}.
  The former exhales permissions to all region instances, guards and fields that
  the Voila program declares and to which the current verification state holds
  permissions.
  The later is analogous, but inhales permissions relative to a given label.
  In Viper, accessibility predicate \vl!acc(x.f)! denotes full (\ie write)
  permission to field \vl!x.f!, and is syntactic sugar for \vl!acc(x.f, write)!.
  Viper also supports fractional permissions~\cite{Boyland03}; for such a
  permission $\pi$, the syntax is \vl!acc(x.f, $\pi$)!. Consequently, the last exhale in the definition of \macro{explicit\_frame\_out} instructs Viper to exhale all permission held to a field \vl!x.f!.
  Predicates are supported analogously, but with additional syntactic sugar: the \vl!acc! around a predicate can be omitted, and \vl!R(x)! (for some predicate \vl!R!) abbreviates \vl!acc(R(x))!, and thus \vl!acc(R(x), write)!.
}
\label{fig:encoding_frame_out_in}
\end{figure}

\begin{figure}
\begin{voila}[language=silver]
MAKE_ATOMIC(R(r, l, $\many{\ctext{p}}$), g, s) ->
  label pre_atomic

  exhale g
  exhale R(r, l, $\many{\ctext{p}}$)
  $\foreach{\ctext{R}}{\metafunc*{Regions}} ~ \foreachStart$
    STABILIZE(R)
  $\foreachEnd$  

  label pre_frame
  EXPLICIT_FRAME_OUT
  
  assert $\alevel$ > l
  var $\alevel$_store := $\alevel$
  $\alevel$ := l

  assert !(r in $\actxtDomain$)
  var $\actxtDomain$_store := $\actxtDomain$
  inhale acc(r.R_A) && r.R_A == r.R_X
  $\actxtDomain$ := $\actxtDomain$ union Set(r)
  
  inhale R(r, l, $\many{\ctext{p}}$) && R_state(r, l, $\many{\ctext{p}}$) in r.R_A
  inhale acc(r.diamond)

  s
  
  ACTION_PERMITTED(R, r.R_from, r.R_to, g)
  
  EXPLICIT_FRAME_OUT
  
  inhale R(r, l, $\many{\ctext{p}}$) && (R_state(r, l, $\many{\ctext{p}}$) == r.R_to
  inhale old[pre_atomic](R_state(r, l, $\many{\ctext{p}}$)) == r.R_from
  exhale acc(r.R_from) && acc(r.R_to)
  inhale g
  
  $\actxtDomain$ := $\actxtDomain$_store
  exhale acc(r.R_A)
  
  $\alevel$ := $\alevel$_store
  
  EXPLICIT_FRAME_IN(pre_frame)
\end{voila}
\caption{%
  General definition of macro \macro{make\_atomic}.
  Before statement \ctext{s} (which is to be proven abstractly atomic) is
  executed, all regions are stabilized and all other resources are framed out, levels and contexts are adjusted, and the diamond resource is obtained.
  After the execution of \ctext{s}, validity of the performed atomic update
  is checked, and parts of the pre-state are restored.
}
\label{fig:encoding_make_atomic}
\end{figure}

\begin{figure}
\begin{voila}[language=silver]
PROCEDURE(M($\many{p}$) returns ($\many{r}$), s) ->
  method M($\many{p}$) returns ($\many{r}$) {
    inhale Pre$_\ctext{M}$

    $\foreach{\ctext{R}}{\metafunc*{Regions}} ~ \foreachStart$
      inhale forall r :: r != null ==> acc(r.R_X)
    $\foreachEnd$

    var level: Int 
    $\foreach{l}{\metafunc*{Levels}(\ctext{M})} ~ \foreachStart$
      inhale $\level$ > $l$ 
    $\foreachEnd$
    var alevel: Int := level
    var update: Set[Ref] := Set()

    s

    exhale Post$_\ctext{M}$
  }

ATOMIC_PROCEDURE(M($\many{p}$) returns ($\many{r}$), s) ->
  method M($\many{p}$) returns ($\many{r}$) {
    inhale Pre$_\ctext{M}$

    $\foreach{\ctext{R}}{\metafunc*{Regions}} ~ \foreachStart$
      inhale forall r :: r != null ==> acc(r.R_X)
    $\foreachEnd$
    $\foreach{Q}{\metafunc*{Inter}(\ctext{M})} ~ \foreachStart$
      inhale r$_Q$.R$_Q$_X subset S$_Q$
    $\foreachEnd$

    var level: Int 
    $\foreach{l}{\metafunc*{Levels}(\ctext{M})} ~ \foreachStart$
      inhale $\level$ > $l$ 
    $\foreachEnd$
    var alevel: Int := level
    var update: Set[Ref] := Set()

    s

    exhale Post$_\ctext{M}$
  }
\end{voila}
\caption{%
  General definition of macro \macro{procedure} and \macro{atomic\_procedure}, which are used to encode, and thus prove, procedure specifications. 
  For a non-atomic procedure, \macro{procedure} first inhales the precondition. Next, necessary resources are inhaled, and local variables declared and constrained. Afterwards, the encoded procedure body is executed. Finally, the postcondition is exhaled.
  \macro{atomic\_procedure} expands similarly, but for atomic procedures and their interference clauses, denoted by \metafunc*{Inter}.
  See also \macro{call} and \macro{atomic\_call} in \figref{encoding_call_atomic}.
}
\label{fig:encoding_method}
\end{figure}

\MaybeFloatBarrier
\section{Soundness}
\label{sec:soundness}

\newcommand{\viperHeap} [0] {H}
\newcommand{\viperPermMask} [0] {P}
\newcommand{\viperStore} [0] {S}
\newcommand{\viperVs} [0] {\upsilon}
\newcommand{\initialViperVs} [0] {\viperVs_{\mathsf{zero}}}
\newcommand{\invalidViperVs} [0] {\lightning}
\newcommand{\isValidViperVs} [1] {#1 \neq \invalidViperVs}
\newcommand{\viperLabelMapping} [0] {\mathsf{lbl}}

\newcommand{\toTaDA} [1] {\llparenthesis \, #1 \, \rrparenthesis}
\newcommand{\toViper} [1] {\llbracket \, #1 \, \rrbracket}
\newcommand{\erased} [1] {\llcorner #1 \lrcorner}
\newcommand{\spc} [2] {\mathsf{post}(#1, #2)}

\newcommand{\exMapping} [1] {\phi( #1 )}
\newcommand{\hoareTriple} [3] {\{ #1 \} \, #2 \,  \{ #3 \}}
\newcommand{\outlineStep} [0] {s}
\newcommand{\atomicOutlineStep} [0] {s_{\mathsf{a}}}
\newcommand{\nonAtomicOutlineStep} [0] {s_{\mathsf{na}}}
\newcommand{\pureStmt} [0] {\hat{s}}
\newcommand{\wellDefStates} [0] {\mathbb{I}}
\newcommand{\wellDefStatePairs} [0] {\mathbb{T} }

In this section, we present a soundness argument for our Voila encoding.
Our encoding is sound when the successful verification of an encoded Voila procedure proof outline
implies that the corresponding TaDA procedure satisfies its TaDA specification.
We deduce the latter by 
showing that the procedure specifications are indeed derivable in TaDA.

We argue soundness of our encoding in four steps:
first, we determine invariants on Viper's pre- and post-verification states of encoded Voila outline statements (programming language statements and key rules statements).
Second, we define a \term{judgment mapping}, which maps from a pair $(\viperVs, \outlineStep)$ of Viper verification states $\viperVs$, satisfying our invariants, and Voila outline statements $\outlineStep$ to a TaDA judgment. 
Third, under the assumption of successful verification,
we show by structural induction over Voila outline statements that the judgment mapping maps to derivable TaDA judgments.
Fourth, we show for each encoded Voila procedure
that the judgment mapping, applied to the encoded procedure body and a Viper state satisfying the procedure's precondition, maps to the desired TaDA judgment.
Combining these ingredients, we formally connect verification of an encoded proof outline to derivability of a TaDA proof, resulting in the soundness of our encoding.

For a better overview, we first illustrate our approach in more detail on a simplified version of TaDA.
Afterwards, we instantiate our approach for normal TaDA.
We demonstrate our soundness argument on four particularly challenging steps of our encoding: the handling of calls, triple changes, \vl!make_atomic!, and \vl!update_region!.

\subsection{Approach}
\label{sec:soundness_approach}

For the sake of simplicity, before targeting full TaDA, 
we introduce our approach informally on a simplified version of TaDA.
For this simplified version, assume that TaDA judgments are standard Hoare triples of the form $\vdash \hoareTriple{P}{\pureStmt}{Q}$,
where $P$, $\pureStmt$, and $Q$ are the precondition, triple statement, and postcondition, respectively.
We omit atomic triples, levels, atomicity contexts, interference contexts, and the requirement that pre- or postconditions are stable.
We use $\erased{\outlineStep}$ to reduce a Voila outline statement $\outlineStep$ to its underlying program statement, by stripping away potentially surrounding rule statements.
\Eg the outline statement \vl!update_region using ... { b := CAS(x,0,1) }! is reduced to \vl!b := CAS(x,0,1)!.

To prove soundness, we need a formal connection between an encoded Voila procedure (that successfully verified in Viper) and a TaDA proof.
On the Viper side, we have the state of a Viper program, \ie the verification state,
and the encoding of procedures and outline statements.
Conversely, on the side of TaDA, we have syntactic judgments and proof rules.
To formally connect both, we define a judgment mapping, a mapping from pairs $(\viperVs, \outlineStep)$ of Viper verification state $\viperVs$ and Voila outline statement $\outlineStep$ to syntactic TaDA judgments.
For our simplified version of TaDA, we can define such a judgment mapping as follows:
assume we have a mapping $\exMapping{\viperVs}$ from Viper verification state $\viperVs$ to assertions of TaDA.
Then, a judgment mapping for a Viper verification state $\viperVs$ and a Voila outline statement $\outlineStep$ can be defined as 
$\toTaDA{\viperVs, \outlineStep} = \hoareTriple{\exMapping{\viperVs}}{\erased{\outlineStep}}{\exMapping{\viperVs'}}$, 
where $\viperVs' = \spc{\toViper{\outlineStep}}{\viperVs}$ is the strongest postcondition verification state of the Viper encoding of $\outlineStep$ and the verification state $\viperVs$.

The judgment mapping is only applied to prestates of encoded Voila outline statements
because only these states, together with encoded statement and resulting poststate, are formally connected to triples in a TaDA proof.
In particular, the mapping is not applied to intermediate verification states of a Viper encoding.
We define invariants on Viper prestates so that the judgment mapping has stronger guarantees on the mapped verification states.
\Eg TaDA 
does not allow partial ownership of points-to predicates ($\ctext{x.f} \pointsto v$).
However, such partial permissions are in general possible in Viper states,  
making judgment mappings for such states with partial permissions nonsensical. 
Therefore, for our encoding, we define the invariant that permissions to fields are either full or none.
We then have to show that these invariants on a verification state hold, before we use the verification state in a judgment mapping.
We use $\wellDefStates$ to refer to the set of all verification states satisfying these invariants.

Using the judgment mapping, 
we can verbally state our soundness lemma of the outline statement encoding:
\textit{``Under the assumption of successful Viper verification, 
we show that the judgment mapping maps to derivable TaDA judgments when applied to encoded Voila outline statements and Viper verification states satisfying our state invariants''}. 
Before we can express this property more formally, 
we have to define the meaning of a successful Viper verification.
A successful verification 
entails that all verification states of
the verified Viper program are \term{valid}.
In Viper, a verification state is valid 
when it is not a special error state $\invalidViperVs$.
Therefore, we refine the soundness lemma from above as follows:
\textit{``Forall Voila outline statements $\outlineStep$ and 
Viper verification states $\viperVs \in \wellDefStates$,
a valid strongest poststate
$\isValidViperVs{\spc{\toViper{\outlineStep}}{\viperVs}}$ implies 
that the mapped judgment $\toTaDA{\viperVs, \outlineStep}$ is derivable in TaDA and that the poststate satisfies our state invariants $\spc{\toViper{\outlineStep}}{\viperVs} \in \wellDefStates$''}. 
We first illustrate the purpose of this lemma and
then argue how to prove it.

The lemma aids us to derive that 
a successfully verified Voila procedure implies 
that the corresponding TaDA procedure with its specification is derivable:
Consider a Voila procedure with the specification $\hoareTriple{P}{\mathsf{m}(\dots)}{Q}$ where $\mathsf{m}(\dots)$ is the procedure itself. 
Let $\outlineStep_m$ be its body and let $\viperVs_\mathsf{pre}$ be the verification state before the encoding of its body.
If the procedure is encoded as \ctext{inhale $\toViper{P}$; $\toViper{\outlineStep_m}$; exhale $\toViper{Q}$}, then $\viperVs_\mathsf{pre} = \spc{\ctext{inhale $\toViper{P}$}}{\initialViperVs}$, where $\initialViperVs$ is the initial (empty) verification state.
Assuming $\viperVs_\mathsf{pre}$ satisfies our state invariants ($\viperVs_\mathsf{pre} \in \wellDefStates$) and that $\toTaDA{ \viperVs_\mathsf{pre}, \outlineStep_m }$ maps to $\hoareTriple{P}{\outlineStep_m}{Q'}$ with $Q' \models Q$,
we can apply the lemma to get that $\hoareTriple{P}{\outlineStep_m}{Q}$, and as such
$\hoareTriple{P}{\mathsf{m}(\dots)}{Q}$, is derivable in TaDA.

We can prove soundness of the outline statement encoding by straightforward structural induction over outline statements.
We illustrate a case of the induction at an abstract level.
Consider a compound outline statement $\outlineStep\{\outlineStep'\}$ ($\outlineStep$ is the compound, \eg \vl!update_region!, and $\outlineStep'$ is its body, \eg \vl!CAS(...)!) with an encoding $\toViper{\outlineStep\{\outlineStep'\}} = \ctext{c$_1$;$\toViper{\outlineStep'}$;c$_2$}$, where $\ctext{c$_1$}$ and $\ctext{c$_2$}$ are the Viper statements before and after the encoding of the body, respectively.
There are four Viper verification states of interest:
the prestate of the compound statement $\viperVs_0$, the prestate of its body $\viperVs_1 = \spc{\ctext{c$_1$}, \viperVs_0}$, the poststate of its body $\viperVs_2 = \spc{\toViper{\outlineStep'}, \viperVs_1}$, and the poststate of the compound statement $\viperVs_3 = \spc{\ctext{c$_2$}, \viperVs_2}$. 
From the induction hypothesis, we know that $\toTaDA{\viperVs_1, \outlineStep'} = \hoareTriple{\exMapping{\viperVs_1}}{\erased{\outlineStep'}}{\exMapping{\viperVs_2}}$ is derivable in TaDA 
and we have to show that $\toTaDA{\viperVs_0, \outlineStep\{\outlineStep'\}} = \hoareTriple{\exMapping{\viperVs_0}}{\erased{\outlineStep\{\outlineStep'\}}}{\exMapping{\viperVs_3}}$ is derivable in TaDA.
Showing this derivation corresponds to applying rules to fill the ($?$)-gap in the following proof tree:
\begin{displaymath}
  \Inf[?]{
    \Inf[\textsc{IH}]{\vdots} {
      \hoareTriple{\exMapping{\viperVs_1}}{\erased{\outlineStep'}}{\exMapping{\viperVs_2}}
    }
  }{\hoareTriple{\exMapping{\viperVs_0}}{\erased{\outlineStep\{\outlineStep'\}}}{\exMapping{\viperVs_3}}}
\end{displaymath}
The application of \textsc{IH} denotes using the fact from the induction hypothesis that $\hoareTriple{\exMapping{\viperVs_1}}{\erased{\outlineStep'}}{\exMapping{\viperVs_2}}$ is derivable in TaDA.
The necessary rule applications for the ($?$)-gap are determined by our encoded proof candidate (\secref{proof_candidate}), where we have to argue that their applications are correctly encoded in the outline statement encoding.

In the next sections, 
we first discuss Viper's verification state.
Afterwards, we introduce the judgment mapping and state invariants for all of TaDA, including atomic triples, levels, atomicity context, interference context, and stability requirements.
Last, we argue soundness of the outline statement encoding.

\subsection{Viper Verification State}
\label{sec:soundness_verification_state}

Viper's verification state~\cite{ParkinsonSummers12}
is defined as a set of traces.
Each trace consists of a sequence of \term{state atoms}:
a triple $(\viperHeap, \viperPermMask, \viperStore)$ 
of a heap $\viperHeap$ (mapping \ctext{Ref} and field name pairs, as well as applied functions, to values), a permission mask $\viperPermMask$ (mapping \ctext{Ref} and field name pairs, as well as predicate instances, to permission amounts; these amounts are non-negative rationals, which for fields cannot exceed $1$),
and a variable store $\viperStore$ (mapping variables to values).
Furthermore, a trace consists of a label mapping $\viperLabelMapping$,
mapping Viper labels to their corresponding state atom.
We have a special error verification state $\invalidViperVs$, which is the result of a verification error, \eg the poststate of \ctext{x := 5; assert  x == 4}.
We use $\viperVs$ to range over verification states.
The semantics of the core logic is given in \cite{ParkinsonSummers12}.
In particular, the semantics of heap-dependent expressions
such as fields accesses \ctext{x.f} comes with well-definedness conditions. 
\Eg reading from a field is only allowed in states with a non-zero permission for that field. The semantics of functions and predicates follows~\cite{SummersDS13}.

When a Viper program verifies successfully, 
this implies that all 
\ctext{assert} and \ctext{exhale} (removes the assertion from the verification state, introduced in \secref{viper_encoding})
assert and exhale, respectively, assertions valid in there respective verification states. 
This includes implicit assertions and exhales, 
such as asserting non-zero permission when accessing a field or exhaling preconditions when calling functions.

\subsection{TaDA Judgment Mapping and State Invariants}
\label{sec:soundness_invariants}
\label{sec:soundness_specification_mapping}

\newcommand{\actxt} [0] {\atctx}
\newcommand{\fdomain} [1] {\mathsf{dom}(#1)}
\newcommand{\fimg} [1] {\mathsf{img}(#1)}
\newcommand{\univ} [0] {\mathbb{U}}
\newcommand{\actxtSet} [0] {\cmcal{AS}}
\newcommand{\actxtleq} [0] {\leq}
\newcommand{\ictxtVar} [0] {x}
\newcommand{\ictxtVars} [1] {\many{x{#1}}}
\newcommand{\historyVar} [0] {h}
\newcommand{\historyVars} [1] {\many{h{#1}}}
\newcommand{\historySet} [0] {H}
\newcommand{\ictxt} [0] {X}
\newcommand{\jlevel} [0] {\lambda}

\newcommand{\najudge} [5] {{#1};{#2} ~ \vdash ~ \left\{ ~ {#3} ~ \right\} ~ {#4} ~ \left\{ ~ {#5} ~ \right\}}
\newcommand{\ajudge}  [6] {{#1};{#2} ~ \vdash ~ {#3} \left\langle ~ {#4} ~ \right\rangle ~  {#5} ~ \left\langle ~ {#6} ~ \right\rangle}

\newcommand{\judgepre} [0] {} 
%

\newcommand{\loweratomicitycontext} [1] {\atctx^{\mathsf{lb}}_{#1}}
\newcommand{\upperatomicitycontext} [1] {\atctx^{\mathsf{ub}}_{#1}}
\newcommand{\stateLevel} [1] {\lambda_{#1}}
\newcommand{\stateInterferenceContext} [1] {{\ictxt{}}_{#1}}
\newcommand{\stateHistory} [1] {\historySet_{#1}}

\newcommand{\stateAPre} [1] {\mathsf{Pre}_{#1}}
\newcommand{\stateAPost} [2] {\mathsf{Post}_{#1, #2}}
\newcommand{\stateP} [1] {\mathsf{P}_{#1}}

\newcommand{\stateClosedPre} [3] {\widehat{\mathsf{Pre}_{#1, #2}}}
\newcommand{\stateClosedPost} [3] {\widehat{\mathsf{Post}_{#1, #2}}}

\newcommand{\ssub} [2] {[{#2}/{#1}]}
\newcommand{\sand} [0] {\slstar}

\newcommand{\vcode} [1] {\text{\vl!#1!}}
\newcommand{\returnVars} [0] {\many{\vcode{r}}}
\newcommand{\argVars} [0] {\many{\vcode{z}}}

\newcommand{\stateALevel} [1] {\delta_{#1}}
\newcommand{\aencInv} [1] {\assertionEncInv{#1}}

\newcommand{\assertionEncInv} [1] {\llparenthesis #1 \rrparenthesis}

\newcommand{\vipertriple} [3] {\vdash_V \{ #1 \} #2 \{ #3 \}}
\newcommand{\viperinvariantassertions} [0] {\cmcal{I}}

In the judgment mapping of TaDA, we distinguish between non-atomic and abstract atomic Voila outline statements.
For an abstract atomic outline statement $\atomicOutlineStep$ and a Viper prestate $\viperVs$, the judgment mapping maps to a TaDA judgment of the following shape:
\begin{multline*}
  \toTaDA{\viperVs, \atomicOutlineStep} = \forall \historyVars{} \in \stateHistory{\viperVs}. ~ \forall \actxt \text{ with } \loweratomicitycontext{\viperVs} \actxtleq \actxt \actxtleq \upperatomicitycontext{\viperVs}. \\ 
  \ajudge{\stateLevel{\viperVs}}{\actxt}{\tadaforall \ictxtVars{} \in \stateInterferenceContext{\viperVs}}{\stateAPre{\viperVs}(\historyVars{}, \ictxtVars{})}{\erased{\atomicOutlineStep}}{\stateAPost{\viperVs}{\viperVs'}(\historyVars{}, \ictxtVars{})} \\
  \text{where } \viperVs' = \spc{\toViper{\atomicOutlineStep}}{ \viperVs} 
\end{multline*}
For the sake of brevity, we omit the exact judgment mapping definition; instead, we describe the different components informally:
$\pmb{(\stateHistory{\viperVs})}$ Viper can deduce facts based on knowledge of old state, \eg from the fact that some variable had the value $5$ at a previous Viper label. 
To account for such deductions at the TaDA level, we use a set $\stateHistory{\viperVs}$, the \term{history set}, to capture Viper's knowledge about old state. 
\Eg consider a variable $\ctext{z}$ whose value is one plus its old value from label $\ctext{lbl}$. 
With the history set, this fact is captured as $\forall (..., \historyVar_{\ctext{z}}, ...)  \in \stateHistory{\viperVs}. \, \ctext{z} = \historyVar_{\ctext{z}} + 1$, where $\historyVar_{\ctext{z}}$ binds the part of the history set that tracks $\ctext{z}$'s value from label $\ctext{lbl}$. 
In our TaDA judgment,
we do not map these facts to the pre- or postcondition of a TaDA triple because rules such as \MakeAtomic restricts the shape of pre- and postconditions. This would force us to remove these facts from TaDA's pre- and postconditions, even though these facts remain in Viper's verification state.
$\pmb{(\actxt)}$ As described in \appref{additional_details_normal_form}, a TaDA triple is proven forall atomicity contexts $\actxt$ within a lower bound $\loweratomicitycontext{\viperVs}$ and an upper bound $\upperatomicitycontext{\viperVs}$;
the order on atomicity contexts is defined as follows:
\begin{multline*}
  \actxt_1 \actxtleq \actxt_2 \Leftrightarrow \forall r \in \fdomain{\actxt_1}. ~ r \in \fdomain{\actxt_2} \\ \land \fdomain{\actxt_2(r)} \subseteq \fdomain{\actxt_1(r)} \\ \land  \forall z \in \fdomain{\actxt_2(r)}. ~ \fimg{\actxt_1(r)(z)} \subseteq \fimg{\actxt_2(r)(z)}
\end{multline*}
The operations $\fdomain{\cdot}$ and $\fimg{\cdot}$ denote domain and image, respectively.
We define the order such that a Viper assertion $P$ being stable for an atomicity context $\actxt_1$ implies that $P$ is also stable for all atomicity contexts $\actxt_2$ with $\actxt_1 \actxtleq \actxt_2$.
Therefore, to satisfy stability of a pre or postcondition for all atomicity contexts $\actxt$ that a triple is proven for ($\loweratomicitycontext{\viperVs} \actxtleq \actxt \actxtleq \upperatomicitycontext{\viperVs}$), it is sufficient to satisfy stability of the pre or postcondition for $\loweratomicitycontext{\viperVs}$.
Regarding the mapping, the lower bound $\loweratomicitycontext{\viperVs}$ has an entry for a region instance $r$ only if, in the verification state $\viperVs$, $r$ is contained in the value of the $\ctext{update}$ variable. 
The domain of such an atomicity context entry for $r$ is the value of $r.\ctext{R\_A}$, where \ctext{R} is the region name of $r$. 
The image of an entry for $r$ depends on the poststate $\viperVs'$. If $r.\ctext{R\_from} = z$ and $r.\ctext{R\_to} = f(z)$ are held in $\viperVs'$ (this corresponds to $\trackres(z,f(z))$ in TaDA), then the image of the entry is defined by the function $f$, otherwise the image is the empty set $\emptyset$.
Conversely, the upper bound $\upperatomicitycontext{\viperVs}$ has an entry for region identifier $r$ only if its region level is at least the value of the \ctext{alevel} variable in $\viperVs$. 
The domain for $r$ is $\fdomain{\loweratomicitycontext{\viperVs}}$ if $r$ is an entry of the lower bound $\loweratomicitycontext{\viperVs}$ ($r \in \fdomain{\loweratomicitycontext{\viperVs}}$), otherwise the domain for $r$ in $\upperatomicitycontext{\viperVs}$ is $\emptyset$. 
Again, the image for $r$ depends on the poststate. If $r.\ctext{R\_from} = z$ and $r.\ctext{R\_to} = f(z)$ are held in $\viperVs'$, then the image of the entry is defined by the function $f$, otherwise the image is the set of all values $\univ$.
$\pmb{(\stateLevel{\viperVs})}$ The level of the triple $\stateLevel{\viperVs}$ is the value of the $\ctext{level}$ variable in the prestate $\viperVs$.
$\pmb{(\stateInterferenceContext{\viperVs})}$ The interference context $\stateInterferenceContext{\viperVs}$ is the cartesian product of all values of $\ctext{r.R\_X}$ for which the predicate $\ctext{R(r, $\many{p}$)}$ is held in the prestate $\viperVs$.
$\pmb{(\stateAPre{\viperVs}, \stateAPost{\viperVs}{\viperVs'})}$ The pre and postcondition of the TaDA triple are $\stateAPre{\viperVs}$ and $\stateAPost{\viperVs}{\viperVs'}$, respectively. 
Both can have occurrences of quantifiers bound by the history set and the interference context quantifier.
We use different assertion mappings for pre and postconditions because the interference context is handled for each of them differently, as they have different restrictions in our normal form. 
For the precondition, if a region predicate $\ctext{R($r$,$\lambda$,$\many{p}$)}$ is held in the prestate $\viperVs$, then this is mapped to $R_r^{\lambda}(\many{p}, x_r)$ where $x_r$ is the interference context quantifier for region identifier $r$. This way, we guarantee that the state of regions in the precondition is bound by the interference context.
For the postcondition, we do not have this requirement.
Holding $\ctext{R($r$,$\lambda$,$\many{p}$)}$ in the poststate $\viperVs'$ is mapped to $R_r^{\lambda}(\many{p}, z_r)$ where $z_r$ is a logical variable, additionally introduced for binding the region state.
The region state function $\text{R\_State($r$,$\lambda$,$\many{p}$)}$ is mapped to constraints on $x_r$ and $z_r$ for the pre and postcondition, respectively.
For all other resources the mapping is the same for pre and postconditions. 
The mapping for these resources corresponds to the inverse of the encoding:
\Eg Holding a guard predicate $\ctext{R\_G($r, \many{p}$)}$ in a verification state is mapped to a TaDA guard instance $\guard[G$(\many{p})$]_r$.
Similarly, holding $\ctext{acc(x.f)}$ is mapped to $\text{x.f}\pointsto z$ where $z$ is the value of $\ctext{x.f}$ in the verification state.
For simplicity, we omit assertions with local program variables in the shown TaDA triple. These are mapped to private assertions of atomic triples and can only depend on the history set. No other resource, \eg guards or points-to predicates, are mapped to private assertions of atomic triples.
The handling of local variables in all rule applications is straightforward.

For a non-atomic outline statement $\nonAtomicOutlineStep$ and a Viper prestate $\viperVs$, the judgment mapping maps to a non-atomic TaDA judgment of the following shape:
\begin{multline*}
  \toTaDA{\viperVs, \nonAtomicOutlineStep} = \forall \historyVars{} \in \stateHistory{\viperVs}. ~ \forall \actxt \text{ with } \loweratomicitycontext{\viperVs} \actxtleq \actxt \actxtleq \upperatomicitycontext{\viperVs}. \\ 
  \najudge{\stateLevel{\viperVs}}{\actxt}{\stateP{\viperVs}(\historyVars{})}{\erased{\nonAtomicOutlineStep}}{\stateP{\viperVs'}(\historyVars{})} \\
  \text{where } \viperVs' = \spc{\toViper{\nonAtomicOutlineStep}}{ \viperVs} 
\end{multline*}
The mapping is the same as for abstract atomic outline steps, except that the TaDA judgment has no interference context and thus the same assertion mapping $\stateP{\viperVs}$ can be used for both, pre- and postcondition.

To express stability of the pre or postcondition, we additionally define a closed form of the pre and postcondition, which has no free variables.
The closed form for Viper prestates $\viperVs$ and abstract atomic statements $\atomicOutlineStep$, as well as non-atomic outline statements $\nonAtomicOutlineStep$, are derived from TaDA and defined as follows:
\begin{align*}
  &\stateClosedPre{\viperVs}{\atomicOutlineStep}{\viperVs'} = \forall \historyVars{} \in \stateHistory{\viperVs}. ~ \exists \, \ictxtVars{} \in \stateInterferenceContext{\viperVs}. ~ \stateAPre{\viperVs}(\historyVars{}, \ictxtVars{}) \\
  &\stateClosedPost{\viperVs}{\atomicOutlineStep}{\viperVs'} = \forall \historyVars{} \in \stateHistory{\viperVs}. ~ \forall \, \ictxtVars{} \in \stateInterferenceContext{\viperVs}. ~ \stateAPost{\viperVs}{\viperVs'}(\historyVars{}, \ictxtVars{}) \\
  &\quad\text{ where } \viperVs' = \spc{\toViper{\atomicOutlineStep}}{ \viperVs}  \\
  &\stateClosedPre{\viperVs}{\nonAtomicOutlineStep}{\viperVs'} = \forall \historyVars{} \in \stateHistory{\viperVs}. ~ \stateP{\viperVs}(\historyVars{}) \\
  &\stateClosedPost{\viperVs}{\nonAtomicOutlineStep}{\viperVs'} = \forall \historyVars{} \in \stateHistory{\viperVs}. ~ \stateP{\viperVs'}(\historyVars{})  \\
  &\quad\text{ where } \viperVs' = \spc{\toViper{\nonAtomicOutlineStep}}{ \viperVs} 
\end{align*}

For stronger guarantees in the judgment mapping, we have several invariants on Viper prestate of encoded Voila outline statements:
\begin{enumerate*}
  \item Fields, region predicates, and guard predicates have either none or full permissions.
  In particular, permissions for the interference context field ($\ctext{R\_X}$) is always full, and permissions to the two tracking fields $\ctext{R\_from}$ and $\ctext{R\_to}$ are either both full or both none.
  An exception are guard predicates for fractional guards, which are allowed to have partial permissions because their Viper permission amount maps to a TaDA guard argument.
  \item If permission to the diamond tracking resource field ($\ctext{r.diamond}$) is held, then $r$ is contained in the set of the $\ctext{update}$ variable.
  \item For all region identifiers contained in $\ctext{update}$, the region level is at least the value of the $\ctext{alevel}$ variable.
  \item The other invariants are more technical and required to define the judgment mapping.
\end{enumerate*}

Besides invariants on single Viper verification states, we define invariants on pairs of pre and poststates of an encoded outline statement. We use $\wellDefStatePairs$ to denote the set of verification state pairs that satisfy these invariants. A state pair $(\viperVs, \viperVs')$ is contained in $\wellDefStatePairs$, when their level, interference context, and both atomicity context bounds are equal in the judgment mapping, \ie when $\stateLevel{\viperVs} = \stateLevel{\viperVs'}$, $\stateInterferenceContext{\viperVs} = \stateInterferenceContext{\viperVs'}$, $\loweratomicitycontext{\viperVs} = \loweratomicitycontext{\viperVs'}$, and $\upperatomicitycontext{\viperVs} = \upperatomicitycontext{\viperVs'}$ holds.
For full TaDA, opposed to the simplified version, we need these additional two-state invariants to guarantee that level, interference context, and atomicity context stay consistent for sequential composition.

\subsection{Proof Candidates}
\label{sec:soundness_canonical_proofs}

As discussed in \secref{soundness_approach}, we prove soundness of our outline statement encoding by induction over outline statements.
We use the induction predicate $W$:
\begin{multline*}
  W(\outlineStep) :\equiv \forall \viperVs \in \wellDefStates. ~ 
  \isValidViperVs{\viperVs'}
   \; \land \; \stateClosedPre{\viperVs}{s}{\viperVs'} \textit{ is stable for } \loweratomicitycontext{\viperVs}\Longrightarrow \\
    \toTaDA{\viperVs, \outlineStep} \textit{ is derivable} \; \land \; \viperVs' \in \wellDefStates \; \land \; (\viperVs, \viperVs') \in \wellDefStatePairs \\
    \; \land \; (\outlineStep \textit{ is non-atomic } \Longrightarrow \stateClosedPost{\viperVs}{s}{\viperVs'} \textit{ is stable for} \loweratomicitycontext{\viperVs}) \\
    \text{ where } \viperVs' = \spc{\toViper{s}}{\viperVs}
 \end{multline*}
The additional properties about stability can be included in our invariants on pre and post Viper verification states $\wellDefStatePairs$ (by making the invariants dependent on the encoded Voila outline statement). We explicitly state the stability properties for clarity. Our normal form is captured in our soundness argument as a combination of the invariants $\wellDefStates$ and $\wellDefStatePairs$, the condition on stability in $W$, and the shape of TaDA judgments in the image of our judgment mapping.

To streamline the proof argument, we add to Voila an outline statement \ctext{atomic\{\outlineStep\}}, which changes the atomicity of a triple from non-atomic to atomic. 
Without this additional outline statement, for cases such as loops, we have to make a case distinction whether the body is abstract atomic or non-atomic.
By introducing the outline statement, it is guaranteed that the atomicity of the body is the atomicity of the non-bridge rules' premise.

For the induction proof, we focus on the cases for atomic calls, \vl!atomic!, \vl!update_region!, and \vl!make_atomic!.
These are particular challenging steps of our encoding.
In our presentation of the induction cases, for the sake of brevity, we reason about Viper code at a higher, more abstract, level to focus on the proofs themselves. 
In particular, 
we take as a lemma that the poststate $\viperVs'$ after the macro \macro{stabilize} (See \figref{encoding_stabilization}) is stable for $\loweratomicitycontext{\viperVs}$ when mapped to TaDA, where $\viperVs$ is the prestate of the macro.
A proof argument about a similar encoding of stabilization was provided in \cite{DinsdaleYoungPAB17}.

\paragraph{Atomic Call.} 

\newcommand{\stmtA} [0] {\many{\text{\vl!y!}}\text{ \vl!:= M(!}\many{\text{\vl!e!}}\text{\vl!)!}}

\figref{soundness_proof_call} shows the filled out TaDA proof tree for the encoding of an abstract atomic call $\stmtA$, where \vl!M!, $\many{\text{\vl!e!}}$, and $\many{\text{\vl!y!}}$ are the called procedure, the arguments, and the return variables, respectively. 
The encoding of atomic calls is given in \figref{encoding_call_atomic}. Let $\viperVs$ and $\viperVs'$ be Viper's pre and poststate of the encoded Voila statement, respectively.
As seen in the definition of the judgment mapping, the TaDA judgment is proven for every $\historyVars{} \in \stateHistory{\viperVs}$ and every atomicity context $\actxt$ between $\loweratomicitycontext{\viperVs}$ and $\upperatomicitycontext{\viperVs}$. 
The important steps of the proof snippet go as follows (from the bottom of the tree to the top): 
Firstly, the current judgment level $\stateLevel{\viperVs}$ is reduced to the level of the called procedure, denoted by $\jlevel'$. 
The side condition of $\LvlWeak$ ($\stateLevel{\viperVs} \geq \jlevel'$) is satisfied, since in the encoding we assert that the $\ctext{level}$ variable is larger than every level in \vl!M!'s precondition and as such is larger than $\jlevel'$, which is one plus the maximum level in \vl!M!'s precondition.
Secondly, the mapped verification state that is not part of the procedure's precondition $R(\historyVars{},\ictxtVars{})$, is weakened to a stabilized version $R'(\historyVars{},\ictxtVars{})$ (by \Consequence), and then framed off.
The stability of the frame $R'(\historyVars{},\ictxtVars{})$ is a consequence from the use of the macro \macro{stabilize} in the encoding.
Furthermore, we know that only the TaDA pre and postcondition of the procedure remain in the proof state since their Viper counterparts are asserted and everything else is framed off. We denote the procedure's pre and postcondition with $P'(\historyVars{},\ictxtVars{})$ and $Q'(\historyVars{},\ictxtVars{})$, respectively.
Thirdly, the current interference context $\stateInterferenceContext{\viperVs}$ is widened to the interference context of the procedure, denoted as $X'$, by applying $\Substitution$. This widening is justified since in the encoding we assert that $\stateInterferenceContext{\viperVs}$ is a subset of $X'$ for the relevant interference context parts.
Lastly, we apply the call rule. We already know that the level, interference context, and pre- and postcondition match. It remains to argue that the current atomicity context $\actxt$ is contained in the set of atomicity contexts handled by \vl!M!, \ie that $\actxt$ is between the lower and upper bound of \vl!M! as defined by the judgment mapping, which we denote by $\loweratomicitycontext{\vcode{M}}$ and $\upperatomicitycontext{\vcode{M}}$, respectively. The inclusion of the lower bound is trivial since $\loweratomicitycontext{\vcode{M}}$ is empty. The upper bound is satisfied since in the encoding we check that the value of the $\ctext{alevel}$ variable is larger or equal to the \vl!M!'s level, which is equal to the the value of $\ctext{alevel}$ initially set for \vl!M!, hence entailing $ \upperatomicitycontext{\viperVs}  \actxtleq \upperatomicitycontext{\vcode{M}}$.

Non-atomic calls are similar, except that interference contexts are not present.

\begin{figure}
\scalebox{0.87}{
\Inf[\textsc{AWeak3}]{
  \Inf[\textsc{Cons}]{
    \Inf[\textsc{Cons}]{
      \Inf[\textsc{Frame}]{
        \Inf[\textsc{Subst}]{
         \Inf[\textsc{Call}] {\vdots}
          { \judgepre{}\ajudge{\jlevel'}{\actxt}{\tadaforall ~\many{\ictxtVar'} \in \ictxt'.~}{{P'(\historyVars{},\ictxtVars{'})}\ssub{{\argVars}}{\many{\vcode{e}}}}{\stmtA}{{Q'(\historyVars{},\ictxtVars{'})}\ssub{{\argVars}}{\many{\vcode{e}}}\ssub{\returnVars}{\many{\vcode{y}}}} }
        }{ \judgepre{}\ajudge{\jlevel'}{\actxt}{\tadaforall ~\many{\ictxtVar} \in \stateInterferenceContext{\viperVs}.~}{{P'(\historyVars{},\ictxtVars{})}\ssub{{\argVars}}{\many{\vcode{e}}}}{\stmtA}{{Q'(\historyVars{},\ictxtVars{})}\ssub{{\argVars}}{\many{\vcode{e}}}\ssub{\returnVars}{\many{\vcode{y}}}} }
      }{ \judgepre{}\ajudge{\jlevel'}{\actxt}{\tadaforall ~\many{\ictxtVar} \in \stateInterferenceContext{\viperVs}.~}{{R'(\historyVars{},\ictxtVars{})} \sand {P'(\historyVars{},\ictxtVars{})}\ssub{{\argVars}}{\many{\vcode{e}}}}{\stmtA}{{R'(\historyVars{},\ictxtVars{})} \sand {Q'(\historyVars{},\ictxtVars{})}\ssub{{\argVars}}{\many{\vcode{e}}}\ssub{\returnVars}{\many{\vcode{y}}}} }
    }{ \judgepre{}\ajudge{\jlevel'}{\actxt}{\tadaforall ~\many{\ictxtVar} \in \stateInterferenceContext{\viperVs}.~}{{R(\historyVars{},\ictxtVars{})} \sand {P'(\historyVars{},\ictxtVars{})}\ssub{{\argVars}}{\many{\vcode{e}}}}{\stmtA}{{R'(\historyVars{},\ictxtVars{})} \sand {Q'(\historyVars{},\ictxtVars{})}\ssub{{\argVars}}{\many{\vcode{e}}}\ssub{\returnVars}{\many{\vcode{y}}}} }
  }{ \judgepre{}\ajudge{\jlevel'}{\actxt}{\tadaforall ~\many{\ictxtVar} \in \stateInterferenceContext{\viperVs}.~}{{\stateAPre{\viperVs}(\historyVars{},\ictxtVars{})}}{\stmtA}{{\stateAPost{\viperVs}{\viperVs'}(\historyVars{},\ictxtVars{})}} }
}{ \judgepre{}\ajudge{\stateLevel{\viperVs}}{\actxt}{\tadaforall ~\many{\ictxtVar} \in \stateInterferenceContext{\viperVs}.~}{{\stateAPre{\viperVs}(\historyVars{},\ictxtVars{})}}{\stmtA}{{\stateAPost{\viperVs}{\viperVs'}(\historyVars{},\ictxtVars{})}} }}
\caption{
  Proof snippet for the encoding of abstract atomic calls $\stmtA$, where \vl!M!, $\many{\text{\vl!e!}}$, and $\many{\text{\vl!y!}}$ are the called procedure, the arguments, and the return variables, respectively. The encoding of atomic calls is given in \figref{encoding_call_atomic}.
  Note that we use shortened TaDA rule names.
}
\label{fig:soundness_proof_call}
\end{figure}

\paragraph{Atomicity Change.}

\newcommand{\stmtB} [0] {\text{\vl!s!}'}

The corresponding proof tree is shown in \figref{soundness_proof_atomic}, where $\stmtB$ is the TaDA statement reduced from $\ctext{atomic\{\outlineStep\}}$ (\ie $\stmtB = \erased{\ctext{atomic\{\outlineStep\}}}$).
The encoding of atomicity changes is given in \figref{encoding_atomicity_change}. 
Let $\viperVs_0$ and $\viperVs_3$ be Viper's pre and poststate of the encoded Voila statement $\ctext{atomic\{\outlineStep\}}$, respectively. Similarly, let $\viperVs_1$ and $\viperVs_2$ be Viper's pre and poststate of the encoded Voila statement $\ctext{s}$, respectively.
Again, let $\historyVars{} \in \stateHistory{\viperVs_0}$ and an atomicity context $\actxt$ between $\loweratomicitycontext{\viperVs_0}$ and $\upperatomicitycontext{\viperVs_0}$ be arbitrary.
As seen in \secref{proof_candidate},
in TaDA, the atomicity of the triple is changed by applying $\Consequence$ to stabilize the postcondition,
$\TripleWeak$ to switch the triple kind,
$\AExists$ to establish the interference context,
where $\ictxtVars{} \in \stateInterferenceContext{\viperVs_1}$ binds all region states in 
$P'(\historyVars{},\ictxtVars{})$ and the corresponding region states from the linearization point in $Q'(\historyVars{},\ictxtVars{})$.
As described in the definition of our judgment mapping, $\Exists$ is applied to move Viper facts about old state out of the triple. Afterwards, the induction hypothesis can be applied.

\begin{figure}
  \scalebox{0.87}{
\Inf[\textsc{Cons}]{
  \Inf[\textsc{AWeak1}]{
    \Inf[\textsc{AExists}]{
      \Inf[\textsc{Cons}]{
        \Inf[\textsc{Exists}]{
          \Inf[\textsc{IH}] {\vdots}
           { \judgepre{}\ajudge{\stateLevel{\viperVs_1}}{\actxt}{\tadaforall ~\many{\ictxtVar} \in \stateInterferenceContext{\viperVs_1}.~}{{\stateAPre{\viperVs_1}(\historyVars{}, \historyVars{''},\ictxtVars{})}}{\stmtB}{{\stateAPost{\viperVs_1}{\viperVs_2}(\historyVars{}, \historyVars{''},\ictxtVars{})}} }
        }{ \judgepre{}\ajudge{\stateLevel{\viperVs_1}}{\actxt}{\tadaforall ~\many{\ictxtVar} \in \stateInterferenceContext{\viperVs_1}.~}{\exists \, \many{\historyVar{}''} \in \historySet{}''. ~  {\stateAPre{\viperVs_1}(\historyVars{}, \historyVars{''},\ictxtVars{})}}{\stmtB}{\exists \, \many{\historyVar''} \in \historySet''. ~{\stateAPost{\viperVs_1}{\viperVs_2}(\historyVars{}, \historyVars{''},\ictxtVars{})}} }
      }{ \judgepre{}\ajudge{\stateLevel{\viperVs_0}}{\actxt}{\tadaforall ~\many{\ictxtVar} \in \stateInterferenceContext{\viperVs_0}.~}{{P'(\historyVars{},\ictxtVars{})}}{\stmtB}{{Q'(\historyVars{},\ictxtVars{})}} }
    }{ \judgepre{}\ajudge{\stateLevel{\viperVs_0}}{\actxt}{}{\exists \, \many{\ictxtVar} \in \stateInterferenceContext{\viperVs_0}.~{P'(\historyVars{},\ictxtVars{})}}{\stmtB}{\exists \, \many{\ictxtVar} \in \stateInterferenceContext{\viperVs_0}.~{Q'(\historyVars{},\ictxtVars{})}} }
  }{ \judgepre{}\najudge{\stateLevel{\viperVs_0}}{\actxt}{\exists \, \many{\ictxtVar} \in \stateInterferenceContext{\viperVs_0}.~{P'(\historyVars{},\ictxtVars{})}}{\stmtB}{\exists \, \many{\ictxtVar} \in \stateInterferenceContext{\viperVs_0}.~{Q'(\historyVars{},\ictxtVars{})}} }
}{ \judgepre{}\najudge{\stateLevel{\viperVs_0}}{\actxt}{{\stateP{\viperVs_0}(\historyVars{})}}{\stmtB}{{\stateP{\viperVs_3}(\historyVars{})}} }}
\caption{
  Proof snippet for the encoding of \vl!atomic!, which switches from the non-atomic triples to the atomic triples. The statement $\stmtB$ is equal to $\erased{\ctext{atomic\{\outlineStep\}}}$. The encoding of atomicity changes is given in \figref{encoding_atomicity_change}.
  Note that we use shortened TaDA rule names.
}
\label{fig:soundness_proof_atomic}
\end{figure}

\paragraph{Update-Region.}

\newcommand{\stmtC} [0] {\text{\vl!s!}'}
\newcommand{\ractxtDomain} [0] {D}
\newcommand{\ractxtImage} [0] {I}
\newcommand{\regionExA} [1] {\text{\region{R}^{\jlevel}}(\many{p}, {#1})}
\newcommand{\openRegionExA} [1] {\interp(\text{\region{R}^{\jlevel}}(\many{p}, {#1}))}
\newcommand{\openRegionExAConstraint} [1] {Z(r,\jlevel,\many{p},{#1},\many{\ictxtVar')}}
\newcommand{\regionExALeftState} [0] {\ictxtVar^{\circ}}

The corresponding proof tree is shown in \figref{soundness_proof_update}, where again $\stmtC$ is the reduced Viper statement ($\stmtC = \erased{\ctext{update\_region using ... \{\outlineStep\}}}$). 
Again, let $\viperVs_0$ and $\viperVs_3$ be Viper's pre and poststate of the encoded Voila statement $\ctext{update\_region using ... \{\outlineStep\}}$, respectively.
Furthermore, let $\viperVs_1$ and $\viperVs_2$ be Viper's pre and poststate of the encoded Voila statement $\ctext{s}$, respectively.
Let $\historyVars{} \in \stateHistory{\viperVs_0}$ and an atomicity context $\actxt$ between $\loweratomicitycontext{\viperVs_0}$ and $\upperatomicitycontext{\viperVs_0}$ be arbitrary.
The encoding of \vl!update_region! is given in \figref{encoding_update_region}.
The important steps of the proof snippet go as follows (from the bottom of the snippet to the top): 
Firstly, as for the atomic call, the judgment level is reduced. 
We denote the new level $\stateLevel{\viperVs_1}$ as $\jlevel$ to not clutter the proof tree with subscripts. 
Again, the encoding asserts explicitly that the new level ($\jlevel + 1$) is smaller or equal to the current level ($\stateLevel{\viperVs_0}$).
Secondly, as also seen before, $\Consequence$ is used to get the pre- and postcondition into the right shape such that $\textsc{UpdateRegion}$ can be applied next. 
All specified resources are justified since their encoding is explicitly asserted in the encoding.
In the updated region $\regionExA{\regionExALeftState}$, we use $\regionExALeftState$ to denote the region's interference context quantifier from the sequence of all interference context quantifiers $\many{\ictxtVar}$. 
Thirdly, $\textsc{UpdateRegion}$ is applied, where $\ractxtDomain$ and $\ractxtImage$ are the domain and image of the atomicity context entry for $r$, respectively.
Splitting the atomicity context is justified, because the encoding tests explicitly that an entry for $r$ exists in the atomicity context. 
Recall from the judgment mapping, that we define images of atomicity context entries such that they coincide with the target of the tracking resource $\trackres(\regionExALeftState, w)$. Therefore, $W$ and $\ractxtImage$ agree on whether or not an update happened.  
In the encoding, the region instance is opened by unfolding the region predicate, which matches the definition of our resource mapping.
Fourthly, as discussed in \secref{proof_candidate}, the interference contexts of regions contained in $\openRegionExA{\regionExALeftState}$, denoted as $\many{\ictxt'}$, is added to the current interference context $\stateInterferenceContext{\viperVs_0}$. Formally, we entail $\openRegionExAConstraint{\regionExALeftState}$, which denotes the assertion that is equivalent to $\openRegionExA{\regionExALeftState}$, except that the region state of regions is explicitly bound by $\many{\ictxtVar'}$. 
Lastly, as seen for \vl!atomic!, surplus old Viper state is removed by applying $\Exists$, so that the invariant can be applied.

The cases for \vl!open_region! and \vl!use_atomic! are similar.

\begin{figure}
\begin{adjustbox}{%
  addcode={%
    \begin{minipage}{\width}%
    \captionsetup{width=.8\textwidth}%
  }{%
    \caption{%
      Proof snippet for the encoding of \vl!update_region!, where $\regionExA{\regionExALeftState}$ is the updated region. The statement $\stmtB$ is the reduced TaDA statement ($\erased{\ctext{update\_region using ... \{\outlineStep\}}}$). The encoding of \vl!update_region! is given in \figref{encoding_update_region}. Note that we use shortened TaDA rule names.%
    }%
    \label{fig:soundness_proof_update}
    \end{minipage}%
  },rotate=90,center}
\scalebox{0.87}{
\Inf[\textsc{AWeak3}]{
  \Inf[\textsc{Cons}]{
    \Inf[\textsc{UpdReg}]{
      \Inf[\textsc{Cons}]{
        \Inf[\textsc{AExists}]{
          \Inf[\textsc{Cons}]{
            \Inf[\textsc{Exists}]{
              \Inf[\textsc{IH}] {\vdots}
                { \judgepre{}\ajudge{\jlevel}{\actxt'}{\tadaforall ~(\many{\ictxtVar},\many{\ictxtVar'}) \in \stateInterferenceContext{\viperVs_1}.~}{ {\stateAPre{\viperVs_1}(\historyVars{}, \historyVars{''},\ictxtVars{})}}{\stmtC}{ {\stateAPost{\viperVs_1}{\viperVs_2}(\historyVars{}, \historyVars{''},\ictxtVars{})}}  }
            }{ \judgepre{}\ajudge{\jlevel}{\actxt'}{\tadaforall ~(\many{\ictxtVar},\many{\ictxtVar'}) \in \stateInterferenceContext{\viperVs_1}.~}{\exists \, \many{\historyVar{}''} \in \historySet{}''. ~ {\stateAPre{\viperVs_1}(\historyVars{}, \historyVars{''},\ictxtVars{})}}{\stmtC}{\exists \, \many{\historyVar{}''} \in \historySet{}''. ~ {\stateAPost{\viperVs_1}{\viperVs_2}(\historyVars{}, \historyVars{''},\ictxtVars{})}} }
          }{ \judgepre{}\ajudge{\jlevel}{\actxt'}{\tadaforall ~\many{\ictxtVar} \in \stateInterferenceContext{\viperVs_0}, ~\many{\ictxtVar'} \in \ictxt'.~}{{P'(\historyVars{},\ictxtVars{})} \sand  \openRegionExAConstraint{\regionExALeftState} \sand \trackres<D>}{\stmtC}{
           \exists w \in W. ~ 
           \begin{array}{c}
              (\regionExALeftState \neq w) ~ ? ~ \openRegionExA{w} \sand \trackres(\regionExALeftState, w) \sand {Q_1(\historyVars{},\ictxtVars{}, w)} \\
               : \openRegionExAConstraint{\regionExALeftState} \sand \trackres<D> \sand {Q_2(\historyVars{},\ictxtVars{})}
           \end{array}
           } 
          }
        }{ \judgepre{}\ajudge{\jlevel}{\actxt'}{\tadaforall ~\many{\ictxtVar} \in \stateInterferenceContext{\viperVs_0}.~}{{P'(\historyVars{},\ictxtVars{})} \sand \exists \, \many{\ictxtVar'} \in \ictxt'. ~ \openRegionExAConstraint{\regionExALeftState} \sand \trackres<D>}{\stmtC}{
           \exists w \in W. ~ 
           \begin{array}{c}
              (\regionExALeftState \neq w) ~ ? ~ \openRegionExA{w} \sand \trackres(\regionExALeftState, w) \sand {Q_1(\historyVars{},\ictxtVars{}, w)} \\
               : \exists \, \many{\ictxtVar'} \in \ictxt'. ~ \openRegionExAConstraint{\regionExALeftState} \sand \trackres<D> \sand {Q_2(\historyVars{},\ictxtVars{})}
           \end{array}
           } 
        }
      }{ \judgepre{}\ajudge{\jlevel}{\actxt'}{\tadaforall ~\many{\ictxtVar} \in \stateInterferenceContext{\viperVs_0}.~}{{P'(\historyVars{},\ictxtVars{})} \sand  \openRegionExA{\regionExALeftState} \sand \trackres<D>}{\stmtC}{
         \exists w \in W. ~ 
         \begin{array}{c}
            (\regionExALeftState \neq w) ~ ? ~ \openRegionExA{w} \sand \trackres(\regionExALeftState, w) \sand {Q_1(\historyVars{},\ictxtVars{}, w)} \\
             : \openRegionExA{\regionExALeftState} \sand \trackres<D> \sand {Q_2(\historyVars{},\ictxtVars{})}
         \end{array}
         } 
       }
    }{ \judgepre{}\ajudge{\jlevel + 1}{r: z \in \ractxtDomain \rightarrow \ractxtImage(z) , \actxt'}{\tadaforall ~\many{\ictxtVar} \in \stateInterferenceContext{\viperVs_0}.~}{{P'(\historyVars{},\ictxtVars{})} \sand \regionExA{\regionExALeftState} \sand \trackres<D>}{\stmtC}{
       \exists w \in W. ~ 
       \begin{array}{c}
          (\regionExALeftState \neq w) ~ ? ~ \regionExA{w} \sand \trackres(\regionExALeftState, w) \sand {Q_1(\historyVars{},\ictxtVars{}, w)} \\
           : \regionExA{\regionExALeftState} \sand \trackres<D> \sand {Q_2(\historyVars{},\ictxtVars{})}
       \end{array}
       } 
     }
  }{ \judgepre{}\ajudge{\jlevel + 1}{\actxt}{\tadaforall ~\many{\ictxtVar} \in \stateInterferenceContext{\viperVs_0}.~}{{\stateAPre{\viperVs_0}(\historyVars{},\ictxtVars{})}}{\stmtC}{{\stateAPost{\viperVs_0}{\viperVs_3}(\historyVars{},\ictxtVars{}})} }
}{ \judgepre{}\ajudge{\stateLevel{\viperVs_0}}{\actxt}{\tadaforall ~\many{\ictxtVar} \in \stateInterferenceContext{\viperVs_0}.~}{{\stateAPre{\viperVs_0}(\historyVars{},\ictxtVars{})}}{\stmtC}{{\stateAPost{\viperVs_0}{\viperVs_3}(\historyVars{},\ictxtVars{})}} }
}
\end{adjustbox}
\end{figure}

\paragraph{Make-Atomic.}

\newcommand{\stmtD} [0] {\text{\vl!s!}'}

The corresponding proof tree is shown in \figref{soundness_proof_make}, where as before $\stmtD$ is the reduced TaDA statement ($\stmtD = \erased{\ctext{make\_atomic using ... \{\outlineStep\}}}$). 
Again, let $\viperVs_0$ and $\viperVs_3$ be Viper's pre and poststate of the encoded Voila statement $\ctext{make\_atomic using ... \{\outlineStep\}}$, respectively.
Furthermore, let $\viperVs_1$ and $\viperVs_2$ be Viper's pre and poststate of the encoded Voila statement $\ctext{s}$, respectively.
Let $\historyVars{} \in \stateHistory{\viperVs_0}$ and an atomicity context $\actxt$ between $\loweratomicitycontext{\viperVs_0}$ and $\upperatomicitycontext{\viperVs_0}$ be arbitrary.
The encoding of \vl!make_atomic! is given in \figref{encoding_make_atomic}.
The important steps of the proof snippet go as follows (from the bottom of the proof tree to the top): 

Firstly, similar to calls, the verification state is split into resources required for the \vl!make_atomic! and the frame $R(\historyVars{},\ictxtVars{})$, where again $R'(\historyVars{},\ictxtVars{})$ is the stabilized version that is framed off to the postcondition.
Afterwards, $\textsc{MakeAtomic}$ is applied.
The new atomicity context for the updated region is $z \in \stateInterferenceContext{\viperVs_0}^{\circ} \rightarrow \ractxtImage(z)$, where $\stateInterferenceContext{\viperVs_0}^{\circ}$ is the projection of $\stateInterferenceContext{\viperVs_0}$ onto the interference context for region identifier $r$. 
The image $\ractxtImage$ of the atomicity context entry for $r$ is chosen according to our judgment mapping.
Similar to the case for \vl!update_region!, $\ractxtImage$ coincides with the target of the tracking resource, thus coincides with $W$.
We can guarantee that $r$ was not in the atomicity context before, since the encoding explicitly checks that $r$ is not in the set of \ctext{update} and that the region's level is smaller than the value of the \ctext{alevel} variable. 
Again, old facts are removed by applying $\Exists$.
However, before we can use the induction hypothesis, 
we have to guarantee that the new atomicity context ($r: z \in \stateInterferenceContext{\viperVs_0}^{\circ} \rightarrow \ractxtImage(z), \actxt$) is between $\loweratomicitycontext{\viperVs_1}$ and $\upperatomicitycontext{\viperVs_1}$. The lower bound follows straight forwardly from $r$ getting added to the atomicity context and $\actxt$ being between $\loweratomicitycontext{\viperVs_0}$ and $\upperatomicitycontext{\viperVs_0}$. The upper bound follows from the value of variable $\ctext{alevel}$ in $\viperVs_0$ being larger then in $\viperVs_1$, which in the encoding is enforced by first asserting that the new value of \ctext{alevel} is lower than the current one and then assigning the new value to the \ctext{alevel} variable.

\begin{figure}
\begin{adjustbox}{%
  addcode={%
    \begin{minipage}{\width}%
    \captionsetup{width=.8\textwidth}%
  }{%
    \caption{%
      Proof snippet for the encoding of \vl!make_atomic!, where $\regionExA{\regionExALeftState}$ is the updated region and $\guard[G]$ the for the updated used guard. The statement $\stmtB$ is equal to $\erased{\ctext{make\_atomic using ... \{\outlineStep\}}}$. The encoding of \vl!make_atomic! is given in \figref{encoding_make_atomic}. Note that we use shortened TaDA rule names.%
    }%
    \label{fig:soundness_proof_make}
    \end{minipage}%
  },rotate=90,center}
\Inf[\textsc{Cons}]{
  \Inf[\textsc{Cons}]{
  \Inf[\textsc{Frame}]{
    \Inf[\textsc{MakeAtomic}]{
      \Inf[\textsc{Cons}]{
        \Inf[\textsc{Exists}]{
         \Inf[\textsc{IH}] { \vdots }
          { \judgepre{}\najudge{\stateLevel{\viperVs_1}}{r: z \in \stateInterferenceContext{\viperVs_0}^{\circ} \rightarrow \ractxtImage(z), \actxt}{ {\stateP{\viperVs_1}(\historyVars{}, \historyVars{''},\ictxtVars{})}}{\stmtC}{ {\stateP{\viperVs_2}(\historyVars{}, \historyVars{''},\ictxtVars{})}} }
        }{ \judgepre{}\najudge{\stateLevel{\viperVs_1}}{r: z \in \stateInterferenceContext{\viperVs_0}^{\circ} \rightarrow \ractxtImage(z), \actxt}{\exists \, \many{\historyVar{}''} \in \historySet{}''. ~ {\stateP{\viperVs_1}(\historyVars{}, \historyVars{''},\ictxtVars{})}}{\stmtC}{\exists \, \many{\historyVar{}''} \in \historySet{}''. ~ {\stateP{\viperVs_1}(\historyVars{}, \historyVars{''},\ictxtVars{})}} }
      }{ \judgepre{}\najudge{\stateLevel{\viperVs_0}}{r: z \in \stateInterferenceContext{\viperVs_0}^{\circ} \rightarrow \ractxtImage(z), \actxt}{  \exists \regionExALeftState \in \stateInterferenceContext{\viperVs_0}^{\circ}.~ \regionExA{\regionExALeftState} \sand \trackres<D>}{\stmtD}{\exists \regionExALeftState \in \stateInterferenceContext{\viperVs_0}^{\circ}, ~ w \in W. ~ \trackres(\regionExALeftState, w) } }
    }{ \judgepre{}\ajudge{\stateLevel{\viperVs_0}}{\actxt}{\tadaforall ~\many{\ictxtVar} \in \stateInterferenceContext{\viperVs_0}.~}{ \regionExA{\regionExALeftState} \sand \guard[G]}{\stmtD}{\exists w \in W. ~ \regionExA{w} \sand \guard[G] } }
  }{ \judgepre{}\ajudge{\stateLevel{\viperVs_0}}{\actxt}{\tadaforall ~\many{\ictxtVar} \in \stateInterferenceContext{\viperVs_0}.~}{{R'(\historyVars{},\ictxtVars{})} \sand \regionExA{\regionExALeftState} \sand \guard[G]}{\stmtD}{{R'(\historyVars{},\ictxtVars{})} \sand \exists w \in W. ~ \regionExA{w} \sand \guard[G] } }
  }{
    \judgepre{}\ajudge{\stateLevel{\viperVs_0}}{\actxt}{\tadaforall ~\many{\ictxtVar} \in \stateInterferenceContext{\viperVs_0}.~}{{R(\historyVars{},\ictxtVars{})} \sand \regionExA{\regionExALeftState} \sand \guard[G]}{\stmtD}{{R'(\historyVars{},\ictxtVars{})} \sand \exists w \in W. ~ \regionExA{w} \sand \guard[G] }
  }
}{ \judgepre{}\ajudge{\stateLevel{\viperVs_0}}{\actxt}{\tadaforall ~\many{\ictxtVar} \in \stateInterferenceContext{\viperVs_0}.~}{{\stateAPre{\viperVs_0}(\historyVars{},\ictxtVars{})}}{\stmtD}{{\stateAPost{\viperVs_0}{\viperVs_3}(\historyVars{},\ictxtVars{})}} }
\end{adjustbox}
\end{figure}

\MaybeFloatBarrier
\section{Complete Encoding of our Running Example}
\label{sec:complete_encoding}

An overview and excerpt of the encoding of our \vl!lock! running example was shown in \figref{lock_enc_overview_ext_disc}; below we show the full encoding atomicity contexts, interference contexts and levels. The encoding uses the macros defined in \appref{macro_definitions_general} (see also their example-specific definitions in \appref{macro_definitions}).




\begin{figure}
\begin{voila}[language=silver]
method lock(r: Ref, lvl: Int, cell: Ref) {
  // Encoded precondition
  inhale forall c: Ref :: c != null ==> acc(c.Lock_X) 
  inhale r.Lock_X == Set(0,1)
  inhale Lock(r, lvl, x) && Lock_state(r, lvl, x) in r.Lock_X
  inhale Lock_G(r)
  
  // Initialize levels
  var level: Int 
  inhale level > lvl
  var alevel: Int := level
  var update: Set[Ref] := Set()

  var b: Bool

  MAKE_ATOMIC(Lock(r, lvl, cell), Lock_G(r), {
    DO_WHILE({
      ATOMIC({
        UPDATE_REGION(Lock(r, lvl, cell), {
          CALL(b := CAS_val(x, 0, 1))
        })
      })
    }, !b, INV)
  })  

  // Encoded postcondition
  exhale Lock(r, lvl, x) && Lock_state(r, lvl, x) == 1 
  exhale Lock_G(r)
  exhale old(Lock_state(r, lvl, x)) == 0
}

@\codenote{where \ctext{INV} is the encoded source invariant:}@
  Lock(r, lvl, cell) &&
  (!b ==> acc(r.diamond)) &&
  ( b ==> acc(r.Lock_from) && acc(r.Lock_to) &&
          r.Lock_from == 0 && r.Lock_to == 1)
\end{voila}
\caption{
  Viper encoding of procedure \vl!lock! from our running example, with macros
  not yet expanded.
}
\label{fig:full_lock_encoding_macros}
\end{figure}

\end{document}